\documentclass[twocolumn]{autart}
\usepackage[usenames,dvipsnames,svgnames,table]{xcolor}
\usepackage{amssymb,latexsym,amsfonts,amsmath,bbm}
\usepackage{mathrsfs}
\usepackage{dsfont}
\usepackage{mathtools}
\usepackage{graphicx}
\usepackage{eso-pic}
\usepackage{epsfig}
\usepackage{mathtools}
\usepackage{tikz}
\usepackage{circuitikz}
\usetikzlibrary{positioning,calc}
\usetikzlibrary{decorations.markings,shapes,arrows,decorations.pathmorphing}
\usetikzlibrary{automata}
\usetikzlibrary{shapes}
\usetikzlibrary{calc,shapes,arrows}
\usetikzlibrary{arrows.meta}
\usepackage{mdwlist}
\usepackage{refcount}
\usepackage{comment}
\usepackage{ragged2e}
\usepackage{optidef}
\usepackage{xcolor}
\usepackage{graphicx}
\usepackage{subcaption}
\DeclareCaptionSubType*[arabic]{figure}

\DeclareCaptionSubType{figure} % revert setting of figureSeries.sty
\usepackage{natbib}

\usepackage{xcolor}
\definecolor{frenchblue}{rgb}{0.0, 0.45, 0.73}
\definecolor{dotcolor}{rgb}{1.0, 0.4, 0.7}
\usepackage{url}

\newcounter{corollary}

\newtheorem{theorem}{Theorem}

\newtheorem{openproblem}{Research Avenue}

\newtheorem{remark}{Remark}
\newtheorem{definition}{Definition}
\newtheorem{assumption}{Assumption}

\usepackage[many]{tcolorbox}
\usetikzlibrary{calc}
\tcbuselibrary{skins}

\usepackage{algorithm}
\usepackage{algorithmic}

\newtcolorbox{resp}[1][]{%
	enhanced jigsaw,%
	colback=gray!5!white,%
	colframe=gray!80!black,%
	size=small,%
	boxrule=1pt,%
	halign title=flush center,%
	coltitle=black,%
	breakable,%
	drop shadow=black!50!white,%
	attach boxed title to top left={xshift=1cm,yshift=-\tcboxedtitleheight/2,yshifttext=-\tcboxedtitleheight/2},%
	minipage boxed title=3cm,%
	boxed title style={%
		colback=white,%
		size=fbox,%
		boxrule=1pt,%
		boxsep=2pt,%
		underlay={%
			\coordinate (dotA) at ($(interior.west) + (-0.5pt,0)$);
			\coordinate (dotB) at ($(interior.east) + (0.5pt,0)$);
			\begin{scope}[gray!80!black]
				\fill (dotA) circle (2pt);
				\fill (dotB) circle (2pt);
			\end{scope}
		}%
	},%
	#1%
}
\makeatletter
\edef\endfrontmatter{%
	\unexpanded\expandafter{\endfrontmatter}% original code
	\noexpand\endNoHyper                  % add this to re-enable links
}
\makeatother
\DeclareRobustCommand{\barriericon}{%
	\tikz[baseline=-0.6ex]{
		\draw[blue, dashed, thick] (0,0) ellipse (0.18cm and 0.1593cm);
	}%
}

\DeclareRobustCommand{\initialicon}{%
	\tikz[baseline=-0.6ex]{
		\draw[fill=green!30, draw=green!50!black, thick]
		plot [smooth cycle, tension=0.8] coordinates {
			(1.8/6,  0.4/6)
			(2.6/6,  1.2/6)
			(3.2/6,  1.0/6)
			(3.4/6,  0.1/6)
			(2.9/6, -0.8/6)
			(2.4/6, -0.4/6)
		};
	}%
}

\DeclareRobustCommand{\unsafeicon}{%
	\tikz[baseline=-0.6ex]{
		\draw[fill=red!30, draw=red!70!black, thick]
		plot [smooth cycle, tension=0.8] coordinates {
			(-2.6/7,  0.5/7)
			(-2.1/7,  1.4/7)
			(-1.3/7,  1.0/7)
			(-1.4/7, -0.1/7)
			(-1.9/7, -1.1/7)
			(-3.0/7, -0.5/7)
		};
	}%
}

\DeclareFontFamily{U}{stix2bb}{}
\DeclareFontShape{U}{stix2bb}{m}{n} {<-> stix2-mathbb}{}

\NewDocumentCommand{\stixbbdigit}{m}{%
	\text{\usefont{U}{stix2bb}{m}{n}#1}%
}
\newcommand{\Zero}{\stixbbdigit{0}}

\newcommand{\R}{{\mathbb{R}}}
\newcommand{\Rpz}{{\mathbb{R}_{\geq 0}}}
\newcommand{\Rp}{{\mathbb{R}_{> 0}}}
\newcommand{\N}{{\mathbb{N}}}
\newcommand{\Np}{{\mathbb{N}_{\geq 1}}}
\newcommand{\I}{{\mathbb{I}}}

\DeclareRobustCommand{\circnum}[1]{%
	\tikz[baseline=(char.base)]{
		\node[
		circle,
		draw=blue,
		fill=gray!10,
		line width=.7pt,
		inner sep=0pt,
		minimum size=1em,
		text height=1.5ex,
		text depth=.25ex,
		font=\fontsize{7.5}{7.5}\selectfont
		] (char) {#1};
	}%
}
                       
\begin{document}
	
	\begin{frontmatter}
		
		\title{Data-Driven Formal Methods for Complex Dynamical Systems: A Survey\thanksref{footnoteinfo}} 
		
		\thanks[footnoteinfo]{Corresponding author: Behrad Samari}
		
		\author[Newcastle]{Behrad Samari}\ead{b.samari2@newcastle.ac.uk},    
		\author[Oxford]{Alessandro Abate}\ead{alessandro.abate@cs.ox.ac.uk},               
		\author[CNRS]{Antoine Girard}\ead{antoine.girard@centralesupelec.fr}, 
		\author[Boulder]{Majid Zamani}\ead{majid.zamani@colorado.edu}, 
		\author[Newcastle]{Amy Nejati}\ead{amy.nejati@newcastle.ac.uk},
		\author[Newcastle]{Abolfazl Lavaei}\ead{abolfazl.lavaei@newcastle.ac.uk}
		
		\address[Newcastle]{School of Computing, Newcastle University, United Kingdom}  
		\address[Oxford]{Department of Computer Science, University of Oxford, United Kingdom}        
		\address[CNRS]{Universit\'e Paris-Saclay, CNRS, CentraleSup\'elec, Laboratoire des signaux et syst\`emes, 91190, Gif-sur-Yvette, France}       
		\address[Boulder]{Department of Computer Science, University of Colorado Boulder, United States}

		
		\begin{abstract}                          
			Data-driven approaches with formal guarantees have recently emerged as a powerful means for the verification and controller synthesis of complex dynamical systems. Interest in these methods is rapidly growing, as system models are often unavailable in practice, and challenges such as nonlinear behavior, uncertainty, and the curse of dimensionality typically render accurate modeling infeasible. These difficulties motivate leveraging \emph{limited data} collected from the system while still providing \emph{formal guarantees} on its overall behavior. The community has therefore proposed a few hundred articles on the development of data-driven frameworks that enable the formal verification and synthesis of dynamical systems without explicit models, addressing complex specifications beyond stability. Despite this rapid growth, existing results remain scattered and lack a coherent organization, limiting a clear understanding of their principles, distinctions, and practical potential.
			This survey fills this gap by providing a comprehensive overview of these data-driven methods for both deterministic and stochastic dynamical systems. We structure the literature around three main methodological pillars in formal methods: (in)finite-abstraction-based techniques, functional certificate approaches, such as control barrier certificates, and compositional methods. For each of these approaches, we classify the resulting data-driven guarantees into three main categories: \emph{(i)} statistical guarantees grounded in probably approximately correct and scenario-based frameworks, \emph{(ii)} guarantees derived from Lipschitz continuity, and \emph{(iii)} guarantees exploiting structural properties, \emph{e.g.}, data-parameterized system representations or monotonicity. While the literature on deterministic systems is considerably richer, we also devote particular attention to the stochastic counterpart, highlighting the inherent differences and challenges that arise compared to the deterministic case. Throughout the survey, we aim to facilitate the entry of younger researchers into this thriving and diverse field by providing a clear overview of key challenges and solutions while highlighting current limitations and research avenues to guide future developments and community growth.
		\end{abstract}
		
	\end{frontmatter}
	
	\section{Introduction}\label{Sec: Introduction}
	
	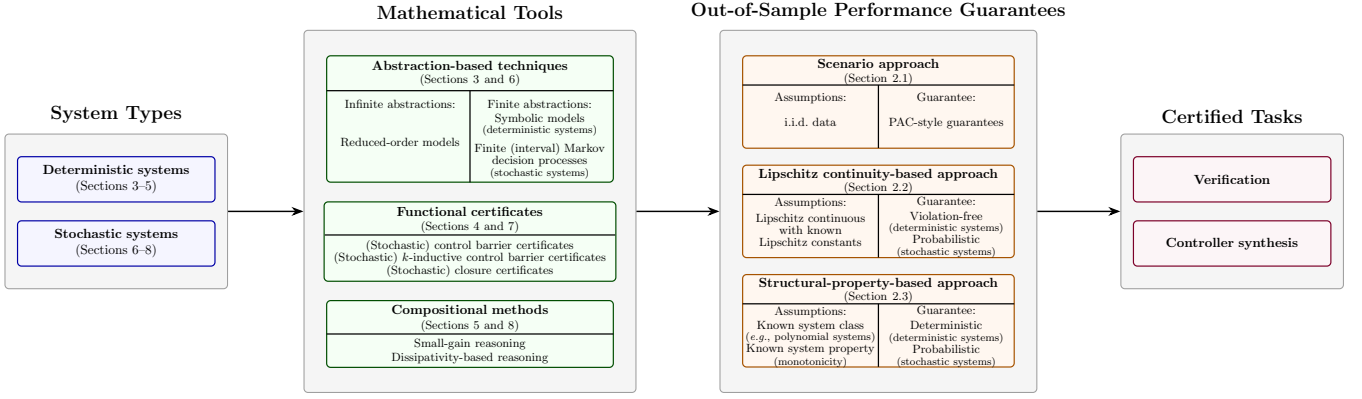
\begin{figure*}[t]
		\centering
		\resizebox{\textwidth}{!}{%
			\begin{tikzpicture}[
				font=\rmfamily,
				outerbox/.style={
					draw=gray!75,
					fill=gray!8,
					line width=0.8pt,
					rounded corners=3pt
				},
				innerboxsystem/.style={
					draw=blue!65!black,
					fill=blue!4,
					line width=0.9pt,
					rounded corners=3pt,
					align=center,
					text width=5.0cm,
					minimum height=1.2cm
				},
				innerboxtools/.style={
					draw=green!30!black,
					fill=green!4,
					line width=0.9pt,
					rounded corners=3pt,
					align=center,
					text width=7.2cm,
					minimum height=1.2cm
				},
				innerboxguarantee/.style={
					draw=orange!65!black,
					fill=orange!6,
					line width=0.9pt,
					rounded corners=3pt,
					align=center,
					text width=6.6cm,
					minimum height=1.4cm
				},
				innerboxtask/.style={
					draw=purple!65!black,
					fill=purple!4,
					line width=0.9pt,
					rounded corners=3pt,
					align=center,
					text width=5cm,
					minimum height=1.35cm
				},
				connectarrow/.style={
					-{Stealth[length=3.2mm,width=2.4mm]},
					line width=1.1pt,
					draw=black
				}
				]
				
				\node[
				outerbox,
				minimum width=5.9cm,
				minimum height=4.1cm
				] (systemblock) at (0,0) {};
				
				\node[
				font=\rmfamily\bfseries\large,
				text=black,
				above=0.15cm of systemblock
				] {System Types};
				
				\node[
				innerboxsystem
				] (det) at ([yshift=0.85cm]systemblock.center) {
					\textbf{Deterministic systems}\\[1pt]
					{(\normalfont Sections~\ref{Sec: Deterministic Setting_ABA}--\ref{Sec: Deterministic Setting_CT})}
				};
				
				\node[
				innerboxsystem
				] (sto) at ([yshift=-0.85cm]systemblock.center) {
					\textbf{Stochastic systems}\\[1pt]
					{(\normalfont Sections~\ref{Sec: Stochastic Setting_ABA}--\ref{Sec: Stochastic Setting_CT})}
				};
				
				\node[
				outerbox,
				minimum width=8.8cm,
				minimum height=9.6cm,
				right=2.0cm of systemblock
				] (toolblock) {};
				
				\node[
				font=\rmfamily\bfseries\large,
				text=black,
				above=0.15cm of toolblock
				] {Mathematical Tools};
				
				\node[
				innerboxtools,
				minimum width=7.6cm,
				minimum height=3.35cm
				] (abs) at ([yshift=2.45cm]toolblock.center) {};
				
				\node[
				font=\rmfamily\bfseries,
				text=black,
				align=center,
				text width=7.0cm
				] at ([yshift=1.2cm]abs.center) {
					Abstraction-based techniques\\[-1pt]
					{\normalfont\footnotesize
						(Sections~\ref{Sec: Deterministic Setting_ABA} and~\ref{Sec: Stochastic Setting_ABA})}
				};
				
				\draw[black, line width=0.95pt]
				([xshift=-3.8cm,yshift=0.75cm]abs.center) --
				([xshift= 3.8cm,yshift=0.75cm]abs.center);
				
				\draw[black, line width=0.95pt]
				([xshift=0cm,yshift=0.75cm]abs.center) --
				([xshift=0cm,yshift=-1.68cm]abs.center);
				
				\node[
				font=\rmfamily\small,
				align=center,
				text width=3.1cm
				] at ([xshift=-1.85cm,yshift=0.48cm]abs.center) {
					Infinite abstractions:
				};
				
				\node[
				font=\rmfamily\footnotesize,
				align=center,
				text width=3.5cm
				] at ([xshift=-1.85cm,yshift=-0.55cm]abs.center) {
					Reduced-order models
				};
				
				\node[
				font=\rmfamily\small,
				align=center,
				text width=3.5cm
				] at ([xshift=1.85cm,yshift=0.48cm]abs.center) {
					Finite abstractions:
				};
				
				\node[
				font=\rmfamily\footnotesize,
				align=center,
				text width=3.3cm
				] at ([xshift=1.85cm,yshift=-0.085cm]abs.center) {
					Symbolic models\\[-1pt]
					{\scriptsize (deterministic systems)}
				};
				
				\node[
				font=\rmfamily\footnotesize,
				align=center,
				text width=3.5cm
				] at ([xshift=1.85cm,yshift=-1.05cm]abs.center) {
					Finite (interval) Markov\\[-1pt]
					decision processes\\[-1pt]
					{\scriptsize (stochastic systems)}
				};
				
				\node[
				innerboxtools,
				minimum width=7.6cm,
				minimum height=1.9cm,
				text width=7.5cm
				] (cert) at ([yshift=-0.8cm]toolblock.center) {
					\textbf{Functional certificates} {\footnotesize (Sections~\ref{Sec: Deterministic Setting_FCA} and~\ref{Sec: Stochastic Setting_FCA})} \\[4pt]
					{\small (Stochastic) control barrier certificates\\[-1pt]
						(Stochastic) \(k\)-inductive control barrier certificates\\[-1pt]
						(Stochastic) closure certificates}
				};
				
				\draw[black, line width=0.95pt]
				([xshift=-3.85cm,yshift=-3.075cm]abs.center) --
				([xshift= 3.85cm,yshift=-3.075cm]abs.center);
				
				\node[
				innerboxtools,
				minimum width=7.6cm,
				minimum height=1.7cm
				] (comp) at ([yshift=-3.25cm]toolblock.center) {
					\textbf{Compositional methods} {\footnotesize (Sections~\ref{Sec: Deterministic Setting_CT} and~\ref{Sec: Stochastic Setting_CT})}\\[4pt]
					{\small Small-gain reasoning\\[-1pt]
						Dissipativity-based reasoning}
				};
				
				\draw[black, line width=0.95pt]
				([xshift=-3.8cm,yshift=-5.7cm]abs.center) --
				([xshift= 3.8cm,yshift=-5.7cm]abs.center);
				
				\node[
				outerbox,
				minimum width=8.4cm,
				minimum height=9.6cm,
				right=2.2cm of toolblock
				] (guarblock) {};
				
				\node[
				font=\rmfamily\bfseries\large,
				text=black,
				above=0.15cm of guarblock
				] {Out-of-Sample Performance Guarantees};
				
				\node[
				innerboxguarantee,
				minimum width=7.2cm,
				minimum height=2.45cm
				] (pac) at ([yshift=2.9cm]guarblock.center) {};
				
				\node[
				font=\rmfamily\bfseries,
				text=black,
				align=center,
				text width=7.0cm
				] at ([yshift=0.8cm]pac.center) {
					Scenario approach\\[-1pt]
					{\normalfont\footnotesize
						(Section~\ref{Subsec: Scenario Approach})}
				};
				
				\draw[black, line width=0.95pt]
				([xshift=-3.60cm,yshift=0.45cm]pac.center) --
				([xshift= 3.60cm,yshift=0.45cm]pac.center);
				
				\draw[black, line width=0.95pt]
				([xshift=0cm,yshift=0.45cm]pac.center) --
				([xshift=0cm,yshift=-1.22cm]pac.center);
				
				\node[
				font=\rmfamily\small,
				align=center,
				text width=2.9cm
				] at ([xshift=-1.8cm,yshift=0.18cm]pac.center) {
					Assumptions:
				};
				
				\node[
				font=\rmfamily\footnotesize,
				align=center,
				text width=2.9cm
				] at ([xshift=-1.8cm,yshift=-0.45cm]pac.center) {
					i.i.d.\ data
				};
				
				\node[
				font=\rmfamily\small,
				align=center,
				text width=2.9cm
				] at ([xshift=1.8cm,yshift=0.2cm]pac.center) {
					Guarantee:
				};
				
				\node[
				font=\rmfamily\footnotesize,
				align=center,
				text width=3.5cm
				] at ([xshift=1.8cm,yshift=-0.5cm]pac.center) {
					PAC-style guarantees
				};
				
				\node[
				innerboxguarantee,
				minimum width=7.2cm,
				minimum height=2.45cm
				] (lip) at (guarblock.center) {};
				
				\node[
				font=\rmfamily\bfseries,
				text=black,
				align=center,
				text width=7.0cm
				] at ([yshift=0.8cm]lip.center) {
					Lipschitz continuity-based approach\\[-1pt]
					{\normalfont\footnotesize
						(Section~\ref{Subsec: Lipschitz Approach})}
				};
				
				\draw[black, line width=0.95pt]
				([xshift=-3.60cm,yshift=0.45cm]lip.center) --
				([xshift= 3.60cm,yshift=0.45cm]lip.center);
				
				\draw[black, line width=0.95pt]
				([xshift=0cm,yshift=0.45cm]lip.center) --
				([xshift=0cm,yshift=-1.22cm]lip.center);
				
				\node[
				font=\rmfamily\small,
				align=center,
				text width=2.9cm
				] at ([xshift=-1.8cm,yshift=0.3cm]lip.center) {
					Assumptions:
				};
				
				\node[
				font=\rmfamily\footnotesize,
				align=center,
				text width=3.9cm
				] at ([xshift=-1.8cm,yshift=-0.45cm]lip.center) {
					Lipschitz continuous\\[-1pt] with known\\[-1pt]
					Lipschitz constants
				};
				
				\node[
				font=\rmfamily\small,
				align=center,
				text width=2.9cm
				] at ([xshift=1.8cm,yshift=0.35cm]lip.center) {
					Guarantee:
				};
				
				\node[
				font=\rmfamily\footnotesize,
				align=center,
				text width=3.9cm
				] at ([xshift=1.8cm,yshift=-0.55cm]lip.center) {
					Violation-free\\[-1pt]
					{\scriptsize (deterministic systems)}\\[-1pt]
					Probabilistic\\[-1pt]
					{\scriptsize (stochastic systems)}
				};
				
				\node[
				innerboxguarantee,
				minimum width=7.2cm,
				minimum height=2.45cm
				] (struc) at ([yshift=-2.9cm]guarblock.center) {};
				
				\node[
				font=\rmfamily\bfseries,
				text=black,
				align=center,
				text width=7.0cm
				] at ([yshift=0.8cm]struc.center) {
					Structural-property-based approach\\[-1pt]
					{\normalfont\footnotesize
						(Section~\ref{Subsec: Structural Approach})}
				};
				
				\draw[black, line width=0.95pt]
				([xshift=-3.60cm,yshift=0.45cm]struc.center) --
				([xshift= 3.60cm,yshift=0.45cm]struc.center);
				
				\draw[black, line width=0.95pt]
				([xshift=0cm,yshift=0.45cm]struc.center) --
				([xshift=0cm,yshift=-1.22cm]struc.center);
				
				\node[
				font=\rmfamily\small,
				align=center,
				text width=2.9cm
				] at ([xshift=-1.8cm,yshift=0.3cm]struc.center) {
					Assumptions:
				};
				
				\node[
				font=\rmfamily\footnotesize,
				align=center,
				text width=3.8cm
				] at ([xshift=-1.8cm,yshift=-0.55cm]struc.center) {
					Known system class\\[-1pt]
					{\scriptsize (\emph{e.g.}, polynomial systems)}\\[-1pt]
					Known system property\\[-.4pt]
					{\scriptsize (monotonicity)}
				};
				
				\node[
				font=\rmfamily\small,
				align=center,
				text width=2.9cm
				] at ([xshift=1.8cm,yshift=0.35cm]struc.center) {
					Guarantee:
				};
				
				\node[
				font=\rmfamily\footnotesize,
				align=center,
				text width=3.5cm
				] at ([xshift=1.8cm,yshift=-0.55cm]struc.center) {
					Deterministic\\[-1pt]
					{\scriptsize (deterministic systems)}\\[-1pt]
					Probabilistic\\[-1pt]
					{\scriptsize (stochastic systems)}
				};
				
				\node[
				outerbox,
				minimum width=5.9cm,
				minimum height=4.1cm,
				right=2.2cm of guarblock
				] (taskblock) {};
				
				\node[
				font=\rmfamily\bfseries\large,
				text=black,
				above=0.15cm of taskblock
				] {Certified Tasks};
				
				\node[
				innerboxtask,
				minimum width=5.0cm,
				minimum height=1.2cm
				] (verify) at ([yshift=0.85cm]taskblock.center) {
					\textbf{Verification}
				};
				
				\node[
				innerboxtask,
				minimum width=5.0cm,
				minimum height=1.2cm
				] (synthesis) at ([yshift=-0.85cm]taskblock.center) {
					\textbf{Controller synthesis}
				};
				
				\draw[connectarrow] (systemblock.east) -- (toolblock.west);
				\draw[connectarrow] (toolblock.east) -- (guarblock.west);
				\draw[connectarrow] (guarblock.east) -- (taskblock.west);
				
			\end{tikzpicture}%
		}
		\caption{Roadmap of the survey, providing a summary of the considered system types, mathematical tools studied, and primary out-of-sample performance guarantees, collectively leading to formal verification and policy synthesis from data.}
		\label{fig:roadmap}
	\end{figure*}
	
	Formal verification and controller synthesis for dynamical systems have become a prominent research direction over the past two decades, driven by their essential role in ensuring the correct and dependable operation of safety-critical systems. These systems arise across a wide range of domains, including aerospace, automotive and transportation, robotics, chemical and industrial processes, critical infrastructure, energy networks, and healthcare technologies~\citep{dutertre2002formal,sun2005challenges,coogan2017formal,kress2018synthesis,schwarting2018planning,akram2018formal}. Such systems operate in environments where malfunction or unanticipated behavior can lead to catastrophic consequences, including loss of life, severe injuries, environmental damage, or substantial economic disruption~\citep{baier2008principles,leveson2016engineering,mcgregor2017analysis}. These risks underscore the growing need for rigorous, mathematically grounded methodologies to ensure both safety and performance in modern autonomous and cyber-physical systems. Moreover, modern autonomous systems increasingly require formal guarantees with respect to high-level temporal specifications that go beyond classical notions of stability, \emph{e.g.}, those expressed as (linear) temporal logic formulae~\citep{pnueli1977temporal,baier2008principles}. A common example is the reach-while-avoid task~\citep{summers2010verification, fan2018controller}, which captures scenarios in which an autonomous system, such as a vehicle, should reach a target region while avoiding obstacles along the way.\vspace{-0.1cm}
   
   While both verification and controller synthesis provide provable guarantees for such specifications, they differ in scope and usage~\citep{belta2017formal}. Specifically, given a deterministic dynamical system and a property of interest, formal verification aims to rigorously determine whether or not the system satisfies the desired specification. For stochastic dynamical systems, this task naturally extends to computing or tightly characterizing the probability that the system satisfies the specification. In contrast, a synthesis problem concerns dynamical systems with control inputs, where the goal is to formally design a controller (also referred to as a policy or strategy), typically in the form of a state-feedback law, that enforces the desired property. This paradigm is also referred to as \emph{correct-by-construction} control design, since each step of the synthesis procedure is accompanied by a formal guarantee. In stochastic settings, this task translates into synthesizing a controller that \emph{maximizes} the probability of satisfying the given specification. Due to their intrinsic soundness, formal-methods-based approaches eliminate the need for costly, exhaustive, and potentially unsuccessful post hoc validation often required in safety-critical real-world applications.\vspace{-0.1cm}
	
	Despite their advantages, formal verification and controller synthesis have typically been developed under the assumption that an accurate mathematical model of the underlying dynamical system is available. In practice, however, such models are often unavailable; when available, they are typically either simplified and therefore incomplete or too complex to be of any effective use. Consequently, classical model-based techniques are often unsuitable for the verification and control design of complex systems with unknown dynamics, making data-driven approaches essential for enabling formal reasoning in such settings. Within this context, this survey provides a structured overview of data-driven formal verification and controller synthesis, as illustrated in Fig.~\ref{fig:roadmap}.\vspace{-0.1cm}
	
	\subsection{Data-Driven Frameworks}\label{Subsec:I-VS-D}
\begin{figure}[t!]
		\centering
		
		\begin{subfigure}[t]{0.4\columnwidth}
			\centering
			\includegraphics[width=0.7\columnwidth]{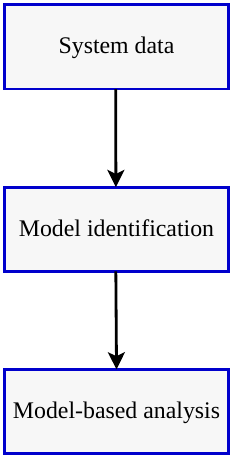}
			\caption{Indirect approach}
			\label{subfig:scheme_indirect}
		\end{subfigure}
		\hfill
		\begin{subfigure}[t]{0.45\columnwidth}
			\centering
			\includegraphics[width=0.87\linewidth]{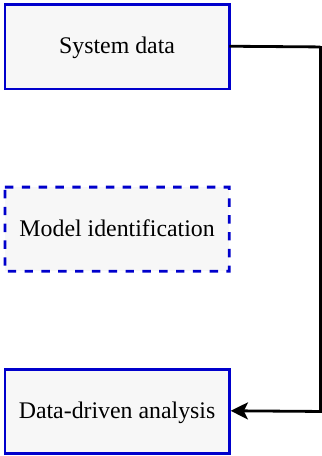}
			\caption{Direct approach}
			\label{subfig:scheme_direct}
		\end{subfigure}
		
		\caption{Conceptual comparison of indirect and direct data-driven approaches. (a) The indirect approach relies on model identification
		followed by model-based analysis, while (b) the direct approach operates on raw data without an explicit model-identification step.
        }
		\label{fig:scheme_direct_vs_indirect}
	\end{figure}

\vspace{-0.1cm}
	To address the aforementioned challenge, the literature has developed two complementary and well-established data-driven paradigms: indirect and direct approaches (cf. Fig.~\ref{fig:scheme_direct_vs_indirect}). In indirect methods, the primary objective is to perform model identification so as to construct approximate models of unknown dynamical systems~\citep{campi2002finite,hjalmarsson2005experiment,hou2013model,pillonetto2014kernel,chiuso2019system,dorfler2022bridging}. These approaches, including methods based on model-based reinforcement learning~\citep{recht2019tour}, neural networks~\citep{zhou2022neural}, and Gaussian process regression~\citep{lederer2019uniform}, aim to reconstruct system dynamics from data, thereby enabling the use of classical model-based approaches once a sufficiently accurate model is obtained.\vspace{-0.1cm}
	
	Nevertheless, the practical effectiveness of such methods ultimately depends on the accuracy of the underlying identification procedure. While an active line of research studies the discrepancy between an identified model and the true system dynamics~\citep{oymak2019non,lederer2019uniform}, this mismatch remains difficult to quantify. Moreover, although well-established methods exist for linear model identification, current approaches for nonlinear systems still face notable limitations~\citep{kerschen2006past}. This constrains the applicability of indirect approaches mainly to linear systems or to specific classes of nonlinear systems. In addition, indirect approaches inherently involve \emph{two-stage} workflows~\citep{haesaert2017data,sadraddini2018formalHSCC}, in which a model is first constructed and then employed for model-based techniques, whereas practitioners often prefer end-to-end methods that bypass this intermediate step and its associated complexity~\citep{11312267}.\vspace{-0.1cm}
	
	Direct data-driven approaches, in contrast, follow a fundamentally different paradigm. Rather than constructing an explicit model, they bypass the identification step and leverage recorded data directly for formal verification or controller synthesis of dynamical systems (cf. Fig.~\ref{subfig:scheme_direct}). This methodology, however, introduces its own challenges. In particular, the finite nature of available data prevents capturing all possible system behaviors, making it challenging to provide guarantees over both observed and unobserved evolution. Consequently, a central challenge lies in providing out-of-sample performance guarantees; that is, formal guarantees, typically in a worst-case or distributional sense, ensuring that conclusions drawn from finite data (\emph{i.e.}, limited observed scenarios) remain provably valid for unseen data (\emph{i.e.}, unobserved scenarios).\vspace{-0.1cm}
	
	This survey places primary emphasis on direct data-driven approaches. For the sake of completeness, we also review selected indirect data-driven approaches and their associated guarantees that address the formal verification and controller synthesis of complex dynamical systems. To provide a thorough discussion of the underlying guarantees, we should first introduce the foundational tools for formal verification and synthesis considered throughout this survey.\vspace{-0.1cm}
	
	\subsection{Abstraction-Based Techniques}\label{Subsec:A-BT}
	The first methodology we consider for formal verification and synthesis of complex dynamical systems is abstraction-based techniques, in which original systems (\emph{a.k.a.} concrete systems) are approximated by simpler ones of either lower dimensions (infinite abstractions) or finite state spaces (finite abstractions). The main advantage of these approaches is that they provide a rigorous quantification of the mismatch between the behavior of the concrete system and that of its abstraction. Thus, one can conduct the analysis and synthesis over simpler abstractions and subsequently transfer them back to concrete systems with formal correctness guarantees~\citep{reissig2016feedback,calbert2024classification}. As noted above, abstraction-based techniques broadly fall into two categories: \emph{(i)} infinite abstractions and \emph{(ii)} finite abstractions. In the sequel, we first discuss these approaches in deterministic settings, and then extend the discussion to stochastic frameworks.\vspace{-0.1cm}
	
	In the realm of infinite abstractions, also referred to as model order reduction in the control literature~\citep{antoulas2005approximation,astolfi2010model}, the aim is to alleviate scalability challenges associated with the formal analysis and design of dynamical systems, especially those with rather high dimensions. More precisely, the core idea in this context is to construct a reduced-order model (ROM) with fewer state variables than the concrete system, enabling analysis and policy synthesis on this simpler model, with the results then translated back to the concrete system while quantifying the closeness of the behavior of the two systems. It is important to note that while such methodologies help address scalability challenges, the resulting ROMs typically retain uncountable state and input sets, which can still lead to computational complexity.\vspace{-0.1cm}
	
	To address this issue, finite abstractions, which are also known as symbolic models~\citep{pola2019control}, provide approximations of either concrete systems or their associated ROMs, which are endowed with finite state sets. Such abstractions (for non-stochastic systems) are broadly divided into \emph{complete} and \emph{sound} abstractions~\citep{tabuada2009verification}. Specifically, complete abstractions provide \emph{necessary and sufficient} guarantees, ensuring that a controller satisfies a desired specification on the abstraction \emph{if and only if} it does so on the original system. In contrast, sound abstractions offer only \emph{sufficient} guarantees; thus, failure to synthesize a controller on the abstraction does not preclude its existence for the original system.\vspace{-0.1cm}
	
	\begin{figure*}
		\centering
		\includegraphics[width=0.85\linewidth]{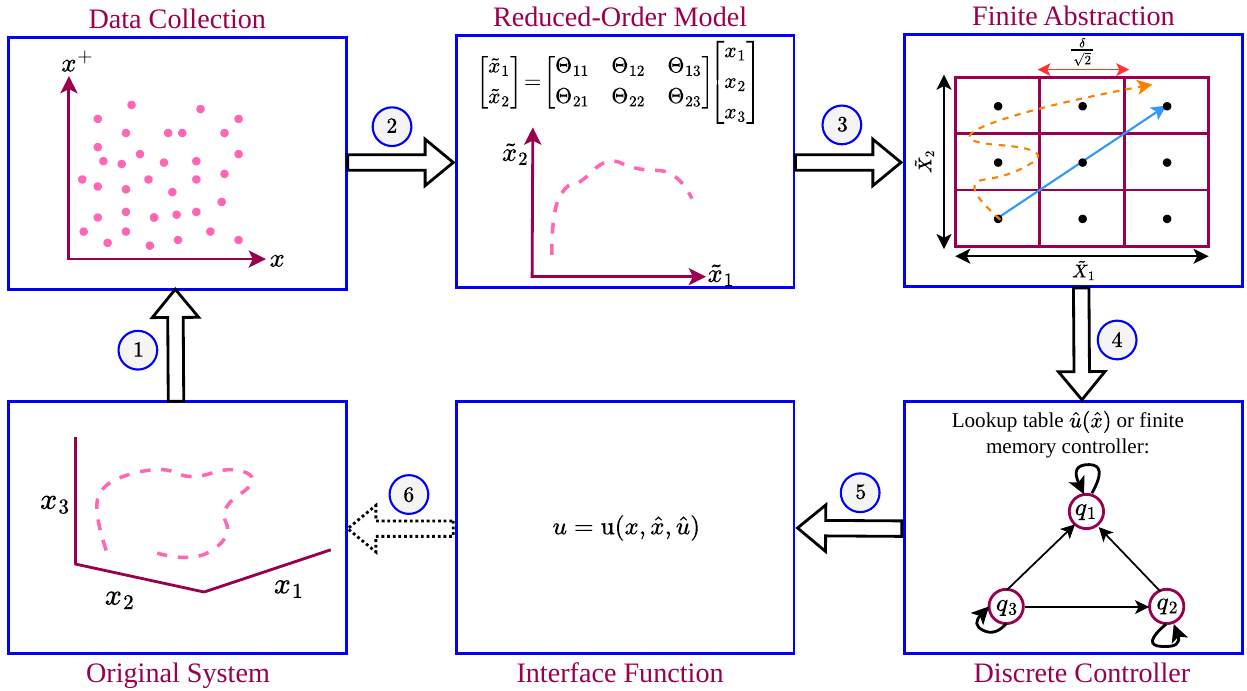}
		\caption{General schematic of data-driven abstraction-based approaches, where symbols associated with the ROM are denoted with a tilde (\emph{e.g.}, $\tilde{x}$), while those associated with the finite abstraction are indicated using a hat (\emph{e.g.}, $\hat{x}$). As depicted, the first step (\protect\circnum{1}) is to collect data from the system, which can typically be carried out in three main ways, each yielding a distinct type of formal guarantee (cf. Sections~\ref{Subsec: Scenario Approach}--\ref{Subsec: Structural Approach}). Then, a ROM for the original dynamical system is constructed (Step \protect\circnum{2}) if it is relatively high-dimensional; otherwise, this step can be skipped.
			Subsequently, for the underlying system, either the original system of low dimensions or the constructed ROM of a high-dimensional original system, a finite abstraction is constructed (Step \protect\circnum{3}), upon which a discrete controller can be synthesized (Step \protect\circnum{4}) using tools introduced in the formal methods community~\citep{calbert2024dionysos, mouelhi2013cosyma, mathiesen2024intervalmdp, mazo2010pessoa, roy2011pessoa,wooding2024impact, van2025syscore}, satisfying a desired specification of interest. This discrete controller can then be refined back to the underlying system through the so-called interface function (Step \protect\circnum{5}), ensuring that the underlying system also satisfies the same specification, albeit with a guaranteed error bound (Step \protect\circnum{6}).}
		\label{fig:data_abstraction-based}
	\end{figure*}
	
	Given a discretization parameter, a finite abstraction is generally constructed by partitioning the state (and input) set so that each discrete state (and input) corresponds to a set of continuous states of the underlying system (similarly for the input). Since the resulting abstractions are finite in this case, one can directly leverage a wide range of algorithmic tools from computer science~\citep{baier2008principles} to perform model checking or synthesize controllers that optimize performance criteria or satisfy complex specifications, including those expressed as temporal logic formulae~\citep{lavaei2022automated}. A central component of finite-abstraction methodologies is the establishment of formal guarantees on the state (or output) relation between the concrete system and its abstract model. Such guarantees ensure that verification outcomes or synthesized strategies obtained on the abstraction model can be transferred back to the original system.\vspace{-0.1cm}
	
	While the general idea of (in)finite abstractions remains the same in both deterministic and stochastic settings, intrinsic differences arise due primarily to the presence of stochasticity in the latter case. In particular, in stochastic settings, the mismatch between the behavior of the original system and that of its infinite abstraction can only be guaranteed with a certain probability, which one typically aims to maximize. This implies that the obtained ROM preserves the relevant behavioral properties of the concrete stochastic system only with a certain probability. The impact of stochasticity becomes even more pronounced when constructing finite abstractions of dynamical systems.
	In particular, in stochastic systems, transitions are inherently probabilistic, so that under the same state and input, the successor state may fall into multiple partitions due to the randomness in the dynamics.
	This is precisely why finite abstractions of stochastic dynamical systems are typically in the form of finite Markov decision processes (MDPs)~\citep{puterman2014markov}, which are generally described by transition probability matrices. Each entry of such a matrix specifies the probability of transitioning from one partition to another, conditioned on the discrete state and applied discrete input, while capturing the inherent randomness of the underlying system. Alternatively, when the transition probabilities between partitions lie within specified intervals, finite abstractions can be constructed in the form of interval Markov decision processes (IMDPs)~\citep{givan2000bounded}. We note that, if the underlying system has no control input, finite MDPs and IMDPs reduce, respectively, to finite Markov chains (MCs) and interval Markov chains (IMCs).\vspace{-0.1cm}
	
	Regardless of whether the interest is to construct infinite or finite abstractions, their construction ultimately hinges on precise knowledge of the underlying dynamics. Specifically, constructing an infinite abstraction of a concrete system requires knowledge of its model to \emph{(i)} derive its ROM and \emph{(ii)} formally relate the two systems by quantifying the closeness of their behaviors. Likewise, accurate model knowledge is also required for finite abstractions. However, as discussed above, in practice, accurate models are often unavailable or too complex to be useful, motivating the development of data-driven abstraction-based approaches for the analysis and policy synthesis of dynamical systems, which typically follow the hierarchy illustrated in Fig.~\ref{fig:data_abstraction-based}.\vspace{-0.1cm}
	
	\subsection{Functional Certificate Techniques}\label{Subsec: MB-FCA}\vspace{-0.1cm}
	While constructing ROMs of dynamical systems mitigates scalability challenges, building finite abstractions of either concrete systems or their associated ROMs still requires the discretization (\emph{i.e.}, gridding or partitioning) of the state and input sets. This leads to the well-known curse of dimensionality, whereby constructing a symbolic model for a dynamical system typically incurs computational complexity that grows \emph{exponentially} with the state and input dimensions. Motivated by this challenge, the literature has developed functional certificate approaches, such as (control) barrier certificates~\citep{prajna2004safety,prajna2006barrier,prajna2007framework,wieland2007constructive,ames2016control,ames2019control,luo2020multi,santoyo2021barrier,wang2017safety,xiao2023safe,clark2024semi,laurenti2025unifying,11373823,mazouz2022safetyNEURIPS}, \(k\)-inductive (control) barrier certificates~\citep{bak2018t,anand2021safetyCDC,11134332,zhi2024unifying,lewis2024verification}, and (control) closure certificates~\citep{murali2024closure}, which can be employed for formal verification and synthesis to enforce diverse complex specifications without requiring state-space discretization.  \vspace{-0.1cm}
	
	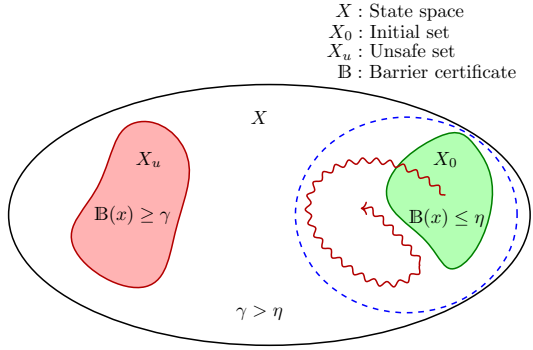
\begin{figure}[t!]
		\centering
		\resizebox{0.82\linewidth}{!}{\begin{tikzpicture}[scale=1.2, line join=round]
				
				\draw[thick] (0,0) ellipse (4cm and 2cm);
				\node at (-0.15,-1.5) {$\gamma > \eta$};
				\node at (-0.15,1.5) {$X$};

				\draw[fill=red!30, draw=red!70!black, thick]
				plot [smooth cycle, tension=0.8] coordinates {
					(-2.6,  0.5)
					(-2.1,  1.4)
					(-1.3,  1.0)
					(-1.4, -0.1)
					(-1.9, -1.1)
					(-3.0, -0.5)
				};
				\node at (-1.85, 0.85) {$X_u$};
				\node at (-2.1, 0.0) {$\mathds{B}(x) \geq \gamma$};
				
				\draw[dashed, thick, blue] (2.1,0) ellipse (1.7cm and 1.5cm);

				\draw[fill=green!30, draw=green!50!black, thick]
				plot [smooth cycle, tension=0.8] coordinates {
					(1.8,  0.4)
					(2.6,  1.2)
					(3.2,  1.0)
					(3.4,  0.1)
					(2.9, -0.8)
					(2.4, -0.4)
				};
				\node at (2.7, 0.85) {$X_0$};
				\node at (2.72, -0.05) {$\mathds{B}(x) \leq \eta$};

				\draw[->, thick, red!70!black,
				decoration={snake, amplitude=0.5mm, segment length=2.5mm},
				decorate]
				(2.7,0.3)
				.. controls (2.1, 0.9) and (1.3, 0.9) ..
				(0.9, 0.6)
				.. controls (0.5, 0.3) and (0.5,-0.3) ..
				(1.2,-0.8)
				.. controls (1.7,-1.2) and (2.1,-1.2) ..
				(2.4,-0.8)
				.. controls (2.0,-0.3) and (1.6, 0.1) ..
				(1.4, 0.1);
				
				\node at (2,3.1) {$X:$ State space};
				\node at (1.84,2.8) {$X_0:$ Initial set};
				\node at (1.87,2.5) {$X_u:$ Unsafe set};
				\node at (2.42,2.2) {$\mathds{B}:$ Barrier certificate};
		\end{tikzpicture}}
		\caption{Illustration of a barrier certificate: The blue dashed ellipse \barriericon\ represents the level set $\mathds{B}(x) = \eta$, which separates the initial set \initialicon\ from the unsafe set \unsafeicon\ $\!\!$.}
		\label{Figure: BC_Scheme}
	\end{figure}
	
	In particular, a barrier certificate is a real-valued function, analogous to a Lyapunov function, defined on the state space of a dynamical system and required to satisfy a set of inequalities involving both the function itself and the dynamics governing the system. If a suitable level of this function is chosen so that the corresponding sublevel set contains the prescribed initial conditions and is disjoint from the unsafe region (cf. Fig.~\ref{Figure: BC_Scheme}), the barrier-certificate inequalities ensure safety. In deterministic settings, trajectories starting from the initial set avoid the unsafe set, while in stochastic settings, the probability of reaching the unsafe set can be suitably bounded. While barrier certificates enable system safety verification, \emph{control} barrier certificates extend this framework to controller-synthesis settings, where the objective is to design a controller that ensures the system state remains within the designated safe region, either deterministically or with a (maximized) quantified probability. We highlight that (control) barrier certificates typically provide \emph{sufficient} results for nonlinear systems; consequently, if a valid certificate cannot be found, no conclusion can be drawn regarding system safety. \vspace{-0.1cm}
	
	It is important to note that the value of a (control) barrier certificate is typically required to be non-increasing along system trajectories, either in the deterministic setting (cf. Definition~\ref{def:cbc} with $c = 0$) or in expectation in the stochastic setting. This requirement may restrict the applicability of (control) barrier certificates in certain scenarios. To mitigate this issue, the literature, building on the notion of $k$-induction for safety verification of finite-state systems and software programs~\citep{sheeran2000checking,de2003bounded,donaldson2011software}, has introduced $k$-inductive (control) barrier certificates to relax the standard non-increasing requirement of (control) barrier certificates~\citep{anand2021safetyCDC}. This approach relaxes the aforementioned constraint by allowing up to \(k-1\) one-step increases (in expectation in the stochastic setting), each bounded by a prescribed threshold, while requiring the non-increasing property over every \(k\) consecutive steps. By allowing such relaxations, the likelihood of obtaining \(k\)-inductive (control) barrier certificates is improved, which in turn increases the chance of being able to conduct formal verification or policy synthesis of dynamical systems where conventional (control) barrier certificates fail to exist.\vspace{-0.1cm}
	
	\begin{figure*}
		\centering
		\begin{tikzpicture}
			\node[anchor=south west, inner sep=0] (img) at (0,0)
			{\includegraphics[width=.55\linewidth]{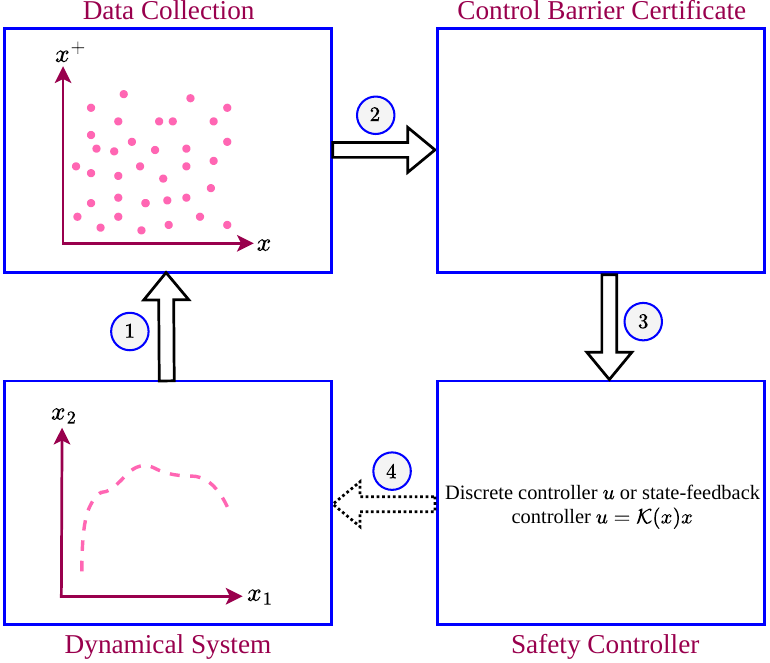}};
			
			\begin{scope}[x={(img.south east)}, y={(img.north west)}]
			
				\node[anchor=south west, inner sep=0] at (0.585,.63)
				{\includegraphics[width={0.22\linewidth}]{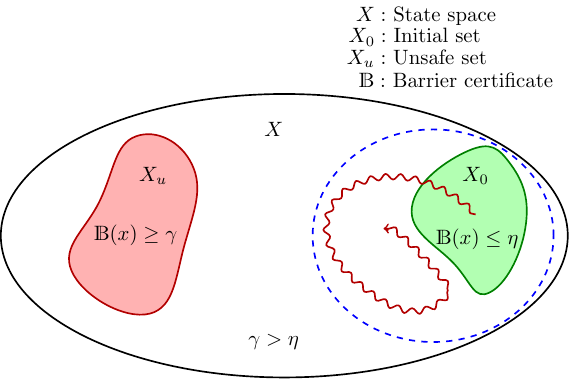}};
				
			\end{scope}
		\end{tikzpicture}
		\caption{Overall diagram of data-driven functional certificate approaches, illustrated using control barrier certificates as a representative example. Similar to data-driven abstraction-based approaches, the first step (\protect\circnum{1}) involves collecting data from the system, which can be carried out in three different ways, each associated with distinct formal guarantees (cf. Sections~\ref{Subsec: Scenario Approach}--\ref{Subsec: Structural Approach}). Leveraging the collected data, control barrier certificates are synthesized (Step \protect\circnum{2}), followed by the design of the corresponding safety controllers (Step \protect\circnum{3}), which are then applied to the system (Step \protect\circnum{4}) to ensure safety.}
		\label{fig:data_barrier-based}
	\end{figure*}
	
	Of paramount importance is that while (control) barrier certificates are mainly employed for guaranteeing the safety of a dynamical system, they have also been adapted to address richer temporal specifications, including those described by $\omega$-regular languages~\citep{baier2008principles,anand2024compositional}. However, such extensions often become conservative since barrier certificates reason only about individual states and their immediate successors. In contrast, $\omega$-regular objectives, such as recurrence, hinge on whether certain states can be visited infinitely often, a question that requires reasoning over the transitive closure of the transition relation rather than over local, one-step behavior. This challenge has led to the development of (control) closure certificates~\citep{murali2024closure}, which extend (control) barrier certificates from state invariants to transition invariants. More precisely, (control) closure certificates are defined over pairs of states, enabling them to characterize an overapproximation of the transitive closure of the transition relation. In this way, they capture the multi-step reachability structure required for reasoning about $\omega$-regular properties. As a result, they offer a more expressive and less conservative foundation for verifying and synthesizing controllers for such specifications.\vspace{-0.1cm}
	
  Despite these advantages, a key challenge in adopting functional certificate approaches lies in their reliance on precise knowledge of the underlying dynamics, which is often unavailable in real-world scenarios. This challenge has motivated the development of data-driven functional certificate approaches, whose typical hierarchy is illustrated in Fig.~\ref{fig:data_barrier-based} for the synthesis of control barrier certificates and the corresponding safety controllers.\vspace{-0.1cm}
   
   \subsection{Compositional Techniques}\vspace{-0.1cm}
  Both the abstraction-based techniques outlined in Section~\ref{Subsec:A-BT} and the functional certificate approaches discussed in Section~\ref{Subsec: MB-FCA} are primarily well suited to dynamical systems of moderate dimensionality. However, an important class of real-world systems comprises large-scale interconnected networks, \emph{e.g.}, power grids. Regrettably, constructing (in)finite abstractions and functional certificates for such large-scale networks in a monolithic manner is intractable, primarily due to the high dimensionality involved. To address this challenge, a promising approach is to view a large-scale network as an interconnection of smaller subsystems. Regarding abstraction-based settings, \emph{compositional techniques} can then be employed to construct an (in)finite abstraction of the overall network by first deriving abstractions of the subsystems and establishing formal behavioral relations between each subsystem and its abstraction. These subsystem abstractions, along with their relations, are then compositionally combined into a global abstraction of the network, which yields the corresponding global behavioral relation.\vspace{-0.1cm}
   
   On the other hand, to construct functional certificates for large-scale interconnected networks, compositional techniques can likewise be utilized, whereby a global functional certificate (together with a controller) for the entire network is constructed by leveraging subsystem-level certificates (and controllers). We note that compositional techniques are largely founded on small-gain or dissipativity-based reasoning, each of which has traditionally been applied to \emph{stability} analysis of large-scale interconnected networks~\citep{dashkovskiy2010small, arcak2016networks,mironchenko2023input}. \vspace{-0.1cm}
   
   While these approaches offer high scalability for handling large-dimensional systems, similar to the monolithic case, constructing (in)finite abstractions or functional certificates using compositional techniques still requires precise knowledge of the network dynamics, which is often unavailable, thereby highlighting the need for data-driven counterparts.\vspace{-0.1cm}
	
	\subsection{Data-Driven Formal Guarantees}\label{Subsec: different_data}\vspace{-0.1cm}
	As discussed previously, the lack of precise mathematical models for real-world systems, together with the widespread availability of inexpensive sensors enabling extensive data collection, has driven significant interest in developing abstraction-based and functional certificate methods that rely on data rather than explicit models. A central challenge in such data-driven approaches, however, is providing out-of-sample performance guarantees, as mentioned in Section~\ref{Subsec:I-VS-D}. Throughout this survey, and across both deterministic and stochastic settings, we classify the principal frameworks for addressing this challenge into three overarching categories. The first comprises statistical guarantees derived via the scenario approach or within probably approximately correct (PAC) frameworks. The second includes guarantees based on Lipschitz continuity to relate observed data to unobserved behaviors. The third encompasses guarantees that exploit structural properties of dynamical systems, such as data-parameterized representations or monotonicity, to infer validity beyond the sampled data.\vspace{-0.1cm}
	
	\textbf{Scenario Approach.} In the first category, the survey considers the scenario approach~\citep{calafiore2006scenarioTAC,campi2009scenarioARC,esfahani2014performanceTAC,campi2008exactSIAM,campi2011samplingJOTA,margellos2014road,romao2022exact,campi2023compressionJMLR,berger2025pac}, which is a probabilistic framework that provides formal guarantees for optimization and control problems with constraints depending on unknown or uncertain quantities, making it particularly well-suited to settings where system models are not explicitly available. More precisely, in both abstraction-based and functional certificate approaches, the required conditions can be generally formulated as a robust optimization program whose constraints should hold over the uncountable state (and input) set, resulting in infinitely many constraints. Moreover, since these constraints also depend on unknown system dynamics, the resulting robust optimization program becomes intractable. The scenario approach, therefore, constructs a tractable counterpart of the optimization problem using a finite number of samples collected from the system, referred to as scenarios. The solution to this sampled problem is then accompanied by a quantifiable guarantee that, with a prescribed confidence level, the probability of constraint violation on unseen realizations does not exceed a user-specified tolerance. In other words, these probabilistic guarantees are typically expressed in terms of violation and confidence levels, which together formalize the likelihood that the solution remains feasible outside the sampled data (cf. Theorem~\ref{thm:scenario_approach_general}).\vspace{-0.1cm}
	
	\begin{figure*}
		\begin{subfigure}[t]{.6\columnwidth}
			\centering
			\resizebox{\columnwidth}{!}{\begin{tikzpicture}[x=1cm,y=1cm,
					axis/.style={
						line width=1.5pt,
						draw=purple!80!black,
						-{Triangle[length=6pt,width=7pt,fill=purple!80!black]}
					},
					datapt/.style={fill=dotcolor, circle, inner sep=1pt},
					dashdot/.style={dash pattern=on 3pt off 1pt on 1pt off 1pt, red, thick},
					]
					
					\pgfmathsetseed{2025}
					
					% Scatter
					\foreach \i in {1,...,120} {
						\pgfmathsetmacro\x{2.5+rand*2.36}
						\pgfmathsetmacro\y{1.7+rand*1.6}
						\node[datapt] at (\x,\y) {};
					}
					
					% Axes with Stealth tips
					\draw[axis] (-0.0231,0) -- (5.5,0) node[right] {$x$};
					\draw[axis] (0,0) -- (0,3.7) node[above] {$x^+$};
			\end{tikzpicture}}
			\caption{Scenario approach}
			\label{subfig:scenario_data}
		\end{subfigure}
		\hfill
		\begin{subfigure}[t]{0.6\columnwidth}
			\centering
			\resizebox{\columnwidth}{!}{\begin{tikzpicture}[x=1cm,y=1cm,
					axis/.style={
						line width=1.5pt,
						draw=purple!80!black,
						-{Triangle[length=6pt,width=7pt,fill=purple!80!black]}
					},
					datapt/.style={fill=dotcolor, circle, inner sep=1pt},
					]
					
					% --- random seed (change this number to get a different "random" figure) ---
					\pgfmathsetseed{2025}
					
					% --- parameters for random x^+ sampling ---
					\def\Nsamp{10}          % how many x^+ samples per x
					\def\ymin{0.15}        % lower bound for random x^+
					\def\ymax{3.45}        % upper bound for random x^+
					\pgfmathsetmacro{\yrange}{\ymax-\ymin}
					
					\draw[gray!30, line width=0.8pt] (0.2,0) -- (0.2,3.7);
					\draw[gray!30, line width=0.8pt] (5.2,0) -- (5.2,3.7);
					
					\draw[gray!30, line width=0.8pt] (0.7,0) -- (0.7,3.7);
					\draw[gray!30, line width=0.8pt] (4.7,0) -- (4.7,3.7);
					
					\draw[gray!30, line width=0.8pt] (1.2,0) -- (1.2,3.7);
					\draw[gray!30, line width=0.8pt] (1.7,0) -- (1.7,3.7);
					
					\draw[gray!30, line width=0.8pt] (2.2,0) -- (2.2,3.7);
					\draw[gray!30, line width=0.8pt] (2.7,0) -- (2.7,3.7);
					
					\draw[gray!30, line width=0.8pt] (3.2,0) -- (3.2,3.7);
					\draw[gray!30, line width=0.8pt] (3.7,0) -- (3.7,3.7);
					
					\draw[gray!30, line width=0.8pt] (4.2,0) -- (4.2,3.7);
					
					% --- grid centers (initial states) ---
					\foreach \xc in {0.2,0.7,1.2,1.7,2.2,2.7,3.2,3.7,4.2,4.7,5.2} {
						
						% initialization point (kept deterministic, as you had it)
						\node[datapt] at (\xc,0.25) {};
						
						% corresponding x^+ samples (random / continuous along y)
						\foreach \k in {1,...,\Nsamp} {
							\pgfmathsetmacro{\yc}{\ymin + \yrange*rnd}
							\node[datapt] at (\xc,\yc) {};
						}
					}
					
					% Axes
					\draw[axis] (-0.0231,0) -- (5.5,0) node[right] {$x$};
					\draw[axis] (0,0) -- (0,3.7) node[above] {$x^+$};
					
			\end{tikzpicture}}
			\caption{Lipschitz continuity-based approach}
			\label{subfig:Lip-based_data}
		\end{subfigure}
		\hfill
		\begin{subfigure}[t]{0.5\columnwidth}
			\centering
			\resizebox{1\columnwidth}{!}{\begin{tikzpicture}[scale=1,
					axis/.style={
						draw=purple!80!black,
						line width=1.5pt,
						-{Triangle[length=6pt,width=7pt,fill=purple!80!black]}
					}
					]
					%── Enlarged lower axes (expanded by 0.5 from each end)
					\draw[axis] (0,-1.25) -- (-1.4, -2.25);
					\draw[axis] (0,-1.25) -- (1.9,-1.25);
					\draw[axis] (0,-1.25) -- (0,0.2);
					
					% Updated axis labels (positions shifted outward)
					\node[below right] at (1.85, -1.05) {$x_{_1}$};
					\node[above left ] at (-0.9, -2.7) {$x_{_2}$};
					\node[above      ] at (0, 0.1) {$x_{_3}$};
					
					%── Curve
					\draw[dotcolor,dotted,thick,smooth]
					plot[samples=200,domain=-0.5:2.2]
					({0.5*(-1.5 + 4*sin(0.3*deg(\x))*0.2 + 2*\x)},
					{0.5*(-2.5 + 1.5*cos(5*deg(\x)) + 0.6*\x)});
					
					%── Points
					\foreach \t in {-0.4,-0.05,0.3,0.65,1.0,1.35,1.65,1.95,2.2} {
						\pgfmathsetmacro{\X}{0.5*(-1.5 + 4*sin(0.3*deg(\t))*0.2 + 2*\t)}
						\pgfmathsetmacro{\Y}{0.5*(-2.5 + 1.5*cos(5*deg(\t)) + 0.6*\t)}
						\fill[dotcolor] (\X,\Y) circle[radius=1.8pt];
					}
			\end{tikzpicture}}
			\caption{Structural-property-based method (data-parameterized system representation)}
			\label{subfig:single_data}
		\end{subfigure}
		\caption{Schematic illustration of three main data collection approaches associated with the data-driven frameworks providing formal out-of-sample performance guarantees considered in this survey. As shown in Fig.~\ref{subfig:scenario_data}, leveraging the scenario approach and its guarantees requires i.i.d. sampling, whereas approaches based on Lipschitz continuity may utilize grid-based data collection to ensure coverage and provide guarantees (without violation and confidence parameters in deterministic settings), as illustrated in Fig.~\ref{subfig:Lip-based_data}.
		In particular, the vertical gray lines in Fig.~\ref{subfig:Lip-based_data}, which discretize the horizontal axis, indicate the discrete set of states at which the system is initialized for data collection, corresponding to the centers of the grid cells induced by the Lipschitz continuity-based framework. For each fixed initial state $x$, multiple successor states $x^{+}$ (vertical axis in Fig.~\ref{subfig:Lip-based_data}) can be observed since the system is repeatedly initialized at the same grid point under distinct control inputs, yielding different successor states.
    	Fig.~\ref{subfig:single_data} illustrates methods that derive data-parameterized system representations, typically relying on a single time-series dataset collected from the system.}\vspace{-0.1cm}
	\end{figure*}
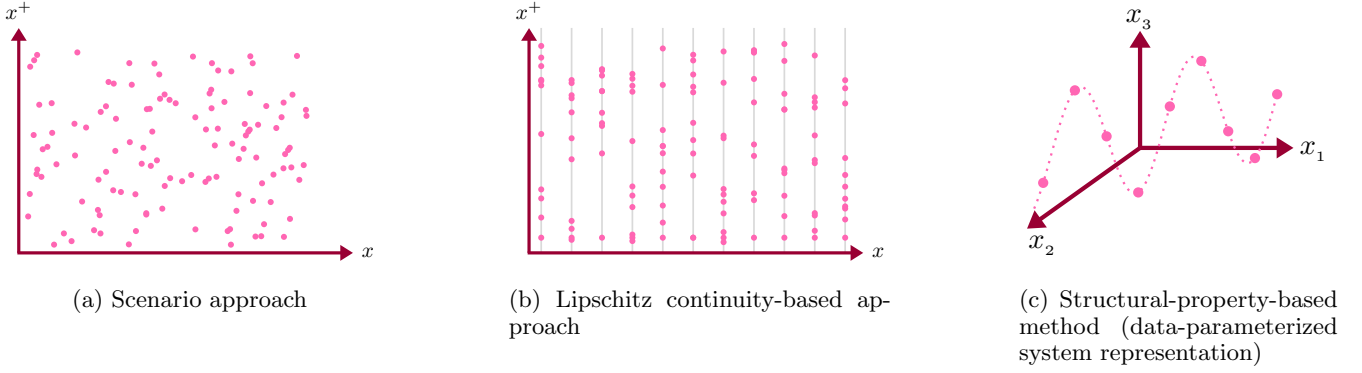
	
	Conceptually, the scenario approach closely aligns with the principles of PAC learning introduced by~\citet{valiant1984theory}, as both frameworks aim to translate empirical observations into out-of-sample performance guarantees with quantifiable statistical confidence. In PAC learning, one seeks solutions that are probably approximately correct, where “probably” denotes the confidence level of the guarantee and “approximately correct” reflects the allowable probability of constraint violation. These two notions are directly analogous to the confidence and violation parameters that characterize the guarantees provided by the scenario approach. Due to this conceptual correspondence, guarantees obtained through scenario-based methods in data-driven abstraction-based and functional certificate approaches are often referred to as PAC-style guarantees. It is worth noting that related finite-sample probabilistic guarantees have also been developed through \emph{conformal prediction}, which has recently been used in formal verification and control to quantify uncertainty, construct prediction regions, design safe controllers, and support offline and online verification of learning-enabled autonomous systems. While standard conformal prediction provides marginal coverage guarantees under exchangeability assumptions, calibration-conditional variants can yield guarantees with a structure close to that of PAC-style guarantees. We refer the interested reader to the recent survey by~\citet{11274485} for more detailed discussions on conformal prediction and its formal guarantees.\vspace{-0.1cm}
	
	\textbf{Lipschitz Continuity-Based Approach.} Despite the advantages of data-driven frameworks grounded in the scenario approach, they yield probabilistic correctness guarantees even in the deterministic setting. This limitation has motivated the development of alternative frameworks whose guarantees rely on certain Lipschitz continuity conditions. More concretely, frameworks based on the scenario approach typically require the collected data to be independent and identically distributed (i.i.d.); see Fig.~\ref{subfig:scenario_data}. In contrast, frameworks relying on Lipschitz continuity impose a grid over the state and input spaces, requiring samples that cover each cell, \emph{i.e.}, by initializing the system at corresponding states and inputs and collecting one-step-ahead data (cf. Fig.~\ref{subfig:Lip-based_data}). Once this coverage is ensured, Lipschitz continuity conditions can be invoked to extend the validity of the results from the sampled points within each cell to the entire cell, and consequently to the entire sample space. As a result, in the deterministic setting, this procedure yields deterministic out-of-sample guarantees in contrast to the inherently probabilistic guarantees provided by the scenario approach. We note that even when an i.i.d. sampling approach is employed, invoking Lipschitz continuity conditions can eliminate the violation parameter from the guarantee; however, the resulting out-of-sample performance guarantee is then expressed in terms of a confidence level due to the random nature of the sampling process.\vspace{-0.1cm}
	
	As discussed above, frameworks based on the scenario approach typically require i.i.d. data, implying that each sample should stem from a distinct, independent trajectory of the system. In practice, this may not be viable and is generally undesirable, as it is time-consuming, costly, and operationally disruptive. Approaches based on Lipschitz continuity do not alleviate this limitation; in fact, they typically require even more samples due to the grid-based sampling, which makes the sample complexity grow \emph{exponentially} with the dimension of the sampling space. Consequently, both approaches are primarily suited to simulator-based settings, where generating multiple independent runs is feasible.\vspace{-0.1cm}
	
	\textbf{Structural-Property-Based Approach.} The above-mentioned challenges have led to growing interest in data-driven frameworks that provide formal guarantees while exploiting structural properties of the system to potentially reduce sample complexity. As a subcategory of these frameworks, recent data-driven methodologies leverage data-parameterized system representations. Specifically, inspired by Willems et al.'s fundamental lemma~\citep{willems2005note} (see also the studies by~\citet{markovsky2021behavioral,van2020willems,shakouri2025new}) and subsequent developments in stability analysis and controller design~\citep{berberich2020data,de2019formulas,van2020noisy,bisoffi2022data,rotulo2022online,van2020data,van2023quadratic,de2023learning,monshizadeh2025versatile,li2026controller,11312975,bianchi2025data,9120225,10810729}, a complementary line of direct data-driven approaches has emerged that often requires only data from a single system trajectory, thereby significantly simplifying data collection (cf. Fig.~\ref{subfig:single_data}).\vspace{-0.1cm}
	
	More concretely, these approaches typically require a single set of non-i.i.d. time-series data collected from the system during a finite-time experiment. Subsequently, if the data is sufficiently rich, typically characterized by a rank condition on the collected data (cf. Assumption~\ref{assump:full-row-rank-data}), it contains adequate information about the system, enabling the development of data-driven abstraction-based and functional certificate approaches with formal out-of-sample performance guarantees. These methodologies provide deterministic formal guarantees for non-stochastic systems, similar in spirit to approaches based on Lipschitz continuity, while requiring substantially less data by exploiting system structural properties.\vspace{-0.1cm}
	
	Another important structural property that can facilitate the development of data-driven methods is \emph{monotonicity}. In broad terms, a monotone system preserves a partial order along its trajectories: if two initial conditions are ordered, then the associated trajectories remain ordered for all future times~\citep{1235373}. In the linear case, such systems are commonly referred to as positive systems~\citep{8310901,8822957}. The importance of this property lies in the additional structure it imposes on system behavior, which typically facilitates analysis based on comparison principles, extremal trajectories, and interval bounds. Consequently, monotonicity enables reasoning based on order relations rather than unstructured exploration, thereby reducing conservatism and improving scalability in data-driven settings. To be more precise, in data-driven settings, this feature is particularly valuable as it allows sparse observations to be exploited more effectively and can omit the dependence on stronger assumptions commonly used to obtain formal out-of-sample guarantees; for instance, for certain classes of order-preserving systems and order-compatible specifications, monotonicity can reduce or eliminate the need for Lipschitz-type assumptions.
	It is worth noting that monotone systems arise in a wide range of application domains, including biological systems~\citep{klipp2005systems}, chemical reaction systems~\citep{leenheer2007monotone}, and transportation networks~\citep{kim2016directed}.\vspace{-0.1cm}
	
	\subsection{Contributions and Organization of the Survey}\vspace{-0.1cm}
	To the best of our knowledge, this is the first survey that systematically organizes data-driven formal verification and synthesis for dynamical systems through the combined lenses of abstraction-based approaches, functional certificates, compositional techniques, and formal out-of-sample guarantees. The survey strives for comprehensiveness while focusing on key developments and situating related methodologies within dedicated sections that provide additional technical detail. Beyond curating the relevant contributions, the survey is structured to help newcomers to the field understand and navigate the key challenges and solutions in this rapidly evolving field. The relevant literature is examined through the lenses of abstraction-based and functional certificate approaches, encompassing both monolithic and compositional manners, applied to both deterministic and stochastic settings (cf.~Fig.~\ref{fig:roadmap}). The resulting data-driven formal guarantees are grouped into three categories: \emph{(i)} statistical guarantees based on PAC and scenario-based frameworks, \emph{(ii)} guarantees derived from Lipschitz continuity, and \emph{(iii)} guarantees exploiting structural properties, \emph{e.g.}, data-parameterized representations or monotonicity. Throughout the survey, several research avenues are raised, thereby paving the way for future advances.\vspace{-0.1cm}
	
	We remark that while recent survey papers by~\citet{martin2023guarantees} and~\citet{de2023learningSurvey} overview data-driven guarantees for nonlinear dynamical systems, their main emphasis is on \emph{stability} analysis and designing stabilizing controllers from data, employing approaches such as polynomial approximations, kernel methods, Koopman operator techniques, and feedback linearization. This survey, however, differs in two fundamental aspects. First, the mathematical tools we consider here, \emph{e.g.}, finite and infinite abstractions, as well as functional certificates, are capable of addressing complex properties beyond stability~\citep{lavaei2022automated}, including those expressed as linear temporal logic (LTL) formulae~\citep{pnueli1977temporal,baier2008principles}. Second, while~\citet{martin2023guarantees} and~\citet{de2023learningSurvey} study only deterministic systems, we consider both deterministic and stochastic settings. These two distinctions necessitate the exploration of different methodologies and frameworks; accordingly, this survey expands the scope of data-driven control with formal guarantees.\vspace{-0.1cm}

   It is worth emphasizing that, while the literature contains many studies that assume known system dynamics while using data such as neural-network-based verification and synthesis~\citep{NNARCH_COMP25_Category_Report}, this survey focuses on settings where the dynamics are (partially) unknown, in line with Section~\ref{Subsec:I-VS-D} and practical application requirements. Furthermore, we note that a substantial body of work has addressed data-driven reachability analysis, where reachable sets are computed from data as over-approximations of all possible system trajectories. Such reachable-set constructions can subsequently be used for several objectives, including, but not limited to, safety verification, the synthesis of safety controllers, and control design under signal temporal logic specifications. While promising, data-driven numerical techniques whose primary objective is to compute possibly tight over-approximations of reachable sets fall outside the scope of this survey and are not reviewed here, unless they are explicitly used for abstraction construction, controller synthesis, or certificate synthesis. This exclusion does not imply that reachability concepts are absent from the discussion; indeed, sublevel sets of barrier certificates can naturally provide over-approximations of reachable sets and play an important role in many verification frameworks. Nevertheless, given the breadth and maturity of data-driven numerical reachability methods, we believe they are more appropriately covered in a dedicated survey.\vspace{-0.1cm}
	
	The survey paper is organized as follows. To facilitate navigation and provide a general overview of the survey, we note that readers primarily interested in formal guarantee mechanisms may focus on Section~\ref{Sec: NaP}, while those interested in deterministic and stochastic systems may proceed directly to Sections~\ref{Sec: Deterministic Setting_ABA}--\ref{Sec: Deterministic Setting_CT} and Sections~\ref{Sec: Stochastic Setting_ABA}--\ref{Sec: Stochastic Setting_CT}, respectively. In more detail, Section~\ref{Sec: NaP} presents the main notations and preliminaries employed throughout the survey, along with formal out-of-sample performance guarantees considered in the paper. Section~\ref{Sec: Deterministic Setting_ABA} is allocated to discussing data-driven abstraction-based techniques for deterministic systems, covering both infinite and finite abstractions. To complement these methodologies, in Section~\ref{Sec: Deterministic Setting_FCA}, we present different data-driven functional certificate approaches for deterministic systems, which do not require discretizing state (and input) spaces. The data-driven approaches outlined in Sections~\ref{Sec: Deterministic Setting_ABA} and~\ref{Sec: Deterministic Setting_FCA} are specifically tailored to monolithic systems and do not scale well to large-scale interconnected networks. To address this, Section~\ref{Sec: Deterministic Setting_CT} provides an overview of data-driven compositional techniques for both abstraction-based and functional certificate approaches.\vspace{-0.1cm}
	
	While deterministic systems are widely applicable, safety-critical systems often exhibit stochastic dynamics, adding complexity to their formal analysis. To cover this, in Section~\ref{Sec: Stochastic Setting_ABA}, we provide a comprehensive overview of data-driven abstraction-based techniques for stochastic dynamical systems. Section~\ref{Sec: Stochastic Setting_FCA} is dedicated to complementing the preceding section by discussing data-driven functional certificate approaches for stochastic systems. Akin to the deterministic setting, Section~\ref{Sec: Stochastic Setting_CT} is allocated to surveying the literature on data-driven compositional techniques, encapsulating both abstraction-based and functional certificate methodologies. Finally, the survey paper is concluded in Section~\ref{Sec: Conclusion}.\vspace{-0.1cm}
	
	\section{Notations and Preliminaries}\label{Sec: NaP}\vspace{-0.1cm}
	We employ the following notation throughout the survey. We denote by $\R$ the set of real numbers, whereas $\Rpz$ and $\Rp$ represent the sets of non-negative and positive real numbers, respectively. Moreover, the sets of non-negative and positive integers are, respectively, given by $\N = \{0,1,2,\ldots\}$ and $\Np = \{1,2,\ldots\}$. The empty set is represented by $\emptyset$.
	The identity matrix of size $n \times n$ is denoted by $\I_n$, while
	$\Zero_n$ represents the zero vector of dimension $n$.
	Given vectors $x_i \in \R^{n_i}$ for $i \in \{1, \ldots, N\}$, we define the stacked vector $x \coloneq [x_1;\ldots;x_N]$ as their vertical concatenation, yielding a column vector of dimension $\sum_{i=1}^N n_i$.
	For any $\bar{p}, \bar{w} \in \R^n$ and relational operator $\simeq \in \{\leq,<,=,>, \geq\}$, where $\bar{p}=[p_1 ; \ldots ; p_n]$ and $\bar{w}=[w_1 ; \ldots ; w_n], \bar{p} \simeq \bar{w}$ is interpreted as $p_l \simeq w_l$ for every $l \in\{1,2, \ldots, n\}$, \emph{i.e.}, component-wise comparison.
	Assuming $\bar{p}<\bar{w}$, then the compact hyper-interval $[\bar{p}, \bar{w}]$ is given as $[p_1, w_1] \times \cdots \times[p_n, w_n]$.
	Furthermore, given $c=[c_1 ; \ldots ; c_n] \in \R^n$, we define the sum $\oplus$ as $c \,\oplus[\bar{p}, \bar{w}] \coloneq [p_1+c_1, c_1+w_1] \times \cdots \times [p_n+c_n, c_n+w_n]$.
	Notation $\vert c \vert$ implies the entry-wise absolute value of $c \in \R^n$, \emph{i.e.}, $[\vert c_1 \vert; \ldots; \vert c_n \vert]$.
	Moreover, $[x_1\;\ldots\;x_N]$ and $[A_1\;\ldots\;A_N]$ represent the horizontal stacking of vectors $x_i \in \R^n$ and matrices $A_i \in \R^{n \times m_i}$ for $i \in \{1, \ldots, N\}$, respectively, forming matrices of sizes $n \times N$ and $n \times \sum_{i=1}^N m_i$.
	Given a \emph{symmetric} matrix $P$, we write $P \succ 0$ ($P \succeq 0$) to denote that $P$ is positive (semi)definite, and $P \prec 0$ ($P \preceq 0$) to denote that $P$ is negative (semi)definite.
	For a matrix $A$ of arbitrary dimensions, its transpose is denoted by $A^\top$.
	The Euclidean norm of a vector $x \in \R^n$ is represented by $\Vert x \Vert \coloneq \sqrt{x^\top x}$.
	The rank of a matrix $P$ is denoted by $\operatorname{rank}(P)$.\vspace{-0.1cm}
	
	The Cartesian product of a collection of sets $X_i$, with $i \in \{1,\ldots,N\}$, is denoted by $\prod_{i=1}^{N} X_i$. For two sets $X$ and $Y$, a relation $\mathscr{R} \subseteq X \times Y$ is defined as a subset of  their Cartesian product, where an element $x \in X$ is said to be related to an element $y \in Y$ if $(x,y) \in \mathscr{R}$; this is equivalently written as $x \mathscr{R} y$.
    Given the sets $X$ and $Y$, their relative complement is denoted by $X \backslash Y$.
    The union among the sets $\mathsf{X}_i$, with $i\in\{1,\dots, n_x\}$, is represented as $\cup_i \mathsf{X}_i$.
    We denote the indicator function of a subset $\mathcal{A}$ of a set $X$ by $\boldsymbol{1}_{\mathcal{A}} : X \to \{0, 1\}$, where $\boldsymbol{1}_{\mathcal{A}}(x) = 1$ if and only if $x \in \mathcal{A}$, and $\boldsymbol{1}_{\mathcal{A}}(x) = 0$ otherwise.
    Given three sets $X$, $Y$, and $Z$, and functions $f : X \to Y$ and $g : Y \to Z$, the composition of $f$ and $g$ is denoted by $g \circ f : X \to Z$. For a function $f : X \to X$, we denote by $f^n$, with $n \in \N$, the $n^{\text{th}}$ iterate of $f$, defined recursively by $f^0 \coloneq \mathrm{id}_X$ and $f^n \coloneq f^{n-1} \circ f$ for $n \ge 1$, where $\mathrm{id}_X$ represents the identity mapping on $X$.
    A block-diagonal matrix with diagonal blocks $A_i$, 
	$i \in \{1,\ldots,N\}$, is denoted by $\mathsf{blkdiag}(A_1,\ldots,A_N)$.
	For a system $\mathcal{S}_g$ and a property $\Delta$, the notation $\mathcal{S}_g \vDash \Delta$ signifies that $\mathcal{S}_g$ satisfies $\Delta$.
	For a function $f : \N \to \R^n$, we define $\Vert f \Vert_{\infty} \coloneq \sup_{k \in \N} \Vert f(k) \Vert$.
	The regularized incomplete beta function~\citep{calafiore2010randomTH5} is defined as
	\[
	 \mathsf{I} \! : \! (c,a,b) \! \mapsto \! \mathsf{I}(c,a,b) \! = \! \frac{\int_0^c \! t^{a-1}(1 \! - \! t)^{b-1} d t}{\int_0^1 \! t^{a-1}(1 \! - \! t)^{b-1} d t}, \forall a, b, c \in \Rp.
	\]
	We consider the probability space $(\Omega, \mathds{F}_\Omega, \mathds{P}_\Omega)$, where $\Omega$ denotes the underlying sample space, $\mathds{F}_\Omega$ is a sigma-algebra on $\Omega$ including subsets of $\Omega$ as events, and $\mathds{P}_\Omega$ is the probability measure assigning probabilities to those events. We assume that random variables introduced in the survey are measurable functions of the form $X : (\Omega,\mathds{F}_\Omega) \rightarrow (S_X,\mathds{F}_X) $ such that each random variable $X$ induces a probability measure on its space $(S_X,\mathds{F}_X)$. We directly specify probability measures on $(S_X,\mathds{F}_X)$ without explicitly referring to the underlying probability space or the mapping $X$. A topological space $S$ is said to be a Borel space if it is homeomorphic to a Borel subset of a Polish space, \emph{i.e.}, a separable and completely metrizable space. A Borel sigma algebra is denoted by $\mathcal{B}(S)$, and can be generated from any Borel space $S$. The map $f : S \rightarrow Y$ is measurable whenever it is Borel measurable.\vspace{-0.1cm}
	
	To avoid ambiguity, we distinguish between two uses of probability throughout the survey. First, for stochastic dynamical systems, probability is induced by the randomness in the system's evolution, such as process noise or random transitions, and therefore characterizes the distribution of trajectories. Under this trajectory distribution, one can then quantify the probability that the system satisfies a desired specification. Second, in data-driven approaches that provide probabilistic out-of-sample performance guarantees (\emph{e.g.}, PAC-style guarantees in Section~\ref{Subsec: Scenario Approach}), probability hinges on the randomly sampled data used to construct the solution; the corresponding probabilistic guarantee quantifies the probability with which the statement inferred from finite samples remains valid for unseen realizations.\vspace{-0.1cm}
	
	\subsection{Scenario Approach: Procedures and Guarantees}\label{Subsec: Scenario Approach}\vspace{-0.1cm}
	As a substantial body of the literature on data-driven formal verification and policy synthesis relies on the scenario approach and its extensions, we begin by reviewing this methodology and the associated formal out-of-sample performance guarantees. Broadly speaking, many problems in systems and control, including those in abstraction-based frameworks and functional certificate-based methods, can be formulated as robust optimization programs, often convex, where constraints are required to hold over a prescribed set. Such a robust convex program (RCP) is typically of the form
	\begin{mini!}|s|[2]<b>
		{[\mu; d]}{\mu}
		{\label{eq:General_RCP}}{}
		\addConstraint{h(x, d) \leq \mu, \; \forall x \in X }{ \label{eq:General_RCP_ST} }
		\addConstraint{d = [d_1; d_2;\ldots;d_{\mathrm{v}}] \in \R^{\mathrm{v}}, \; \mu \in \R,}{ \notag}
	\end{mini!}
	where \(d\) denotes the vector of decision variables, \(\mu\) is an auxiliary scalar variable to be minimized, and, for each fixed \(x\in X\), \(h(x,\cdot)\) is assumed to be convex in \(d\). Moreover, \(x \in X\) denotes the state vector, where \(X \subset \R^n\) is the state set.
	We denote by \(\mu^\ast_{\mathrm{RCP}}\) the optimal value of the RCP in~\eqref{eq:General_RCP}, and by \(d^\ast_{\mathrm{RCP}}\) its corresponding optimizer.
	Importantly, \(h(x,d)\) in~\eqref{eq:General_RCP_ST} can encode the constraints associated with the construction of either (in)finite abstractions or functional certificates. We note that the specific forms of $h(x,d)$ for abstraction-based and functional certificate approaches are provided in their respective sections, and here we keep the discussion general for clarity.\vspace{-0.1cm}
	
	It is evident that solving the RCP~\eqref{eq:General_RCP} is, in general, intractable, primarily for two reasons. First, $h(x,d)$ incorporates the unknown system dynamics, as it encodes the underlying conditions for constructing either abstractions or functional certificates. Second, since the state set \(X\) is uncountable, the problem involves infinitely many constraints. This is precisely where the scenario approach plays a crucial role in addressing the problem by concentrating attention on $N$ i.i.d. samples drawn from $X$, denoted by $x^{z_i}$ for $i \in \{1, \ldots, N\}$, each of which is referred to as a scenario (cf. Fig.~\ref{subfig:scenario_data}).
	Subsequently, rather than focusing on the RCP~\eqref{eq:General_RCP}, based on the extracted samples $x^{z_i} \in X$ for all $i \in \{1, \ldots, N\}$, one can consider the scenario convex program (SCP)
	\begin{mini!}|s|[2]<b>
		{[\mu; d]}{\mu}
		{\label{eq:General_SCP}}{}
		\addConstraint{h(x^{z_i}, d) \leq \mu, \; \forall x^{z_i} \in X, \; \forall i \in \{1, \ldots, N\}}{ \label{eq:General_SCP_ST} }
		\addConstraint{d = [d_1; d_2;\ldots;d_{\mathrm{v}}] \in \R^{\mathrm{v}}, \; \mu \in \R,}{ \notag}
	\end{mini!}
	which constitutes a scenario-based counterpart of the RCP~\eqref{eq:General_RCP}.
	We denote by \(\mu^\ast_{\mathrm{SCP}}\) the optimal value of the SCP~\eqref{eq:General_SCP}, and by \(d^\ast_{\mathrm{SCP}}\) its corresponding optimizer.\vspace{-0.1cm}
	
	Unlike the RCP~\eqref{eq:General_RCP}, the SCP~\eqref{eq:General_SCP} avoids the two aforementioned difficulties, as it involves only finitely many constraints and allows the unknown dynamics in these constraints to be replaced by quantities inferred from data. However, to provide out-of-sample performance guarantees, a key question remains: if the SCP~\eqref{eq:General_SCP} is solved in place of the RCP~\eqref{eq:General_RCP}, what can be claimed about the satisfaction (or violation) of the remaining constraints associated with unseen realizations (\emph{i.e.}, those not enforced during optimization)? We present the following theorem, which addresses this crucial question~\citep{campi2008exactSIAM}.\vspace{-0.1cm}
	
	\begin{theorem}\label{thm:scenario_approach_general}
		For a chosen violation parameter $\varepsilon_1 \in (0,1]$ and confidence parameter $\varepsilon_2 \in (0,1]$, if $N \geq N(\varepsilon_1, \varepsilon_2)$, where
		\begin{align}
			N(\varepsilon_1, \varepsilon_2) \! \coloneq \! \min \! \Big\{ \! N \! \in \! \Np \,  \Bigl\vert \,  \sum_{i=0}^{\mathrm{v}} \! {N \choose i} \varepsilon_1^i (1 - \varepsilon_1)^{N-i} \! \leq \! \varepsilon_2 \! \Big\} \! , \label{eq:scenario_number_of_data}
		\end{align}
		 and the SCP~\eqref{eq:General_SCP} has $\mathrm{v}+1$ optimization variables, then the solution $[\mu^\ast_{\mathrm{SCP}}; d^\ast_{\mathrm{SCP}}]$ satisfies the constraints over $X$ with violation probability at most $\varepsilon_1$ and confidence at least $1-\varepsilon_2$, {i.e.},
		\begin{align}
			\mathds{P}\Big(\mathds{P}\big(h(x, d^\ast_{\mathrm{SCP}}) > \mu^\ast_{\mathrm{SCP}}\big) \leq \varepsilon_1 \Big) \geq 1 - \varepsilon_2. \label{eq:scenario_guarantee}
		\end{align}
	\end{theorem}
	
	\begin{remark}\label{rem:reformulation_datanumber_scenario}
		As shown by~\citet{campi2009scenarioARC}, a simpler expression than that in~\eqref{eq:scenario_number_of_data} is given by
		\[
		N(\varepsilon_1, \varepsilon_2) \coloneq \min \! \Big\{ \! N \! \in \! \Np \; \Bigl\vert \; N \geq \frac{2}{\varepsilon_1} \big( \! \ln (\frac{1}{\varepsilon_2}) + \mathrm{v} + 1 \! \big) \! \Big\} .
		\]
		This expression shows that the required number of scenarios depends logarithmically on the confidence parameter $\varepsilon_2$ and linearly on $\varepsilon_1^{-1}$. Consequently, $\varepsilon_2$ can be chosen to be very small ({e.g.}, $10^{-10}$) without significantly increasing the required sample size.
	\end{remark}
	
	We proceed with providing insight into Theorem~\ref{thm:scenario_approach_general}, first regarding the violation and confidence parameters, and then its connection to the principle of PAC learning. In broad terms, the parameters $\varepsilon_1$ and $\varepsilon_2$ play complementary roles in quantifying the probabilistic nature of the scenario-based guarantee. Specifically, the violation parameter $\varepsilon_1$, also referred to as a risk parameter, quantifies the fraction of the state space $X$ over which constraint violations are tolerated, \emph{i.e.}, where the constraint~\eqref{eq:General_RCP_ST} may not be satisfied. The confidence parameter $\varepsilon_2$, on the other hand, reflects the reliability of this risk bound by quantifying the probability that, due to the randomness in the finite sampling process, the actual violation probability exceeds $\varepsilon_1$. Equivalently, with confidence at least $1 - \varepsilon_2$, the solution obtained from the SCP~\eqref{eq:General_SCP} exhibits a violation probability no greater than $\varepsilon_1$ when evaluated on unseen realizations.\vspace{-0.1cm}
	
	Moreover, through the parameters $\varepsilon_1$ and $\varepsilon_2$, the probabilistic guarantee in Theorem~\ref{thm:scenario_approach_general} admits a natural interpretation within the framework of PAC learning.
	More precisely, the parameter \(\varepsilon_1\) plays the role of an accuracy (or violation) level, while \(\varepsilon_2\) quantifies the confidence with which this accuracy guarantee holds. This result is distribution-free in the sense that the guarantee does not depend on the underlying probability measure on $X$, and the sample complexity bound depends only on the desired accuracy, the confidence level, and the number of optimization variables. Accordingly, the scenario-based guarantee in Theorem~\ref{thm:scenario_approach_general} can be interpreted as a PAC-style generalization bound for constraint satisfaction, ensuring that a solution learned from finite samples extends to unseen realizations with high probability.\vspace{-0.1cm}
	
	\begin{remark}
		For brevity, the formulation considered here assumes that the constraint function depends only on the state variable \(x\) and the design variables \(d\). When the constraint additionally depends on the control input $u \in U$, where $U \subset \R^m$ denotes the input set, the scenario-based framework can be extended accordingly, with the required modifications depending on the controller structure (e.g., whether the control input takes values only from a finite input set or is generated by a state-feedback law). \vspace{-0.1cm}
	\end{remark}
	
	Despite the advantages offered by the scenario approach, its formal out-of-sample performance guarantee involves two nested probability layers, the inner of which (\emph{i.e.}, violation or risk) can be undesirable in certain safety-critical applications. Specifically, the inner probability layer, associated with the violation parameter, only ensures constraint satisfaction for most realizations, rather than over the entire state space $X$. As a result, rare yet potentially critical constraint violations cannot be excluded, which may not be acceptable in practice. This motivates the development of alternative guarantees that bypass, at least, the inner probabilistic layer and establish a more direct connection between the solution of the SCP~\eqref{eq:General_SCP} and that of the corresponding RCP~\eqref{eq:General_RCP}. In this context, approaches that exploit Lipschitz continuity conditions provide a natural alternative and form the basis of the guarantee presented in the following subsection.\vspace{-0.1cm}
	
	\subsection{Lipschitz Continuity-Based Approach: Procedures and Guarantees}\label{Subsec: Lipschitz Approach}\vspace{-0.1cm}
	Approaches based on Lipschitz continuity assumptions for deriving out-of-sample performance guarantees establish a direct link between the feasibility of the RCP~\eqref{eq:General_RCP} and the solution of the SCP~\eqref{eq:General_SCP}. Depending on the data collection procedure, this link may be deterministic, requiring no confidence parameter, or probabilistic, involving a confidence level. In either case, such guarantees provide a stronger notion of reliability when compared to the PAC-style guarantee in~\eqref{eq:scenario_guarantee}. 
	In particular, the guarantee in~\eqref{eq:scenario_guarantee} inherently allows for constraint violations, as quantified by the violation parameter $\varepsilon_1$. Indeed, as is evident from Remark~\ref{rem:reformulation_datanumber_scenario}, achieving $\varepsilon_1 \rightarrow 0$ requires $N \rightarrow \infty$. Accordingly, within the scenario-based framework discussed in Section~\ref{Subsec: Scenario Approach}, complete avoidance of constraint violation cannot be guaranteed when only finitely many samples are available.\vspace{-0.1cm}
	
	To overcome this limitation, approaches based on Lipschitz continuity are instrumental in enabling violation-free guarantees, even with finitely many samples. To elaborate on these approaches, let us reconsider the RCP in~\eqref{eq:General_RCP}, which may be viewed as a formulation aimed at enforcing \(h(x,d)\leq 0\) over the state set \(X\) through the minimization of the auxiliary scalar \(\mu\); this objective is achieved whenever \(\mu^\ast_{\mathrm{RCP}} \leq 0\). To proceed with approaches that rely on Lipschitz continuity conditions, one can collect data using the i.i.d. sampling employed in Section~\ref{Subsec: Scenario Approach}. Alternatively, one may collect \(N\) samples \(x^{z_i} \in X\), \(i \in \{1,\ldots,N\}\), and consider a ball \(\mathbf{B}_i\) of radius \(\varrho \in \Rp\) around each sampled point \(x^{z_i}\) such that $X \subseteq \cup_i \mathbf{B}_i$ and
	\begin{align}
		\Vert x - x^{z_i} \Vert \leq \varrho, \quad \forall x \in \mathbf{B}_i, \quad \forall i \in \{1,\ldots,N\}, \label{eq:max_dist}
	\end{align}
	where \(\varrho\) denotes the covering radius induced by the collected samples and plays a key role in establishing violation-free guarantees~\citep{10149443}. For simplicity, we refer to this type of data collection as \emph{grid-based} sampling hereafter.
	With the collected data, irrespective of the data collection procedure employed, one can reconstruct the SCP in~\eqref{eq:General_SCP}.\vspace{-0.1cm}
	
	The central assumption underpinning the derivation of a formal \emph{violation-free} guarantee in this framework is that the function \(h(x,d)\) is Lipschitz continuous with respect to the state variable \(x\), with Lipschitz constant \(\mathscr{L}\). Since the analysis is carried out over a compact domain \(X\), the existence of such a constant is guaranteed whenever \(h(x,d)\) is locally Lipschitz in \(x\). This regularity assumption effectively accounts for state realizations that are not explicitly observed during the data collection process. Under the aforementioned Lipschitz continuity assumption, it becomes possible to establish a violation-free out-of-sample performance guarantee, which is formalized in the following theorem, adapted from the contributions of~\citet{10066195} and~\citet{10149443}.\vspace{-0.1cm}
	
		\begin{theorem}\label{thm:Lipschitz_approach_general}
			Let the SCP~\eqref{eq:General_SCP} be solved using $N \in \Np$ collected samples, with the optimal value of $\mu^\ast_{\mathrm{SCP}}$ and solution $d^\ast_{\mathrm{SCP}} = [d_1^\ast; d_2^\ast;\ldots;d_{\mathrm{v}}^\ast]$.
			\begin{itemize}
				\item If the data are collected under the i.i.d. sampling scheme, then for an a priori chosen confidence parameter \(\varepsilon_2 \in (0,1]\), the following statement holds with confidence at least \(1 - \varepsilon_2\): if
				\begin{align}
					\mu^\ast_{\mathrm{SCP}} + \mathscr{L} \kappa^{-1} (\iota) \leq 0, \label{eq:tmp2_thm2}
				\end{align}
				with
				\begin{align}
					\iota \geq \mathsf{I}^{-1} (1 - \varepsilon_2, \mathrm{v}+1, N-\mathrm{v}),\notag
				\end{align}
				where $\mathsf{I}^{-1}$ denotes the inverse of the regularized incomplete beta function~\citep{calafiore2010randomTH5} with respect to its first argument (i.e., $1 - \varepsilon_2$), and $\kappa : \Rpz \to [0, 1]$, which depends on the sampling distribution and the geometry of the uncertainty set $X$~\citep[Remarks~5.4 and~5.5]{10066195}, then
				\begin{align}
					h(x,d^\ast_{\mathrm{SCP}}) \le 0, \quad \forall x\in X. \label{eq:tmp_thm2}
				\end{align}
				\item If the data are collected under the grid-based sampling scheme, and if
				\begin{align}
					\mu^\ast_{\mathrm{SCP}} + \mathscr{L} \varrho \leq 0, \label{eq:LIP_S2}
				\end{align}
				with $\varrho$ as in~\eqref{eq:max_dist}, then~\eqref{eq:tmp_thm2} holds.
			\end{itemize}
		\end{theorem}
		
		We now present two remarks on Theorem~\ref{thm:Lipschitz_approach_general}. First, both guarantees in the theorem are violation-free. In the first case, the guarantee is still associated with a confidence parameter, arising primarily from the use of i.i.d. sampling, whereas in the second case, no confidence parameter appears and the resulting guarantee is deterministic. Second, both conditions~\eqref{eq:tmp2_thm2} and~\eqref{eq:LIP_S2} should be verified a posteriori. More precisely, one should first solve the SCP~\eqref{eq:General_SCP} and then check whether~\eqref{eq:tmp2_thm2} or~\eqref{eq:LIP_S2} is satisfied. This stands in contrast to the approach described in Section~\ref{Subsec: Scenario Approach}, which does not require any a posteriori verification.\vspace{-0.1cm}
		
	    While promising, both the Lipschitz continuity-based approaches discussed in this section and the PAC-style approaches presented in Section~\ref{Subsec: Scenario Approach} require repeated system initializations and corresponding data collection; recall that the function $h(x,d)$ explicitly depends on the system dynamics. In practice, however, this requirement can be both challenging and time-consuming, motivating the development of alternative methodologies that exploit system structural properties to alleviate these challenges; such approaches are the focus of the subsequent subsection.\vspace{-0.1cm}
	    
	\subsection{Structural Property-Based Approaches: Procedures and Guarantees}\label{Subsec: Structural Approach}\vspace{-0.1cm}
	In the frameworks discussed in Sections~\ref{Subsec: Scenario Approach} and~\ref{Subsec: Lipschitz Approach}, the problem of interest is formulated as an optimization program, where $h(x,d)$ encapsulates constraints arising from the construction of either (in)finite abstractions or functional certificates.
	As discussed, solving the scenario-based counterpart of the optimization program requires collecting samples at which $h(x,d)$ is evaluated. Since $h(x,d)$ depends on the dynamics, this entails initializing the system at each sampled state, letting it evolve, and recording the resulting state. From this perspective, no explicit knowledge of system structural properties is required, such as whether the dynamics are linear or nonlinear, the specific form of nonlinearities, whether the system is input-affine, or whether it is monotone. Instead, it suffices to be able to evaluate the system dynamics starting from prescribed initial states.\vspace{-0.1cm}
	
	In contrast, the data-driven approaches to be discussed in this subsection assume that certain structural properties of the system are known a priori, which proves beneficial in certain aspects. For instance, one key benefit of approaches that exploit system structural properties lies in simplifying the data collection procedure. Specifically, approaches based on (generalizations of) Willems et al.'s fundamental lemma~\citep{willems2005note}, viewed here as a subcategory of frameworks that exploit system structural properties, often require only a single set of non-i.i.d. time-series data collected during a finite-time experiment. Based on the collected data, the system dynamics can be equivalently represented in a data-parameterized form; consequently, this representation is incorporated into the constraints arising from the construction of (in)finite abstractions or functional certificates, rather than the unknown dynamics themselves. The satisfaction of these constraints can then be explored via suitable data-driven feasibility conditions. For ease of exposition and to facilitate a clearer illustration of the mechanism of such data-driven approaches, we restrict our attention here to simple discrete-time linear dynamical systems, which are formally defined below.\vspace{-0.1cm}
	
	\begin{definition}\label{def:LTI-system}
		A discrete-time linear time-invariant system $\mathcal{S}_{l}$ evolves according to
		\begin{align}
			\mathcal{S}_{l} : x^+ = A x + B u, \label{eq:LTI_model}
		\end{align}
		where $x^+$ denotes the state vector at the next time step, {i.e.}, $x^+ \coloneq x(k+1)$ for $k \in \N$. The matrices $A \in \R^{n \times n}$ and $B \in \R^{n \times m}$ are the unknown system matrices, while $x \in X$ and $u \in U$ denote the state and control input, respectively, with $X \subset \R^n$ and $U \subset \R^m$ representing the compact state and input sets. For a given initial condition $x_{0} \coloneq x(0) \in X$ and an input sequence $u : \N \to U$\footnote{The symbol $u$ is used to denote both an input value in $U$ and an input sequence $u:\N\to U$; the intended meaning will be clear from the context. We adopt the same notational convention elsewhere.}, the state reached at discrete time instant $k \in \N$ is denoted by $x_{x_{0}u}(k)$.
        It is assumed that all state variables of $\mathcal{S}_{l}$ are directly measurable. \vspace{-0.1cm}
	\end{definition}
	
	Considering the system $\mathcal{S}_{l}$, we now illustrate how the dynamics in~\eqref{eq:LTI_model} can be parameterized using data collected along a single trajectory of the system. Specifically, assuming direct measurability of all system states and inputs, one can perform a finite-horizon experiment on the system over the interval $[0, \, \mathcal T]$, where $\mathcal T \in \Np$ denotes the experiment horizon, and collect input--state data
	\begin{subequations}\label{eq:data_single_trajectory}
		\begin{align}
			\mathcal{O} & \coloneq \begin{bmatrix}
				x(0) & ~~ x(1) & ~~ \ldots & ~~ x(\mathcal T - 1)
			\end{bmatrix} \in \R^{n \times \mathcal T}\!,\\
			\mathcal{I} & \coloneq \begin{bmatrix}
				u(0) & ~~ u(1) & ~~ \ldots & ~~ u(\mathcal T - 1)
			\end{bmatrix} \in \R^{m \times \mathcal T}\!,\\
			\mathcal{O}^+ & \coloneq \begin{bmatrix}
				x(1) & ~~ x(2) & ~~ \ldots & ~~ x(\mathcal T)
			\end{bmatrix} \in \R^{n \times \mathcal T}\!,
		\end{align}
	\end{subequations}
	which we refer to as a single trajectory. Assuming that the samples in~\eqref{eq:data_single_trajectory} are noise-free, it follows directly that
	\begin{align*}
		\mathcal{O}^+ = A \mathcal{O} + B \mathcal{I} = \overbrace{\begin{bmatrix}
				A & ~~ B
		\end{bmatrix}}^{\text{unknown}} \begin{bmatrix}
		\mathcal{O} \\ \mathcal{I}
		\end{bmatrix} \! \! .
	\end{align*}
	This observation motivates the idea of parameterizing the closed-loop counterpart of the dynamics in~\eqref{eq:LTI_model} using the collected data in~\eqref{eq:data_single_trajectory}. To enable such a parameterization, however, the following key assumption is typically required.
	
	\begin{assumption}\label{assump:full-row-rank-data}
		It is assumed that the matrix $\begin{bmatrix}
			\mathcal{O}^\top & \mathcal{I}^\top
		\end{bmatrix}^{\! \! \top}$ has full row rank, {i.e.},
		\begin{align}
			\operatorname{rank}\left( \begin{bmatrix}
				\mathcal{O} \\ \mathcal{I}
			\end{bmatrix} \right) = n + m. \label{eq:rank-condition}
		\end{align}
	\end{assumption}
	
	Assumption~\ref{assump:full-row-rank-data} is, in essence, a data richness condition ensuring that the collected data are sufficiently informative to characterize the system dynamics in a data-driven manner. Under this assumption, which is closely related to the notion of persistency of excitation~\citep{willems2005note}, it is essentially guaranteed that the applied input sequence sufficiently excites all modes of the system. Closely related discussions on persistently exciting signals can be found, for instance, in the studies by~\citet{verhaegen2007filtering, padoan2017geometric, alsalti2023data} and the references therein. We note that, for~\eqref{eq:rank-condition} to be potentially satisfied, the experiment horizon \(\mathcal{T}\) should at least fulfill \(\mathcal{T} \geq n + m\); in practice, however, \(\mathcal{T}\) may need to be chosen sufficiently large to ensure that~\eqref{eq:rank-condition} holds.\vspace{-0.1cm}
	
	Having stated the required rank condition on the data matrices, we now present a theorem that parameterizes the system dynamics using the data in~\eqref{eq:data_single_trajectory}, thereby providing a data-driven representation of the closed-loop dynamics~\citep{de2019formulas}.
	
	\begin{theorem}\label{thm:data-based-representation}
		Consider the system $\mathcal{S}_{l}$ in Definition~\ref{def:LTI-system}, and suppose that Assumption~\ref{assump:full-row-rank-data} holds. Then, under the state-feedback control law $u = \mathcal Kx$, with $\mathcal K \coloneq \mathcal{I} G$, and $G \in \R^{\mathcal T \times n}$ satisfying
		\begin{align*}
			\I_n = \mathcal{O} G,
		\end{align*}
		the system $\mathcal{S}_{l}$ in~\eqref{eq:LTI_model} admits the following data-based closed-loop representation:
		\begin{align}
			x^+ = \mathcal{O}^+ G x. \label{eq:data-based-rep}
		\end{align}
	\end{theorem}
	We now present several remarks on Theorem~\ref{thm:data-based-representation} and its associated formal guarantee.
	First, it is evident that the discussed approach relies on knowledge of certain structural properties of the system (\emph{e.g.}, that the system \(\mathcal{S}_{l}\) has linear dynamics). Indeed, systems with different structures admit different data-based closed-loop representations, which should be derived accordingly. For instance, if the system of interest were an input-affine nonlinear system with polynomial dynamics, then a different data-based closed-loop representation would be obtained (cf. Theorem~\ref{thm:poly-rep}). This contrasts with the data-driven approaches discussed in the previous subsections, which do not require such prior structural knowledge and can therefore be applied more broadly when this information is unavailable.
	Second, if the collected data are corrupted by noise, it is still possible to derive a closed-loop parametrization, albeit with suitable modifications to account for the presence of noise.
	Third, and most importantly, it is essential to observe that the data-based closed-loop representation in~\eqref{eq:data-based-rep} is equivalent to the closed-loop dynamics in~\eqref{eq:LTI_model} under $u = \mathcal Kx$. Consequently, if the constraints induced by the construction of either (in)finite abstractions or functional certificates are satisfied when expressed in terms of the data-based parameterization~\eqref{eq:data-based-rep}, then these constraints are also guaranteed to hold for the original closed-loop system in~\eqref{eq:LTI_model}.
	We also note that the nature of this formal guarantee, deterministic or probabilistic, depends on the underlying system dynamics (non-stochastic or stochastic) and, in the presence of noise in the collected data, on whether the data are affected by unknown-but-bounded noise with known bounds or by stochastic noise with a probabilistic distribution.\vspace{-0.1cm}
	
	As mentioned in Section~\ref{Subsec: different_data}, another structural property that can facilitate the development of data-driven approaches is monotonicity.
	It is important to emphasize that monotonicity can not only facilitate analysis and design by imposing additional structure on system behavior but can also be integrated into other data-driven frameworks, where it may help reduce conservatism and, depending on the method, lower the required sample complexity.\vspace{-0.1cm}
	
	\begin{remark}
		The three classes of data-driven guarantees reviewed in this section should be viewed as complementary frameworks, differing mainly in their data requirements, prior knowledge assumptions, and strength of the resulting out-of-sample performance guarantees. In particular, the scenario approach in Section~\ref{Subsec: Scenario Approach} is the most flexible in terms of system knowledge, since it does not require an explicit model or a prescribed structural property of the dynamics. It is therefore suitable when i.i.d. samples can be generated, particularly in simulator-based settings, and when a small probability of constraint violation is acceptable. While promising, its main limitation is that the resulting guarantee is PAC-style even in deterministic settings; it holds up to prescribed violation and confidence levels rather than certifying violation-free correctness.
		The Lipschitz continuity-based approach in Section~\ref{Subsec: Lipschitz Approach} is more appropriate when such violations cannot be tolerated, provided that valid Lipschitz constants and sufficient sample coverage are available. In deterministic settings, this can yield violation-free guarantees, but at the price of exponential sample complexity with respect to the dimensionality of the sampling space (in both i.i.d. and grid-based sampling). In contrast, the structural-property-based approaches in Section~\ref{Subsec: Structural Approach} are preferable when reliable prior knowledge about the system class or qualitative properties (e.g., monotonicity) is available. These methods are less general, but when the required structural assumptions are satisfied, they offer substantial advantages, such as reducing the data-collection burden while still providing deterministic out-of-sample performance guarantees in deterministic settings.\vspace{-0.1cm}
	\end{remark}
	
	Having reviewed the main data-driven techniques and the formal guarantees they provide, we now turn to a comprehensive overview of data-driven abstraction-based methods for deterministic dynamical systems, with particular emphasis on the challenges involved in their development and on how the previously reviewed techniques can be leveraged to address them.\vspace{-0.1cm}
	
	\section{Deterministic Setting: Data-Driven Abstraction-Based Approaches}\label{Sec: Deterministic Setting_ABA}\vspace{-0.1cm}
	In this section, we provide a comprehensive review of data-driven abstraction-based methodologies, specifically tailored to deterministic dynamical systems. As discussed in Section~\ref{Subsec:A-BT}, abstraction-based techniques can be broadly categorized into infinite and finite abstractions, which we cover in separate subsections.\vspace{-0.1cm}
    \subsection{Infinite Abstractions}\label{Subsec:IA-data}
    Within the context of infinite abstractions, the primary objective is to approximate original (\emph{a.k.a.} concrete) dynamical systems by simpler lower-dimensional models, \emph{i.e.}, reduced-order models (ROMs); see Step~2 in Fig.~\ref{fig:data_abstraction-based}. Such approaches have proven particularly beneficial as the mismatch between the behavior of a concrete system and that of its ROM can be rigorously quantified. This, in turn, enables analysis or controller design to be carried out on the ROM, thereby improving scalability, while still allowing the obtained results to be formally transferred to the concrete system. Specifically, this paradigm is particularly advantageous for synthesizing controllers for high-dimensional systems to satisfy complex specifications, as it enables the use of formal methods tools on the ROM to design controllers that guarantee the satisfaction of the desired specifications. The resulting controller can then be refined to the concrete system via an interface function, ensuring that the specification is satisfied by the original system as well, albeit up to a quantified error bound. Due to this multi-layered design procedure, such approaches are also commonly referred to as \emph{hierarchical} control frameworks~\citep{girard2009hierarchical,863598,4200859,tabuada2005hierarchical,smith2020approximate}.\vspace{-0.1cm}

    In order to proceed with such approaches, the notion of simulation functions (SFs) is typically employed to establish a formal relation between the two systems. To elaborate on this notion, consider the system \(\mathcal{S}_l\) in Definition~\ref{def:LTI-system}, equipped with an output map as
    \begin{align}
        \mathcal{S}_{l}\!: \begin{cases}
            \begin{array}{lll}
                x^+ & \hspace{-0.1cm} = & \hspace{-0.05cm} A x + B u,\\
                y   & \hspace{-0.1cm} = & \hspace{-0.05cm} C x,
            \end{array}
        \end{cases} \label{eq:LTI_model_output}
    \end{align}
    where \(C \in \R^{r \times n}\) denotes the output matrix and \(y \in Y\) represents the system output, with \(Y \subset \R^{r}\) being the compact output set. Notice that Definition~\ref{def:LTI-system} corresponds to the special case in which the full state is taken as the output, \emph{i.e.}, \(r=n\) and \(C=\I_n\). Accordingly, the output of the system at discrete time instant \(k \in \N\) is denoted by \(y_{x_{0}u}(k) = C x_{x_{0}u}(k)\).\vspace{-0.1cm}
    
      In the subsequent definition, we first introduce the ROM of the system $\mathcal{S}_{l}$ in~\eqref{eq:LTI_model_output}, paving the way for presenting the notion of SFs.\vspace{-0.1cm}

    \begin{definition}\label{def:LTI-system-ROM}
		A ROM of the system $\mathcal{S}_{l}$ in~\eqref{eq:LTI_model_output} evolves according to
		\begin{align}
            \hat{\mathcal{S}}_{l}\!: \begin{cases}
                \begin{array}{lll}
                    \hat x^+ & \hspace{-0.1cm} = & \hspace{-0.05cm} \hat A \hat x + \hat B \hat u,\\
                    \hat y   & \hspace{-0.1cm} = & \hspace{-0.05cm} \hat C \hat x,
                \end{array}
            \end{cases} \label{eq:LTI_model_output_ROM}
        \end{align}
		where the matrices $\hat A \in \R^{\hat n \times \hat n}$, $\hat B \in \R^{\hat n \times \hat m}$, and $\hat C \in \R^{r \times \hat n}$ characterize the ROM, with potentially $\hat n \ll n$. Moreover, $\hat x \in \hat X$, $\hat y \in \hat Y$, and $\hat u \in \hat U$ denote the state, output, and control input of the ROM, respectively, where $\hat X \subset \R^{\hat n}$, $\hat Y \subset \R^{r}$, and $\hat U \subset \R^{\hat m}$ represent the corresponding compact state, output, and input sets. For a given initial condition $\hat x_{0} \coloneq \hat x(0) \in \hat X$ and an input sequence $\hat u : \N \to \hat U$, the state of the ROM at discrete time instant $k \in \N$ is denoted by $\hat x_{\hat x_{0}\hat u}(k)$, while $\hat y_{\hat x_{0}\hat u}(k) = \hat C \hat x_{\hat x_{0}\hat u}(k)$ represents the corresponding output.\vspace{-0.1cm}
	\end{definition}

    Having introduced the two systems $\mathcal{S}_{l}$ and $\hat{\mathcal{S}}_{l}$, we now formally define the notion of SFs in the subsequent definition, adapted from the work by~\citet{7857702}, which is used to quantify the proximity between the output trajectories of $\mathcal{S}_{l}$ and $\hat{\mathcal{S}}_{l}$.\vspace{-0.1cm}

    \begin{definition}\label{def:SFs-deterministic}
        Consider the two systems $\mathcal{S}_{l}$ and $\hat{\mathcal{S}}_{l}$ as in~\eqref{eq:LTI_model_output} and~\eqref{eq:LTI_model_output_ROM}, respectively.
		A function $\mathds{S} : X \times \hat{X}\to\Rpz$ is
		called a simulation function (SF) from $\hat{\mathcal{S}}_{l}$  to $\mathcal{S}_{l}$ if there exist constants $\alpha_s, \,\rho_s\in \Rp$, and $0 < \kappa_s < 1 $ such that
		\begin{subequations}
			\begin{itemize}
				\item $\forall x \in  X,\forall \hat x \in \hat{X},$
				\begin{align}
					\alpha_s\Vert Cx-\hat C\hat x\Vert^2\le  \mathds{S}(x,\hat x),\label{eq:con1-def-SF-discrete}
				\end{align}
				\item $\forall x\in X,\forall\hat x\in\hat{X}, \forall \hat u \in \hat{U}, \exists u \in U,$ such that
				\begin{align}
					\mathds{S}(A x + B u,  \hat A \hat x + \hat B \hat u) \leq \kappa_s \mathds{S}(x,\hat x) + \rho_s \,\Vert \hat u \Vert^2. \label{eq:con2-def-SF-discrete}
				\end{align}
			\end{itemize}
		\end{subequations}
    \end{definition}

    The notion of SFs, as outlined in Definition~\ref{def:SFs-deterministic}, intuitively implies that if the outputs of \(\mathcal{S}_{l}\) and \(\hat{\mathcal{S}}_{l}\) are initially sufficiently close (ensured by condition~\eqref{eq:con1-def-SF-discrete}), they remain close over time (enforced by condition~\eqref{eq:con2-def-SF-discrete}). Condition~\eqref{eq:con2-def-SF-discrete} also indicates the existence of a function \(u = \mathrm{u}(x,\hat x,\hat u)\) that, for any \(x\), \(\hat x\), and \(\hat u\), returns a corresponding input \(u\) for which condition~\eqref{eq:con2-def-SF-discrete} holds; this function is referred to as an \emph{interface} function.\vspace{-0.1cm}
    
    The following theorem highlights the significance of the notion of SFs by providing a formal quantification of the error between the output trajectories of $\mathcal{S}_{l}$ and $\hat{\mathcal{S}}_{l}$~\citep{samari2026noisyCDCDLD}.\vspace{-0.1cm}

    \begin{theorem}\label{thm:closeness_det_IA}
        Consider the system $\mathcal{S}_{l}$ in~\eqref{eq:LTI_model_output} and its ROM $\hat{\mathcal{S}}_{l}$ as in~\eqref{eq:LTI_model_output_ROM}. Let $\mathds{S}$ be an SF from $\hat{\mathcal{S}}_{l}$ to $\mathcal{S}_{l}$ with associated constants $\alpha_s, \,\rho_s \in \Rp$, and $0 < \kappa_s < 1$. Then, for any $x_0 \in X$, $\hat{x}_0 \in \hat{X}$, and input sequence $\hat{u} : \N \to \hat{U}$, there exists an input sequence $u : \N \to U$ such that the following inequality holds for all $k \in \N$:
        \begin{align}
            \Vert y_{x_{0}u}(k) - \hat y_{\hat x_{0}\hat u}(k) \Vert \leq \sqrt{ \frac{\mathds{S}(x_0, \hat x_0)}{\alpha_s} + \frac{\phi_s}{\alpha_s (1 - \kappa_s)}}, \label{eq:closeness_guarantee_det_IA}
        \end{align}
        where $ \phi_s = \rho_s \Vert \hat u \Vert^2_\infty$.\vspace{-0.1cm}
    \end{theorem}

    As Theorem~\ref{thm:closeness_det_IA} establishes an explicit bound on the mismatch between the output trajectories of \(\mathcal{S}_{l}\) and \(\hat{\mathcal{S}}_{l}\), the ROM can be used as a basis for enforcing a wide range of specifications on the concrete system, including safety, reachability, and reach-while-avoid. More specifically, one may first synthesize a formal controller for the lower-dimensional ROM \(\hat{\mathcal{S}}_{l}\) in a way that the required specification is fulfilled. Subsequently, through a constructed interface function, this controller can be concretized for the higher-dimensional system \(\mathcal{S}_{l}\), while preserving the bounded discrepancy between the output trajectories of the two systems, as characterized in~\eqref{eq:closeness_guarantee_det_IA}. \vspace{-0.1cm}

   To better highlight the main challenge in developing data-driven approaches for constructing infinite abstractions, we present the following theorem~\citep{8264180,lavaei2022automated}, which provides conditions for constructing the ROM $\hat{\mathcal{S}}_{l}$ and guarantees the existence of a quadratic SF from $\hat{\mathcal{S}}_{l}$ to $\mathcal{S}_{l}$.\vspace{-0.1cm}

    \begin{theorem}\label{thm:quadraticSF_deterministic}
        Consider the system $\mathcal{S}_{l}$ in~\eqref{eq:LTI_model_output} and its ROM $\hat{\mathcal{S}}_{l}$ as in~\eqref{eq:LTI_model_output_ROM}. Suppose there exist matrices $K_s \in \R^{m \times n}$ and $P_s \in \R^{n \times n}$, with $P_s \succ 0$, such that
        \begin{subequations}\label{eq:ROM_S3}
            \begin{align}
                & C^\top C \preceq P_s, \label{eq:SF1_linear_det}\\
                & \kappa_s P_s - (1 + \varpi) \big(\! (A + B K_s)^\top P_s (A + B K_s)  \!\big) \succeq 0\label{eq:SF2_linear_det}
            \end{align}
            hold for some constants $\varpi \in \Rp$ and $0 < \kappa_s < 1$. If, in addition,
            \begin{align}
                & A \Theta + B\Xi = \Theta \hat A, \label{eq:SF3_linear_det}\\
                & C\Theta = \hat C \label{eq:SF4_linear_det}
            \end{align} 
            hold for some matrices $\Theta \in \R^{n \times \hat n}$ and $\Xi \in \R^{m \times \hat n}$, then there exists a quadratic SF from $\hat{\mathcal{S}}_{l}$ to $\mathcal{S}_{l}$ of the form
            \begin{align}
                \mathds{S}(x, \hat x) = (x - \Theta \hat x)^\top P_s (x - \Theta \hat x), \label{eq:quadSF-det}
            \end{align}
            where $\Theta$ is referred to as the reduction matrix.\vspace{-0.1cm}
        \end{subequations}
    \end{theorem}
    
    \begin{remark}
    	If the pair $(A,B)$ is stabilizable, then condition~\eqref{eq:SF2_linear_det} can be satisfied~\citep{smith2020approximate}. Moreover, condition~\eqref{eq:SF3_linear_det} holds provided that geometric condition~(30) by~\citet{7857702} is satisfied. We also note that Theorem~\ref{thm:quadraticSF_deterministic} does not pose any restriction on the matrix $\hat B$, which can therefore be selected arbitrarily. For instance, one may choose $\hat B = \I_{\hat n}$, resulting in a fully actuated ROM, thereby considerably simplifying the controller synthesis problem.\vspace{-0.1cm}
    \end{remark}

    Despite the advantages offered by SFs, constructing such functions requires precise knowledge of the system dynamics. In particular, the system matrices appear explicitly in conditions~\eqref{eq:con2-def-SF-discrete} and~\eqref{eq:ROM_S3}. However, as discussed previously, system models are often unavailable in practical settings, thereby motivating the development of data-driven approaches for constructing infinite abstractions using the notion of SFs.\vspace{-0.1cm}
    
    To develop such methodologies, one can make use of the data-driven approaches described in Section~\ref{Sec: NaP}. Broadly speaking, upon fixing a suitable structure for the SF, with its coefficients treated as decision variables, as well as potentially selecting a candidate for the ROM (\emph{e.g.,} fixing $\hat{A}$ and $\hat{B}$ in the linear case), the conditions required for its construction can be cast as robust optimization programs; in particular, conditions~\eqref{eq:con1-def-SF-discrete} and~\eqref{eq:con2-def-SF-discrete} can be encoded as optimization constraints that are required to hold over the relevant sets. The scenario-based counterpart can then be constructed using samples collected from the system. Subsequently, one can leverage Theorems~\ref{thm:scenario_approach_general} and~\ref{thm:Lipschitz_approach_general} to construct the SF by solving the resulting scenario optimization program with formal guarantees. While the general idea is outlined here, the construction of infinite abstractions, together with the establishment of formal relations with their concrete systems using the scenario-based approaches in Sections~\ref{Subsec: Scenario Approach} and~\ref{Subsec: Lipschitz Approach}, remains an important avenue for future work, as stated below.\vspace{-0.1cm}
    
    \begin{resp}
    	\begin{openproblem}
    		Given an unknown general nonlinear system and a selected ROM candidate, develop a data-driven framework based on the scenario approach to construct an SF and its corresponding interface function, while providing out-of-sample performance guarantees.
    	\end{openproblem}
    \end{resp}
    
    Alternatively, by fixing suitable structures for both the SF and the interface function, one may exploit structural properties of the system to derive a data-parameterized representation of the closed-loop system (see Section~\ref{Subsec: Structural Approach}), which can then be used in place of the unknown system dynamics, thereby bypassing the need for an explicit system model. While the approaches described here can be regarded as representative solution frameworks, we note that other data-driven methodologies have also been proposed in the literature, as reviewed hereafter.\vspace{-0.1cm}

    Within this realm, \citet{10976413} propose a data-driven approach that combines Active Subspace theory and Gaussian process (GP) regression~\citep{rasmussen2003gaussianML} to construct ROMs for partially unknown discrete-time input-affine nonlinear control systems, providing probabilistic guarantees characterized by a confidence parameter (but not a violation parameter). We also note that the approach proposed by~\citet{10976413} relies on an i.i.d. sampling scheme (cf.~Fig.~\ref{subfig:scenario_data}), which necessitates multiple system initializations for data collection. To move beyond probabilistic guarantees, \citet{nadali2024transferNEURALSF}, without explicitly constructing ROMs, propose a neural-network-based data-driven approach for unknown discrete-time dynamical systems that leverages transfer learning techniques~\citep{9134370Survey,ben2010theoryML} to formally transfer a controller from a known abstract model, possibly a ROM, to an unknown system with deterministic guarantees. Specifically, the approach relies on grid-based sampling (cf.~Fig.~\ref{subfig:Lip-based_data}) and employs certain Lipschitz continuity assumptions to establish deterministic out-of-sample correctness guarantees (cf. Section~\ref{Subsec: Lipschitz Approach}).\vspace{-0.1cm}
    
    We now continue by presenting studies that fall within the class of data-driven approaches exploiting system structural properties and yielding data-parameterized system representations, as discussed in Section~\ref{Subsec: Structural Approach}. Unlike the two previous studies, these frameworks do not require multiple system initializations, thereby avoiding the key assumption of those studies. In this context, using the notion of SFs, \citet{11312476} develop a data-driven methodology for constructing ROMs of continuous-time linear dynamical systems, while providing deterministic trajectory-based closeness guarantees. Building on this work, \citet{samari2025dataMORNONLINEAR} propose a data-driven framework for constructing linear ROMs for continuous-time input-affine nonlinear dynamical systems, requiring only noise-corrupted input--state data collected along a single trajectory of the system, in contrast to the previous study, which requires noise-free data, while still providing deterministic closeness guarantees. Although neither of these approaches can be directly applied to systems subject to process disturbances, \citet{samari2026noisyCDCDLD} present a data-driven methodology for constructing ROMs of discrete-time linear dynamical systems in the presence of process disturbances, offering deterministic closeness guarantees.\vspace{-0.1cm}

    \begin{remark}
        It is worth noting that while energy-based approaches, including balanced truncation~\citep{6819789}, and Krylov methods, which rely on interpolation and/or moment-matching techniques~\citep{astolfi2010model}, also lead to the construction of ROMs, a fundamental distinction between these methods and SF-based approaches is that these methods are primarily tailored to ensure stability or input--output behavior preservation. In particular, within these two aforementioned frameworks, the same input is typically applied to both the concrete system and its ROM, and the reduction error is quantified as the discrepancy between their input--output maps. In contrast, SF-based schemes allow different inputs for the two systems, coupled through an interface, while providing explicit a priori trajectory-wise bounds on the output mismatch. This property enables their application in controller synthesis for complex specifications. As the focus of this survey is on such specifications that go beyond mere stability, we restrict our attention to SF-based approaches; however, for data-driven methods within the two aforementioned frameworks, we refer the interested reader to the works by~\citet{10024367,8822957,scarciotti2017data,11021390,11193710,mao2024data} and references therein.\vspace{-0.1cm}
    \end{remark}
    
    While infinite abstractions are effective in addressing scalability challenges and can substantially simplify formal verification and controller synthesis, the resulting ROMs still possess uncountable state and input sets, which may render formal analysis and/or synthesis computationally demanding. Finite abstractions address this challenge by approximating concrete systems\footnote{Here, by concrete systems, we refer either to a relatively low-dimensional dynamical system or to a ROM of a high-dimensional dynamical system.} with models that have finite state and input sets. Such abstractions are the focus of the next subsection.\vspace{-0.1cm}
    
    \subsection{Finite Abstractions}\label{Subsec:FA-data}
    A finite abstraction (\emph{a.k.a.} a symbolic model) provides an approximate representation of a concrete system, where each discrete state corresponds to a set of continuous states of the concrete system (cf. Step~3 in Fig.~\ref{fig:data_abstraction-based}). Owing to their finite nature, such abstractions are particularly advantageous, as they enable the use of algorithmic machinery and existing software tools for systematic controller synthesis (cf. Step~4 in Fig.~\ref{fig:data_abstraction-based}). The synthesized controllers can then be refined to the concrete model, thereby enforcing complex specifications (cf. Step~5 in Fig.~\ref{fig:data_abstraction-based}).
    As discussed in Section~\ref{Subsec:A-BT}, such abstractions can be categorized into complete and sound abstractions. We first focus on complete abstractions and then elaborate on sound abstractions, highlighting their differences.\vspace{-0.1cm}
    
    To proceed, consider $\mathcal{S}_l$ in~\eqref{eq:LTI_model_output} as the concrete system\footnote{Notice that while we consider $\mathcal{S}_l$ in~\eqref{eq:LTI_model_output} as the concrete system, the subsequent discussion extends directly to the case where the concrete system is endowed with nonlinear dynamics, without requiring any substantial modifications.}. Here, with a slight abuse of notation, we use \(\hat{\mathcal{S}}_{l}\) to denote the symbolic model of the system \(\mathcal{S}_{l}\). The construction process for complete abstractions begins by partitioning the state and input sets of the concrete system into finite segments, denoted by \(X = \cup_i \mathsf{X}_i\) and \(U = \cup_i \mathsf{U}_i\), respectively. Representative points \(\hat{x}_i \in \mathsf{X}_i\) and \(\hat{u}_i \in \mathsf{U}_i\) are then selected to serve as abstract states and control inputs. The following definition formally outlines the procedure for constructing a symbolic model~\citep{pola2016symbolicTAC,swikir2019compositionalAUTOMATICA}.\vspace{-0.1cm}
    
    \begin{definition}\label{def:FA_construction}
    	Given the concrete system \(\mathcal{S}_l\), its symbolic model \(\hat{\mathcal{S}}_{l}\) (i.e., complete abstraction) is described by
    	\begin{align}
    		\hat{\mathcal{S}}_{l}\!: \begin{cases}
    			\begin{array}{lll}
    				\hat x^+ & \hspace{-0.1cm} = & \hspace{-0.05cm} \hat f(\hat x, \hat u) = \Pi(A\hat{x} + B\hat{u}),\\
    				\hat y   & \hspace{-0.1cm} = & \hspace{-0.05cm} C \hat x,
    			\end{array}
    		\end{cases} \label{eq:LTI_model_output_FA}
    	\end{align}  
    	where \(\hat x \in \hat X\) and \(\hat u \in \hat U\) represent the abstract state and input, respectively, with
    	\(\hat{X} \coloneq \{\hat{x}_i \mid i = 1, \ldots, n_{\hat{x}}\}\) and \(\hat{U} \coloneq \{\hat{u}_i \mid i = 1, \ldots, n_{\hat{u}}\}\) denoting the finite state and input sets of \(\hat{\mathcal{S}}_{l}\), and
    	$\hat y \in \hat Y$ is the output, with $\hat Y$ representing the finite output set of $\hat{\mathcal{S}}_{l}$. Moreover, $\hat f: \hat X \times \hat U \to \hat X$ is the transition map, where the quantization map \(\Pi: X \to \hat{X}\) assigns to any \(x \in X\) a representative point \(\hat{x} \in \hat{X}\) from the corresponding partition, satisfying
    	\begin{align}\label{eq:quantization_map}
    		\Vert \Pi(x) - x \Vert  \leq \delta, \quad \forall x \in X,
    	\end{align}  
    	where \(\delta \coloneq \sup \big\{ \Vert \mathrm{x} - \mathrm{x}^{\prime} \Vert , \; \mathrm{x}, \, \mathrm{x}^{\prime} \in \mathsf{X}_i, \; i = 1, 2, \ldots, n_{\hat{x}}\big\}\) is the state discretization parameter.\vspace{-0.1cm}
    \end{definition}
    
    Despite the benefits of constructing symbolic models, a key challenge is ensuring that properties established for the symbolic model can be reliably transferred to the corresponding concrete system. To address this, a similarity relation between the output trajectories of the two systems should be established, which is typically achieved using the notion of SFs, as defined in Definition~\ref{def:SFs-deterministic}. We note that when employing SFs between a concrete system and its symbolic model, condition~\eqref{eq:con2-def-SF-discrete} requires a slight modification. In particular, an additional constant term \(\psi_s \in \Rp\) should be included on the right-hand side of condition~\eqref{eq:con2-def-SF-discrete} to capture an inherent approximation error, which depends on the state discretization parameter \(\delta\). Consequently, the corresponding closeness guarantee in~\eqref{eq:closeness_guarantee_det_IA} should be modified accordingly, with $\phi_s$ being defined as $\phi_s \coloneq \rho_s \Vert \hat u \Vert_\infty^2 + \psi_s$. Alternatively, the literature proposes that the closeness between the behavior of a concrete system and that of its symbolic model can also be captured through simulation \emph{relations}. More specifically, using the so-called \(\epsilon\)-approximate simulation relation~\citep[Definition~9.2]{tabuada2009verification}, one can guarantee that the output behaviors of the two systems are \(\epsilon\)-close, as established in model-based frameworks~\citep[Proposition 2.4]{swikir2019compositionalAUTOMATICA} and data-driven methodologies~\citep[Theorem~1]{pmlr-v283-samari25aL4DCSINGLE}.

    In data-driven settings for complete abstractions, two main challenges arise: how to construct the symbolic model directly from data, and how to relate the concrete system to its symbolic model via SFs when the unknown dynamics appear explicitly in the conditions for their construction.
    Regarding the first challenge, we note that while the mathematical model of the concrete system $\mathcal{S}_l$ is unknown, its symbolic model $\hat f(\hat x, \hat u) = \Pi(A\hat{x} + B\hat{u})$ can still be constructed in a data-driven manner. Specifically, assuming direct measurability of all state variables (\emph{i.e.}, $C = \I_n$), for each discrete state $\hat{x}$ and input $\hat{u}$, we initialize the system $\mathcal{S}_l$ at $\hat{x}$ under the input $\hat{u}$, and obtain the corresponding successor state associated with $A\hat{x} + B\hat{u}$. Given a state discretization parameter $\delta$, the quantization map $\Pi$ is applied to determine $\hat f(\hat x, \hat u) = \Pi(A\hat{x} + B\hat{u})$ as the representative point closest to $A\hat{x} + B\hat{u}$ that satisfies condition~\eqref{eq:quantization_map} (cf. Step~3 in Fig.~\ref{fig:data_abstraction-based}). Repeating this procedure for all combinations of discrete states $\hat{x} \in \hat X$ and inputs $\hat{u} \in \hat U$ yields the data-driven symbolic model. This construction is fully aligned with the procedure adopted in the model-based setting.
    
    As discussed in Section~\ref{Subsec:IA-data}, the second challenge, namely the construction of SFs in a data-driven setting, can also be addressed by employing the data-driven frameworks described in Section~\ref{Sec: NaP}. In particular, one may either reformulate the conditions for SF construction as robust optimization programs, construct their scenario-based counterparts using collected samples, and exploit the guarantees presented in Sections~\ref{Subsec: Scenario Approach} and~\ref{Subsec: Lipschitz Approach}, or leverage system structural properties, as described in Section~\ref{Subsec: Structural Approach}, to replace the unknown dynamics with their data-parameterized representations.
    
    Notwithstanding the above discussion, it is important to note that the existence of an SF between a concrete system, possibly with nonlinear dynamics, and its symbolic model is guaranteed under the assumption that the system is incrementally input-to-state stable ($\delta$-ISS). In the case of linear dynamics, this property reduces to conventional stability. We  highlight that this assumption is implicitly reflected in condition~\eqref{eq:con2-def-SF-discrete} through the requirement $0 < \kappa_s < 1$. Yet, for completeness, we formally present the notion of $\delta$-ISS for the system $\mathcal{S}_{l}$ in the following definition~\citep{swikir2019compositionalAUTOMATICA}. Notice that while we present the following definition for the system $\mathcal{S}_l$ with linear dynamics, it extends naturally to nonlinear dynamical systems, as well.
    
    \begin{definition}\label{def:DISS}
    	The system $\mathcal{S}_{l}$~\eqref{eq:LTI_model_output} is incrementally input-to-state stable ($\delta$-ISS) if, for some constants $\underline{\alpha}, \, \overline{\alpha} \, , \overline{\rho} \in \Rp$, and $0 < \overline{\kappa} < 1 $, there exists a function $\mathds{S} : X \times X \to\Rpz$, such that
    	\begin{subequations}
    		\begin{itemize}
    			\item $\forall x, \, x^{\prime}  \in  X,$
    			\begin{align}
    				\underline{\alpha} \Vert x - x^{\prime} \Vert^2 \le  \mathds{S}(x, x^{\prime}) \leq \overline{\alpha} \Vert x - x^{\prime} \Vert^2,\label{eq:con1-def-DISS-discrete}
    			\end{align}
    			\item $\forall x, \, x^{\prime}  \in  X, \forall  u, \, u^{\prime} \in U,$
    			\begin{align}
    				\mathds{S}(A x + B u,  A x^{\prime} + B u^{\prime}) \leq \overline{\kappa} \,  \mathds{S}(x, x^{\prime}) + \overline{\rho}  \,\Vert u - u^{\prime} \Vert^2. \label{eq:con2-def-DISS-discrete}
    			\end{align}
    		\end{itemize}
    	\end{subequations}
    \end{definition}
    The $\delta$-ISS property in Definition~\ref{def:DISS} characterizes the incremental behavior of the system by relating the distance between two trajectories to both their initial separation and the mismatch between their input sequences. In particular, when the two systems are driven by identical inputs ($u=u'$), the trajectory mismatch decays exponentially to zero. More generally, when the inputs differ, the influence of the initial mismatch still decays exponentially, while the state mismatch is bounded by a discounted accumulation of the input mismatch $\Vert u-u'\Vert^2$. The results in~\citet{swikir2019compositionalAUTOMATICA} establish that $\hat{\mathcal{S}}_{l}$, defined in Definition~\ref{def:FA_construction}, is a complete finite abstraction of $\mathcal{S}_{l}$, where the function $\mathds{S}$ in Definition~\ref{def:DISS} serves as an SF from $\hat{\mathcal{S}}_{l}$ to $\mathcal{S}_{l}$ and vice versa.
    	
    Despite the advantages of complete abstractions, particularly in providing necessary and sufficient guarantees on controller design, their construction requires the concrete system to be $\delta$-ISS as in Definition~\ref{def:DISS}, which may limit applicability. 
    Constructing sound abstractions, instead of complete ones, helps relax this restrictive requirement; we now explain how such abstractions are typically constructed. Such construction typically relies on an overapproximation of the system’s reachable sets at each time step. To do so, as in the case of complete abstractions, the uncountable state and input sets are first partitioned into finitely many cells, each represented by a discrete state and input. The key difference, however, arises in how transitions are computed. In particular, for each pair of a discrete state and input, the system is conceptually evolved over an instant in time; however, rather than computing an exact successor, an overapproximation of all reachable states is determined. This overapproximation is commonly obtained using the notion of a growth bound~\citep{reissig2016feedback}, which characterizes how the set of possible next states can expand. Once such a reachable set estimate is obtained, all discrete cells intersecting the reachable set are identified as possible successors. In this way, the abstraction deliberately includes every behavior that the concrete system could exhibit, thereby ensuring soundness. In fact, while this approach does not require incremental stability of the underlying system, the trade-off is that sound abstractions provide only sufficient guarantees. That is, the absence of a controller on the abstraction does not imply that no suitable controller exists for the original system.
    
    To formalize the above discussion, let us consider the system
    \begin{align}
    	\mathcal{S}_{c} : \dot{x} = f_c(x,u), \label{eq:ct_sys}
    \end{align}
    where $X \subset \R^n$ and $U \subset \R^m$ denote the state and input sets, respectively, and $f_c : X \times U \to \R^n$ denotes the vector field, which is assumed to be locally Lipschitz continuous for each $u \in U$. Let $\tau \in \Rp$ be the sampling time. For an initial state $x_0 \in X$, a trajectory of the system over $[0,\tau]$ is defined as an absolutely continuous function $\sigma_{x_0u} : [0,\tau] \to X$ satisfying~\eqref{eq:ct_sys} for all $t \in [0,\tau]$, given a constant input $u \in U$ over the interval $[0,\tau]$; \emph{i.e.}, the input signal is piecewise constant across sampling intervals.
    We recall that the uncountable state and input sets are partitioned as  \(X = \cup_i \mathsf{X}_i (x)\) and \(U = \cup_i \mathsf{U}_i\), respectively, where $\hat{x}_i \in \mathsf{X}_i(x)$ and $\hat{u}_i \in \mathsf{U}_i$ are representative discrete state and input points. Accordingly, the corresponding discrete state and input sets are defined as  \(\hat{X} \coloneq \{\hat{x}_i \mid i = 1, \ldots, n_{\hat{x}}\}\) and \(\hat{U} \coloneq \{\hat{u}_i \mid i = 1, \ldots, n_{\hat{u}}\}\), respectively, where $\delta$ and $\delta_u$ denote the state and input discretization parameters (see Definition~\ref{def:FA_construction}).
    
    As mentioned earlier, to construct sound abstractions, one needs to exploit the notion of growth bound, as formalized below~\citep{ajeleye2023dataLCSSLIP}.
    
    \begin{definition}\label{def:GB}
    	For the system $\mathcal{S}_{c}$ in~\eqref{eq:ct_sys}, let $\hat X$ and $\hat U$ denote its symbolic state and input sets. A function $\chi : \R^n_{\geq 0} \times \hat X \times \hat U \to \R^n_{\geq 0}$ is called a growth bound of $\mathcal{S}_{c}$ if it satisfies
    	\begin{align}
    		\vert \sigma_{x^\prime \hat u}(\tau) - \sigma_{\hat x \hat u}(\tau) \vert \leq \chi (\vert x^\prime - \hat x \vert, \hat x, \hat u),
    		\label{eq:GB}
    	\end{align}
    	for any $\hat x \in \hat X$, $\hat u \in \hat U$, and $x^\prime \in \mathsf{X}_i(\hat x)$. 
    \end{definition}
    
Under the growth bound notion in Definition~\ref{def:GB}, we now present the definition of sound abstractions~\citep{ajeleye2023dataLCSSLIP}.
    
    \begin{definition}\label{def:SFA}
    	Given the system $\mathcal{S}_{c}$ in~\eqref{eq:ct_sys} with a growth bound $\chi$, let $\mathcal{S}_{c}^\tau$ denote the sampled system associated with $\mathcal{S}_{c}$. Then, $\hat{\mathcal{S}}_{c}$ is a sound abstraction of $\mathcal{S}_{c}^\tau$, with the finite state and input sets $\hat X$ and $\hat U$, respectively, and the transition map $\hat f_c : \hat X \times \hat U \rightrightarrows \hat X$ if:
    	\begin{itemize}
    		\item the set $\cup_{\hat x \in \hat X} \mathsf{X}_i(\hat x)$ forms a non-empty cover of $X$;
    		\item for any $\hat x, \hat x^\prime \in \hat X$ and $\hat u \in \hat U$, $( \sigma_{\hat x \hat u}(\tau) \oplus [-p^\prime , p^\prime] ) \cap \mathsf{X}_i(\hat x^\prime) \neq \emptyset \Rightarrow \hat x^\prime \in \hat f_c (\hat x, \hat u)$, where $p^\prime = \chi (\delta, \hat x, \hat u)$.
    	\end{itemize}
    \end{definition}
    
    If a sound abstraction is constructed as described in Definition~\ref{def:SFA}, then a feedback refinement relation~\citep{reissig2016feedback} can be established between the sampled system associated with $\mathcal{S}_{c}$ and its finite abstraction. This enables the synthesis of a controller over the finite abstraction and its subsequent refinement to the original system, while fulfilling the property of interest.
    
    Akin to the case of complete abstractions, constructing sound abstractions in data-driven settings is challenging, primarily due to the unknown system dynamics. In particular, computing a growth bound, as described in Definition~\ref{def:GB}, necessitates model knowledge, as it is typically obtained by bounding the Jacobian of the system dynamics. Accordingly, several data-driven approaches have been proposed for constructing sound abstractions and for establishing formal relations between them and their concrete counterparts, providing various types of formal guarantees, which we review in the subsequent subsection.
    
    \subsection{Literature on Data-Driven Construction of Finite Abstractions (Sound and Complete)}
	As expected, indirect data-driven approaches for constructing finite abstractions of dynamical systems can be regarded as the earliest solution frameworks. In this context, \citet{9303814} propose an indirect data-driven approach to obtain finite abstractions of discrete-time dynamical systems, where the unknown dynamics are learned via  GP regression. The resulting abstraction is subsequently employed for safety verification, providing Bayesian-based probabilistic guarantees.
	By leveraging GP regression, together with certain Lipschitz continuity bounds on the unknown dynamics, \citet{hashimoto2022learningAuto} present an indirect data-driven approach for constructing symbolic models of discrete-time dynamical systems subject to bounded disturbances. Through the notion of alternating simulation relations, the framework enables refinement of a controller designed for the symbolic model into a controller for the original system for arbitrary specifications (not limited to safety, unlike the previous work), while providing deterministic correctness guarantees.
	While the approach by~\citet{hashimoto2022learningAuto} applies to partially unknown dynamical systems, the indirect data-driven method proposed by~\citet{10849583} accommodates fully unknown dynamics and, importantly, stochastic noise in the output measurements (while the system dynamics themselves remain deterministic), providing Bayesian-based probabilistic correctness guarantees in the construction of finite abstractions.
	
	As discussed in Section~\ref{Subsec:I-VS-D}, due to the inherent challenges of indirect data-driven approaches, significant attention has been devoted to developing \emph{direct} data-driven methods for constructing finite abstractions of dynamical systems.
	In this vein, \citet{devonport2021symbolicCDCPAC}, without imposing smoothness or regularity assumptions on the dynamics, propose a data-driven approach that requires only the ability to evaluate successor states under given inputs. The approach relies on i.i.d. sampling (cf. Fig.~\ref{subfig:scenario_data}) and provides PAC guarantees (cf. Section~\ref{Subsec: Scenario Approach}) for the constructed abstractions. Nevertheless, the proposed framework can only be employed to satisfy \emph{finite-horizon} specifications. Building on the works of~\citet{pmlr-v211-banse23a} and~\citet{10383513}, \citet{11048505} propose a data-driven approach for constructing finite abstractions of discrete-time dynamical systems, with a focus on formal verification. The approach incorporates memory to better capture the dynamics in specific regions of the state space.
	
	Based on i.i.d. sampling, \citet{coppola2023dataLCSSPAC} and~\citet{coppola2022dataPACARXIV} propose data-driven frameworks for constructing symbolic abstractions of discrete-time linear and nonlinear dynamical systems for verifying complex logic specifications. The general idea in these studies is to sample finite-length trajectories of an unknown system and construct an abstraction from the observed behaviors, while providing PAC guarantees that the resulting abstraction behaviorally includes the concrete system over both finite and infinite time horizons.
    Building on this line of work, \citet{10590827} extend the framework to general unknown deterministic systems and derive classical PAC guarantees via the scenario approach (cf. Section~\ref{Subsec: Scenario Approach}), as well as finite-sample, distribution-free probabilistic inclusion guarantees based on conformal prediction~\citep{shafer2008tutorial,11274485}.
	
	The application of the aforementioned data-driven approaches to event-triggered control systems is also investigated by~\citet{peruffo2022dataLCSSPAC} and~\citet{10591386}. While the former considers only linear dynamical systems, the latter extends the framework to nonlinear dynamical systems and noisy data; both studies provide PAC-style out-of-sample performance guarantees.
    Within this line of work, \citet{coppola2024dataAUTO} move beyond constructing finite abstractions solely for formal verification and propose a data-driven approach, based on i.i.d. sampling, for controller synthesis. To do so, \citet{coppola2024dataAUTO} rely on the notion of probabilistic alternating simulation and provide PAC guarantees that the constructed symbolic model captures all behaviors of the concrete system.
	
	Despite the benefits of the aforementioned advancements, we recall that PAC-bound guarantees are inherently endowed with a violation parameter (cf.~Section~\ref{Subsec: Scenario Approach}, in particular Theorem~\ref{thm:scenario_approach_general}). This implies that the constructed finite abstraction may be incorrect over a small subset of the state space, whose measure can be reduced at the expense of increased computational effort. In particular, achieving zero violation would require an infinite number of samples, as discussed in Section~\ref{Subsec: Lipschitz Approach}. 
	This limitation motivates the development of direct data-driven approaches for constructing finite abstractions that provide correctness guarantees over the entire state space, albeit with a confidence level (cf. Section~\ref{Subsec: Lipschitz Approach}).
	
	Within this context, \citet{lavaei2022dataLCSSONLYCONFIDENCE} propose a data-driven approach for constructing complete finite abstractions of discrete-time dynamical systems, which, by employing $\epsilon$-approximate alternating bisimulation relations, can be used for formal control synthesis tasks. Although the proposed method relies on i.i.d. sampling, by exploiting certain Lipschitz continuity conditions, the resulting guarantee is not endowed with a violation parameter and involves only a confidence parameter. That is, the constructed abstraction is guaranteed to be correct over the entire state space with a prescribed (potentially high) confidence level.
	In addition, by exploiting Lipschitz continuity conditions, \citet{kazemi2024dataNAHSONLYCONFIDENCE} and \citet{ajeleye2023dataLCSSLIP} propose data-driven approaches for constructing sound finite abstractions of continuous-time dynamical systems subject to process disturbances. While the correctness guarantee by~\citet{kazemi2024dataNAHSONLYCONFIDENCE} is endowed with a confidence parameter, \citet{ajeleye2023dataLCSSLIP} eliminate this, thereby providing deterministic correctness guarantees (cf. Section~\ref{Subsec: Lipschitz Approach}). We note that this distinction primarily arises from the data collection procedure; \citet{kazemi2024dataNAHSONLYCONFIDENCE} rely on i.i.d. sampling, whereas \citet{ajeleye2023dataLCSSLIP} employ grid-based sampling (cf.~Fig.~\ref{subfig:Lip-based_data}).
	It is worth noting that, compared with approaches providing PAC guarantees, the studies discussed here exhibit \emph{exponential} sample complexity with respect to the dimension of the sampling space. This increase in sample complexity reflects the cost of providing formal guarantees without a violation parameter.
	
	A common feature of most studies surveyed above is the lack of explicit exploitation of structural properties in dynamical systems. However, as discussed in Section~\ref{Subsec: Structural Approach}, leveraging such properties can be highly beneficial in various aspects, albeit typically at the cost of restricting the class of dynamical systems.
	Within this context, as an extension of the works by~\citet{pmlr-v144-makdesi21aL4DCMONOTONE} and~\citet{makdesi2021efficientIFACMONOTONE}, \citet{makdesi2023dataTACMONOTONE} propose a data-driven approach for computing set-valued overapproximations of unknown monotone functions subject to additive bounded disturbances. In particular, \citet{makdesi2023dataTACMONOTONE} provide a characterization of a simulating map that provably contains all monotone functions consistent with the collected data, providing deterministic correctness guarantees. The resulting data-driven overapproximations are subsequently employed to construct models of partially unknown systems whose unknown components are monotone. These models can then be used to build finite-state symbolic models suitable for formal control synthesis, as they are related to the corresponding concrete systems via alternating simulation relations.
	We note that, even with i.i.d. sampling, the proposed approach, unlike previously surveyed studies, can provide deterministic guarantees, primarily due to exploiting structural properties of the system.
	Notice also that while the approach by~\citet{makdesi2023dataTACMONOTONE} can be classified as an indirect data-driven method, we present it here because it explicitly exploits the structural properties of the underlying systems.
	
	Another line of work that involves exploiting structural properties of dynamical systems focuses on data-parameterized system representations, as discussed in Section~\ref{Subsec: Structural Approach}. In this context, \citet{pmlr-v283-samari25aL4DCSINGLE} propose a data-driven approach for constructing complete finite abstractions of discrete-time input-affine polynomial dynamical systems, using input--state data collected along only two system trajectories, while providing deterministic guarantees. By employing the notion of $\epsilon$-approximate alternating simulation relations, the framework enables the design of a discrete controller for the symbolic model and its subsequent refinement to the concrete system via a hybrid interface function, ensuring the satisfaction of the desired specification.
	
	\begin{resp}
		\begin{openproblem}\label{OP2}
      Consider a concrete system with general nonlinear dynamics subject to bounded process disturbances. Develop a data-driven approach that exploits structural properties of the system to derive a data-parameterized representation (cf. Section~\ref{Subsec: Structural Approach}) and construct a symbolic model of the concrete system from noisy data, while providing deterministic correctness guarantees for the resulting abstraction.
		\end{openproblem}
	\end{resp}
	
	While abstraction-based approaches offer several benefits, such as being well suited to rich specifications and algorithmic synthesis, the discretization of uncountable state and input sets required for their construction can limit their applicability due to the resulting computational complexity, particularly in higher-dimensional settings. This central challenge motivates the development of discretization-free approaches that avoid gridding while retaining formal guarantees, namely functional certificate approaches. In particular, while certificate-based approaches can be potentially conservative and may often be more difficult to construct for rich specifications, they avoid explicit state-space discretization and can scale more favorably. We review such methodologies in the subsequent section.
	
	\section{Deterministic Setting: Data-Driven Functional Certificate Approaches}\label{Sec: Deterministic Setting_FCA}
	This section is devoted to data-driven functional certificate methodologies for deterministic dynamical systems, including (control) barrier certificates, control barrier functions, $k$-inductive (control) barrier certificates, and (control) closure certificates. These certificates provide a powerful framework for formal verification and synthesis to enforce diverse, complex specifications without resorting to state-space discretization (cf. Fig.~\ref{Figure: BC_Scheme}). The data-driven developments reviewed in this section build upon a rich body of model-based literature discussed in Section~\ref{Subsec: MB-FCA}. One of the early efforts bridging model-based and data-driven perspectives is the work by~\citet{han2015sublinear}, where the multiplicative weights method is employed within a barrier-certificate-based framework for data-driven model validation of dynamical systems, with an emphasis on scalability to large collections of trajectory data. 
	
	Recent years have witnessed substantial progress in the development of data-driven frameworks for the synthesis of functional certificates, which often follow the scheme in Fig.~\ref{fig:data_barrier-based}. In alignment with the overall structure of the survey, this section systematically reviews three principal data-driven approaches, \emph{i.e.}, those grounded in the scenario approach with PAC guarantees, those on the basis of certain Lipschitz continuity conditions, and those exploiting system structural properties. In each part, we discuss the underlying assumptions, the associated synthesis procedures, and, crucially, the proposed formal out-of-sample performance guarantees that ensure the validity of the learned certificates beyond the observed data. To do so, we first present a general deterministic discrete-time dynamical system in the following definition.
	
	\begin{definition}\label{def:dt-GNS}
		A discrete-time nonlinear dynamical system $\mathcal{S}_{g}$ evolves according to
		\begin{align}
			\mathcal{S}_{g} : x^+ = f(x, u), \label{eq:gen_disc_org}
		\end{align}
		where $x \in X$ denotes the system state and $u \in U$ represents the control input, with $X \subset \R^n$ and $U \subset \R^m$, respectively, denoting the compact state and input spaces. The mapping $f : X \times U \rightarrow \R^n$ characterizing the system dynamics is assumed to be unknown. For a given initial condition $x_{0} \coloneq x(0) \in X$ and an input sequence $u : \N \to U$, the state reached at discrete time instant $k \in \N$ is denoted by $x_{x_{0}u}(k)$.
	\end{definition}
	
	The system $\mathcal{S}_g$ in~\eqref{eq:gen_disc_org} is intentionally stated in its general nonlinear  form. While we adopt this level of generality to provide a unified framework, a significant portion of the existing literature focuses on specific subclasses of~\eqref{eq:gen_disc_org}. These include input-affine structures, particular classes of nonlinearities, or purely linear dynamics. Throughout the discussion of related studies, we explicitly indicate whenever such structural restrictions are imposed.
	We further remark that, in relation to the system $\mathcal{S}_g$ introduced in Definition~\ref{def:dt-GNS}, a similar system definition can be formulated in a continuous-time framework. However, for the sake of brevity and to maintain a consistent exposition, we confine our discussion to the discrete-time setting. Nevertheless, we stress that the results surveyed in this work encompass both discrete-time and continuous-time settings, and this distinction is clearly specified whenever relevant.

    Having introduced the system of interest in Definition~\ref{def:dt-GNS}, we now proceed by formally presenting the control barrier certificate notion in the subsequent definition~\citep{prajna2004safety,samari2025dataHSCC}.

    \begin{definition}\label{def:cbc}
        Consider the system $\mathcal{S}_g$ in Definition~\ref{def:dt-GNS}, with the initial set $X_0 \subset X$ and the unsafe set $X_u \subset X$. A function $\mathds{B} : X \to \R$ is considered as a control barrier certificate (CBC) for the system $\mathcal{S}_g$ over the time horizon $[0, T - 1]$, with $T \in \Np$, if there exist $c \in \Rpz$ and $\eta, \gamma \in \R$, with $\eta + cT < \gamma$, such that
        \begin{subequations}\label{eq:cbc}
            \begin{itemize}
                \item $\forall x \in X_0$:
                \begin{align}
                    \mathds{B}(x) \leq \eta,\label{eq:cbc1}
                \end{align}
                \item $\forall x \in X_u$:
                \begin{align}
                    \mathds{B}(x) & \geq \gamma, \label{eq:cbc2}
                \end{align}
                \item $\forall x \in X, \; \exists u \in U,$ such that
                \begin{align}
                    \mathds{B}(f(x, u)) &\leq \mathds{B}(x) + c.\label{eq:cbc3}
                \end{align}
            \end{itemize}
        \end{subequations}
    \end{definition}

    According to Definition~\ref{def:cbc}, CBCs are defined and constructed on the system state space, enforcing a collection of inequality constraints on the CBC itself (cf. conditions~\eqref{eq:cbc1} and~\eqref{eq:cbc2}), as well as on its one-step evolution (cf. condition~\eqref{eq:cbc3}). A suitably chosen level set of the CBC, denoted by $\eta$, acts as a separator that isolates the unsafe region $X_u$ from trajectories originating in the set of admissible initial states $X_0$; see Fig.~\ref{Figure: BC_Scheme} for illustration.

    Before proceeding further, it is worthwhile to note that a barrier certificate (BC) is closely related to a CBC, differing primarily in the underlying objective, namely, verification versus synthesis. Specifically, when the system $\mathcal S_g$ admits no control input, the problem reduces to verifying whether the system satisfies a safety property. In this case, a BC is defined in an analogous way to Definition~\ref{def:cbc}, with the key distinction that the control input $u$ does not appear in condition~\eqref{eq:cbc3}.

    \begin{remark}\label{rem:c0}
        Condition~\eqref{eq:cbc3} does not require the CBC to be non-increasing because of the constant $c \in \Rpz$. Allowing the CBC to increase in this way substantially improves the feasibility of CBC construction, while ensuring safety over a finite time horizon. Moreover, since Definition~\ref{def:cbc} enforces the constraint $T < \frac{\gamma - \eta}{c}$, it follows directly that $T \to \infty$ as $c \to 0$. We further note that, particularly in deterministic settings, condition~\eqref{eq:cbc3} typically does not involve the parameter $c$, {i.e.}, $c = 0$, which yields infinite-horizon safety guarantees.
    \end{remark}

    \begin{remark}
        It is important to note that the sets $X_0$ and $X_u$ should be disjoint in Definition~\ref{def:cbc} (cf. Fig.~\ref{Figure: BC_Scheme}). In particular, we require $\gamma > \eta + cT$, which directly implies $\gamma > \eta$. This relation ensures that the sets $X_0$ and $X_u$ do not intersect, as follows from conditions~\eqref{eq:cbc1} and~\eqref{eq:cbc2}; otherwise, the system is unsafe from the outset.
    \end{remark}

    The following theorem, adapted from the literature~\citep{prajna2007framework,10066195}, employs the notion of CBCs to establish formal safety guarantees.

    \begin{theorem}\label{thm:safety_cbc}
        Given the system $\mathcal S_g$, let $\mathds{B}$ be a CBC for $\mathcal S_g$ in the sense of Definition~\ref{def:cbc}. Then, $\mathcal S_g$ is safe over a finite horizon $T$; that is, $x_{x_0u}(k) \notin X_u$ for all $x_0 \in X_0$ and all $k \in [0, T-1]$, under the control input $u$ (associated with the CBC $\mathds{B}$), provided that $T < \frac{\gamma - \eta}{c}$.
    \end{theorem}

   Theorem~\ref{thm:safety_cbc} applies whenever the system $\mathcal{S}_g$ is equipped with a CBC in the sense of Definition~\ref{def:cbc}, or with a BC when addressing safety verification. Nevertheless, the construction of such certificates inherently depends on the system dynamics, since the mapping $f(x,u)$ appears explicitly in condition~\eqref{eq:cbc3}. This motivates the development of data-driven approaches for synthesizing CBCs or BCs.
    
    For completeness, we note that safety specifications have also been extensively studied using control barrier \emph{functions} (CBFs)~\citep{wieland2007constructive,ames2019control}. This notion, which can be viewed as complementary to the concept of CBCs, offers another means for safety analysis. Indeed, frameworks based on CBFs typically provide online safety filters, where the safety controller is obtained via an optimization problem that aims to account for both safety specifications and desired performance. In such approaches, safety is often characterized as the forward invariance of a user-defined safe set $\mathcal{C}$. Such a safety definition coincides with the one considered in Theorem~\ref{thm:safety_cbc} (see also Definition~\ref{def:cbc}) when $\mathcal{C} \coloneq X \backslash X_u$ and $X_0 \coloneq \mathcal{C}$. The notion of CBFs is formally defined in the subsequent definition~\citep{agrawal2017discrete,cosner2024generative}.

    \begin{definition}\label{def:cbf}
        Consider a system $\mathcal{S}_g$ in~\eqref{eq:gen_disc_org}. Let $\mathds{C} : X \to \R$ be a continuous function, with $\mathcal{C} \subset \R^n$ being its $0$-superlevel set, {i.e.}, $\mathcal{C} \coloneq \{x \in X \mid \mathds{C}(x) \geq 0 \}$. The function $\mathds{C}$ is a control barrier function (CBF) for $\mathcal{S}_g$ if, for some $\alpha \in [0, 1]$ and for each $x \in X$, there exists a corresponding $u \in U$ satisfying
        \begin{align}
            \mathds{C}(f(x, u)) \geq \alpha \mathds{C}(x).\label{eq:cbf_general}
        \end{align}
    \end{definition}\vspace{-0.25cm}
    As shown by~\citet{agrawal2017discrete} and~\cite{cosner2024generative}, the existence of a CBF implies controlled forward invariance of the set $\mathcal{C}$. Nevertheless, one can observe that the unknown dynamics also appear in~\eqref{eq:cbf_general}, similar to condition~\eqref{eq:cbc3}, highlighting the need for the development of data-driven approaches.
   Over the past decade, this challenge has attracted significant attention, leading to a substantial body of literature proposing various data-driven methodologies, which we review in the following subsection. While the focus here is on approaches based on CBCs/CBFs, we also include closely related data-driven methods for constructing safety certificates.

    \subsection{Literature on Data-Driven Design of Safety Certificates}\label{Subsec:DD-CBC}
    Within the research line on synthesizing safety certificates for dynamical systems with (partially) unknown dynamics, indirect data-driven approaches can be viewed as one of the earliest solution paradigms (cf. Fig.~\ref{subfig:scheme_indirect}). In particular, GPs have been employed due to their ability to provide nonparametric models equipped with probabilistic uncertainty quantification~\citep{kocijan2016modelling}, where the resulting guarantees are also probabilistic. In this context, and assuming prior knowledge of CBCs, the works by~\citet{8460471} and~\citet{cheng2019end} address safety-critical learning in continuous- and discrete-time settings, respectively. Specifically, these contributions enable safe online learning of GP models and the safe learning of reinforcement learning policies. In a related direction, \citet{9303847} introduce a two-stage framework for the synthesis of safety controllers for continuous-time dynamical systems. In this approach, GPs are first employed to learn the unknown component of the system dynamics, after which CBC-based safety controllers are designed, while providing safety guarantees with specified confidence levels.

    We note that the aforementioned works focus on \emph{input-affine} nonlinear systems, which constitute a specific subclass of the model introduced in Definition~\ref{def:dt-GNS}. In particular, the transition map is assumed to take the form $f(x,u) = f_a(x) + g(x)u$, where $g : X \to \R^{n \times m}$ is known, while $f_a : X \to \R^{n}$ remains unknown. Furthermore, the presence of an explicit model identification phase renders the overall design procedure inherently two-step, which can be both computationally expensive and time-consuming. We refer interested readers to the studies by~\citet{9658123,8493361,wabersich2023data,wabersich2021predictive,10354078,lederer2024safearXivTAC} for additional insights on the CBC/CBF construction grounded in indirect data-driven approaches. 
    We note that alternative methodologies for safety controller synthesis based on CBCs/CBFs have also been proposed, such as the work by~\citet{qin2022sablas} (see also the survey by~\citet{10015199} for a broader overview). In this survey, however, we restrict our attention to approaches that provide formal guarantees.

    In the vein of direct data-driven approaches, \citet{akella2022barrier} propose a safety verification framework based on barrier functions for continuous-time dynamical systems. The approach focuses on systems whose controllers vary with respect to a parameterized input, \emph{e.g.}, varying obstacle locations. For such systems, \citet{akella2022barrier} provide either a PAC guarantee for system safety, as discussed in Section~\ref{Subsec: Scenario Approach}, or a counterexample indicating a safety violation. Moreover, a data-driven semi-parametric approach for learning CBFs is proposed by~\citet{11059253}. Specifically, using noise-corrupted data, the method simultaneously learns a CBF-based safety certificate  to ensure robust controlled invariance for continuous-time nonlinear systems subject to disturbances, providing PAC-based  guarantees. Inspired by the pick-to-learn framework~\citep{paccagnan2023pick}, \citet{RICKARD2026112798} develop a methodology for synthesizing various certificates, including BCs, for discrete-time dynamical systems, providing PAC-style guarantees. The extension of this framework to the continuous-time setting is also explored by~\citet{11312635}.

    Although the aforementioned studies focus on deterministic systems, the PAC nature of the proposed guarantees introduces not only a confidence level but also a certain degree of risk in the safety guarantees, \emph{i.e.}, the possibility of safety being violated (cf. Section~\ref{Subsec: Scenario Approach}). To mitigate this issue, several studies exploit Lipschitz continuity conditions to eliminate either only the violation parameter or both the violation and confidence parameters (cf. Section~\ref{Subsec: Lipschitz Approach}). Before reviewing these studies, we first outline how such frameworks operate in the synthesis of BCs for clarity of exposition.

    As the first step, the structure of a BC is chosen as $\mathds{B}(x, \mathrm{q}) = \sum_{j = 1}^{\mathrm{r}} \mathrm{q}_j \mathrm{p}_j (x)$, where each $\mathrm{p}_j(x)$ denotes a user-defined (potentially nonlinear) basis function and $\mathrm{q} = [\mathrm{q}_1 ~\ldots ~ \mathrm{q}_{\mathrm{r}}]^\top \in \R^{\mathrm{r}}$ represents the unknown coefficients to be designed. For instance, in the case of polynomial-type BCs, each basis function $\mathrm{p}_j(x)$, for all $j \in \{1, \ldots,\mathrm{r}\}$, is a monomial in $x$.
   Subsequently, the conditions in~\eqref{eq:cbc}, with condition~\eqref{eq:cbc3} adapted for BCs since the control input $u$ no longer appears in~\eqref{eq:cbc3}, can be formulated as the following robust convex program (RCP):
    \begin{mini!}|s|[2]<b>
		{[\mu; d]}{\mu}
		{\label{eq:RCP_BC}}{}
		\addConstraint{ \max_j \{ h_j(x, d) \} \leq \mu, \; j \in \{1,\ldots, 4\},  \; \forall x \in X }{ \label{eq:RCP_BC_ST} }
		\addConstraint{d = [\eta;\gamma;c;\varkappa;\mathrm{q}_1; \mathrm{q}_2;\ldots;\mathrm{q}_{\mathrm{r}}] \in \R^{\mathrm{r}+4}}{ \notag}
        \addConstraint{\varkappa \in \R \backslash \Rpz, \; \eta,\gamma,\mu \in \R, \; c \in \Rpz,}{ \notag}
	\end{mini!}
    where
    \begin{align}
        \begin{split}\label{eq:h_CBC}
            h_1(x, d) & =(\mathds{B}(x, \mathrm{q})-\eta) \boldsymbol{1}_{X_0}(x), \\
            h_2(x, d) & =(-\mathds{B}(x, \mathrm{q})+\gamma) \boldsymbol{1}_{X_u}(x), \\ h_3(x, d) & =\eta+c T-\gamma-\varkappa, \\
            h_4(x, d) & =\mathds{B}(f(x), \mathrm{q})-\mathds{B}(x, \mathrm{q})-c,
        \end{split}
    \end{align}
    with $\varkappa$ incorporated into $h_3(x, d)$ in~\eqref{eq:h_CBC} to ensure that the condition $\gamma > \eta$ is satisfied even when $c = 0$. If $\mu^\ast_{\mathrm{RCP}} < 0$, with $\mu^\ast_{\mathrm{RCP}}$ being the optimal value of the RCP~\eqref{eq:RCP_BC}, then a solution to the RCP~\eqref{eq:RCP_BC} indicates that the conditions in~\eqref{eq:cbc} are satisfied. 

    As discussed previously, solving the RCP~\eqref{eq:RCP_BC} is intractable since it involves the unknown dynamics $f(x)$, which motivates the development of its corresponding scenario convex program (SCP). Based on the approach described in Section~\ref{Subsec: Lipschitz Approach}, the data used to construct the SCP corresponding to the RCP~\eqref{eq:RCP_BC} can typically be obtained in two different ways, which nevertheless share a common structure. In both cases, two consecutive sample points along system trajectories are required, \emph{i.e.}, sample pairs of the form $(x^{z_i}, x^{+^{z_i}})$, or equivalently $(x^{z_i}, f(x^{z_i}))$, where each $x^{z_i}$, for all $i \in \{1, \ldots, N\}$, denotes a scenario; thus, a total of $N$ sample pairs are gathered from the system. The key difference, however, lies in the sampling strategy: either i.i.d. samples are drawn from $X$ (cf. Fig.~\ref{subfig:scenario_data}), or sampling is performed using a grid-based scheme over $X$ (cf. Fig.~\ref{subfig:Lip-based_data}).

    The SCP corresponding to the RCP~\eqref{eq:RCP_BC} is described as
    \begin{mini!}|s|[2]<b>
		{[\mu; d]}{\mu}
		{\label{eq:SCP_BC}}{}
		\addConstraint{ \max_j \{ h_j(x, d), h_4(x^{z_i}, d) \} \leq \mu, \; j \in \{1,2,3\} }{ \label{eq:SCP_BC_ST} }
        \addConstraint{\forall x \in X, \; \forall x^{z_i}   \in   X,  \; \forall i \in \{1, \ldots, N\}}{ \notag}
		\addConstraint{d = [\eta;\gamma;c;\varkappa;\mathrm{q}_1; \mathrm{q}_2;\ldots;\mathrm{q}_{\mathrm{r}}] \in \R^{\mathrm{r}+4}}{ \notag}
        \addConstraint{\varkappa \in \R \backslash \Rpz, \; \eta,\gamma,\mu \in \R, \; c \in \Rpz,}{ \notag}
	\end{mini!}
    with an optimal value $\mu^\ast_{\mathrm{SCP}}$ and  optimizer $d^\ast_{\mathrm{SCP}}$.

    The key assumption enabling the derivation of formal guarantees without introducing a violation parameter is that $\mathds{B}(x, \mathrm{q})$ and $f(x)$ are Lipschitz continuous with respect to $x$, with Lipschitz constants $\mathscr{L}_1$ and $\mathscr{L}_{f}$, respectively. Consequently, as both $\mathds{B}(x, \mathrm{q})$ and $f(x)$ are Lipschitz continuous over the compact state space $X$, $h_4(x,d)$ is also Lipschitz continuous with respect to $x$, with Lipschitz constant $\mathscr{L}_2$.
    In the following theorem, we formally state the guarantees obtained by solving the SCP~\eqref{eq:SCP_BC} under both i.i.d. and grid-based sampling schemes~\citep{10066195,10149443}.

    \begin{theorem}\label{thm:unified_lip_iid_BC}
        Consider the system $\mathcal{S}_g$ as in Definition~\ref{def:dt-GNS} (without the control input $u$), with the initial set $X_0 \subset X$ and the unsafe set $X_u \subset X$. Let the SCP~\eqref{eq:SCP_BC} be solved using $N$ sample pairs, with the optimal value of $\mu^\ast_{\mathrm{SCP}}$ and the solution $d^\ast_{\mathrm{SCP}} = [\eta^\ast;\gamma^\ast;c^\ast;\varkappa^\ast;\mathrm{q}_1^\ast; \mathrm{q}_2^\ast;\ldots;\mathrm{q}_{\mathrm{r}}^\ast]$.
        \begin{itemize}
            \item If the data are collected under the i.i.d. sampling scheme, and if condition~\eqref{eq:tmp2_thm2} holds, with $\mathrm{v} \coloneq \mathrm{r} + 4$ and $\mathscr{L} \coloneq \mathscr{L}_2$,
            then, $\mathcal{S}_g$ is safe with confidence at least $1 - \varepsilon_2$, in the sense of Theorem~\ref{thm:safety_cbc}, over the time horizon $[0, T-1]$, with $T < \frac{\gamma^\ast - \eta^\ast}{c^\ast}$.
            \item If the data are collected under the grid-based sampling scheme, and if condition~\eqref{eq:LIP_S2} holds, with $\mathscr{L} \coloneq \mathscr{L}_2$ and $\varrho$ as in~\eqref{eq:max_dist},
            then $\mathcal{S}_g$ is safe, in the sense of Theorem~\ref{thm:safety_cbc}, over the time horizon $[0, T-1]$, with $T < \frac{\gamma^\ast - \eta^\ast}{c^\ast}$.
        \end{itemize}
    \end{theorem}

    \begin{remark}
     The constraints corresponding to $h_1$ and $h_2$ in the SCP~\eqref{eq:SCP_BC} are required to hold for all $x \in X_0$ and $x \in X_u$, respectively. This requirement arises under the i.i.d. sampling scheme, as the presence of piecewise constant indicator functions prevents the use of Lipschitz continuity arguments. However, under the grid-based sampling scheme, the indicator functions appearing in $h_1$ and $h_2$ can be removed, allowing the corresponding constraints to be imposed only on samples drawn from $X_0$ and $X_u$, respectively. Specifically, in this setting, all constraints can be enforced directly using data, in contrast to the SCP~\eqref{eq:SCP_BC}, where the collected data are used solely for the constraint associated with $h_4$.
     Upon this modification, if condition~\eqref{eq:LIP_S2} holds, with $\mathscr{L} \coloneq \max\{\mathscr{L}_1,\mathscr{L}_2\}$,
     the unknown system is safe, in the sense of Theorem~\ref{thm:safety_cbc}, over the time horizon $[0, T-1]$, for any $T < \frac{\gamma^\ast - \eta^\ast}{c^\ast}$. 
    \end{remark}

    Having described the frameworks based on Lipschitz continuity conditions, we now review their corresponding literature. Within this line of work, using the notion of BCs, \citet{10066195} propose a data-driven framework for safety verification in both discrete- and continuous-time settings. Although the proposed approach eliminates the violation parameter in the safety guarantee, it still involves an a priori confidence parameter arising from the use of i.i.d. sampling (cf. first case in Theorem~\ref{thm:unified_lip_iid_BC}). As discussed by~\citet{10066195} in Remark~5.7, the sample complexity of the proposed approach grows exponentially with respect to the number of state variables of the unknown system, which is generally undesirable. To address this challenge, \citet{aminzadeh2024physics} propose a data-driven framework for safety verification of discrete-time dynamical systems, based on the concept of BCs, which provides both deterministic and probabilistic guarantees by incorporating physical principles of the underlying dynamics to eliminate redundant samples and thereby reduce the required number of samples.
    
    While the aforementioned study focuses on safety verification, the data-driven design of safety controllers has also been investigated in the literature. In this respect, \citet{10149443} propose a data-driven methodology for synthesizing safety controllers for discrete-time unknown dynamical systems with finite input sets, introducing the notion of multiple CBCs. Notably, using grid-based sampling and certain Lipschitz continuity conditions, the proposed framework provides deterministic guarantees that the synthesized controller satisfies the desired safety specification.
    Moreover,~\citet{10540046} propose a data-driven methodology for synthesizing controllers for discrete-time dynamical systems based on grid-based sampling and Lipschitz continuity conditions. The objective is to design controllers that ensure a given region in the state space is visited only finitely often during the system evolution, with the number of visits limited to at most $l$, a property expressed using the so-called $l$-universal co-B\"{u}chi automata.

    Neural-network-based safety certificate synthesis has also attracted significant attention in recent years, even in settings where the system dynamics are known~\citep{zhao2020synthesizing,zhao2021learning,lindemann2024learningOJCS,lindemann2021learningCRL,zhang2023exactNeurIPS}.
    This interest is primarily driven by two factors: \emph{(i)} the training process is entirely data-driven, rendering neural networks well-suited for settings with unknown models, and \emph{(ii)} neural networks possess universal approximation capabilities for continuous functions~\citep{hornik1989multilayerUA}, thereby overcoming the limitations imposed by fixed certificate templates. A central challenge in employing such approaches, however, is the lack of formal out-of-sample performance guarantees. To tackle this, for instance, \citet{zhao2020synthesizing} encode CBC constraints as a satisfiability modulo theory (SMT) problem~\citep{de2011satisfiability} to verify the validity of the trained CBCs, while~\citet{jin2020neural} exploit Lipschitz continuity of the trained certificates for their correctness verification.
    Nevertheless, aside from requiring knowledge of system dynamics, such approaches typically verify the trained CBCs a posteriori, which can be computationally costly.

    To bypass the need for a posteriori verification, as well as knowledge of system dynamics, for discrete-time unknown dynamical systems, \citet{anand2023neuraldata} integrate the training and verification processes to obtain provably correct CBCs within a unified framework. In this approach, both the CBCs and the corresponding control policies are parameterized as neural networks, which are learned jointly. Accordingly, Lipschitz continuity properties of neural networks~\citep{NEURIPS2019_95e1533e}, as well as grid-based sampling, are leveraged to establish deterministic out-of-sample performance guarantees.
    Furthermore, \citet{nadali2025choiceQEST} encode CBC conditions using a mean-squared error loss function to obtain smoother gradients and improve the stability and convergence of neural network training. This approach yields CBC-based controllers for discrete-time unknown dynamical systems that, by exploiting specific Lipschitz continuity properties and grid-based sampling, deterministically guarantee system safety without requiring post hoc verification.
    Moreover, \citet{kashani2026data} explore the synthesis of barrier functions that characterize invariant sets for locally Lipschitz continuous unknown nonlinear systems, embedding safety guarantees directly into the training process while relying on grid-based sampling and Lipschitz continuity arguments to provide deterministic guarantees.

    Another aspect that neural-network-based approaches can address is the question of whether knowledge gained from synthesizing certificates and control policies in a given environment (the so-called source environment) can be transferred to a different but related environment (the so-called target environment).
    To address this question, \citet{nadali2023transferCDC} propose a data-driven framework for unknown discrete-time dynamical systems that leverages tools from transfer learning as a solution paradigm, enabling safety guarantees to be transferred from one system to another related system. This implies that the applicability of the proposed framework is limited to settings in which the source and target domains are sufficiently similar to permit comparable control strategies. However, \citet{nadali2024transferADHS} argue that, in many practical scenarios, two systems may differ enough to preclude the use of identical control, yet still share a common logical control structure, thereby enabling the transfer of safety guarantees through learning-based controller design. In particular, \citet{nadali2024transferADHS} propose incorporating inverse dynamics (\emph{i.e.}, a neural network that suggests the required action given a desired successor state) of the target system into the BC of the source system to provide formal safety guarantees. We note that both of the aforementioned studies rely on conditions grounded in Lipschitz continuity assumptions, together with grid-based sampling, to establish deterministic out-of-sample performance guarantees.

    Almost all studies surveyed in this section assume that all state variables of the system are directly measurable. This assumption may limit their practical applicability, as it does not hold in many real-world systems.
    To overcome this limitation, \citet{jahanshahi2023data} propose a data-driven approach for designing CBC-based safety controllers for unknown \emph{partially observable} discrete-time dynamical systems. In particular, this work assumes the availability of a state estimator with unknown dynamics but a known upper bound on the estimation error. Building upon this assumption and leveraging certain Lipschitz continuity conditions and grid-based sampling, deterministic safety guarantees are subsequently established.

   While most of the data-driven studies reviewed above do not explicitly exploit structural properties of dynamical systems (beyond certain Lipschitz continuity assumptions) and are therefore applicable to systems with general nonlinearities, many other works focus on more specific classes of systems. 
    Before reviewing such approaches, we first outline how these frameworks operate, particularly those based on data-parameterized system representations, by focusing on a widely studied subclass of the system in Definition~\ref{def:dt-GNS}, namely input-affine dynamical systems with polynomial nonlinearities, as described in the following definition.

    \begin{definition}\label{def:poly_general_sys}
        A discrete-time input-affine nonlinear system with polynomial dynamics, denoted by $\mathcal{S}_{p}$, evolves according to
		\begin{align}
			\mathcal{S}_{p} : x^+ = f(x) + Bu, \label{eq:poly_disc_org}
		\end{align}
		where $B \in \R^{n \times m}$ is an unknown input matrix, $x \in X$ denotes the state vector, and $u \in U$ represents the control input. Here, $X \subset \R^n$ and $U \subset \R^m$ denote the compact state and input sets, respectively. The polynomial mapping $f : X \rightarrow \R^n$, satisfying $f(\Zero_n) = \Zero_n$, characterizes the system dynamics and is assumed to be unknown. 
    \end{definition}

    While the mapping $f$ is assumed to be unknown, the related literature typically imposes the following assumption on its structure.

    \begin{assumption}\label{assump:dictionary}
        A vector-valued polynomial function $\mathcal{M} : X \to \R^M$, satisfying $\mathcal{M}(\Zero_n) = \Zero_M$, referred to as a dictionary, is assumed to be known such that $f(x) = \mathrm{A} \mathcal{M}(x)$ for some unknown constant matrix $\mathrm{A} \in \R^{n \times M}$.
    \end{assumption}

    \begin{remark}\label{remark:on_dictionary}
        Assumption~\ref{assump:dictionary} is motivated by the observation that, in many practical scenarios ({e.g.}, electrical and mechanical systems), structural information about the system dynamics can often be derived from first principles, which aligns naturally with the dictionary $\mathcal{M}(x)$. However, the system parameters, captured by the unknown matrices $\mathrm{A}$ and $B$, may remain unknown. This assumption is satisfied either when a dictionary $\mathcal{M}(x)$ is available that contains all nonlinear terms appearing in the system dynamics (possibly including additional terms), or when an upper bound on the maximum degree of $f(x)$ is known, enabling the construction of $\mathcal{M}(x)$ by including all monomials up to that degree. Notice that, since $\mathcal{M}(\Zero_n) = \Zero_M$, one can, without loss of generality, express $\mathcal{M}(x)$ as $\mathcal{M}(x) = \aleph(x)x$, where $\aleph : X \to \R^{M \times n}$ denotes a matrix-valued polynomial function, enabling all derivations to be carried out in terms of the state $x$ instead of $\mathcal{M}(x)$, thereby simplifying the analysis~\citep{10804185}.
    \end{remark}
    
    \begin{remark}
    	It is worth highlighting that, through Assumption~\ref{assump:dictionary}, $f(x)$ is represented in a higher-dimensional feature space induced by the dictionary $\mathcal{M}(x)$, yielding a representation that is linear in the lifted features. A similar lifting procedure has also been extensively explored in the literature through approaches grounded in Koopman operator theory~\citep{10565947,9516947,9674041,10153400} and immersion-based methods~\citep{wang2020data,wang2023computationAUTOMATICA}.
    \end{remark}

    Upon Assumption~\ref{assump:dictionary} and Remark~\ref{remark:on_dictionary}, the dynamics in~\eqref{eq:poly_disc_org} can be rewritten as
    \begin{align}
		\mathcal{S}_{p} : x^+ = \mathrm{A} \aleph(x)x + Bu. \label{eq:poly_disc_after_dic}
	\end{align}
    At the same time, by collecting a single set of input--state data from $\mathcal{S}_{p}$ (referred to as a single trajectory) as described in~\eqref{eq:data_single_trajectory}, and using the dictionary $\mathcal{M}(x)$, one can construct the data matrix
    \begin{align}
        \mathcal{D} \coloneq \begin{bmatrix}
				\mathcal{M}(x(0)) &  \mathcal{M}(x(1)) &  \ldots &  \mathcal{M}(x(\mathcal T - 1))
			\end{bmatrix}  \in  \R^{M \times \mathcal T}, \label{eq:data_dict}
    \end{align}
    which is assumed to be of full row rank.
    Having introduced the required preliminaries, we now present a theorem that parameterizes the system dynamics using the data in~\eqref{eq:data_single_trajectory} and~\eqref{eq:data_dict}, thereby providing a data-driven representation that serves as a replacement for the closed-loop system $\mathcal{S}_{p}$~\citep{10804185}.

    \begin{theorem}\label{thm:poly-rep}
        Given the system $\mathcal{S}_{p}$ in Definition~\ref{def:poly_general_sys}, let Assumption~\ref{assump:dictionary} hold. If there exists a matrix-valued polynomial function $\mathds{G} : X \to \R^{\mathcal T \times n}$ such that
		\begin{align*}
			\aleph(x) = \mathcal{D} \mathds{G}(x),
		\end{align*}
		then, under the state-feedback control law $u = \mathcal{K}(x) x$, with $\mathcal{K}(x) \coloneq \mathcal{I} \mathds{G}(x)$, the system $\mathcal{S}_{p}$ admits the following data-based closed-loop representation:
		\begin{align}
			x^+ = \mathcal{O}^+ \mathds{G}(x) x. \label{eq:data-based-rep-poly}
		\end{align}
    \end{theorem}\vspace{-0.25cm}
    Since the data-based closed-loop representation in~\eqref{eq:data-based-rep-poly} can be used on the left-hand side of~\eqref{eq:cbc3}, many works in the literature have exploited approaches that parameterize the system dynamics directly from collected data to derive safety guarantees. An immediate advantage of leveraging such approaches, compared with the previously discussed scenario-based ones, is that input–state data collected along a single trajectory of the system typically suffices, thereby eliminating the need to repeatedly reset the system to gather multiple datasets. 

    In a related context, the work by~\citet{ahmadi2020safe} represents one of the early studies proposing a data-driven methodology for designing finite-horizon safety controllers for unknown continuous-time dynamical systems using input--state data collected from a single trajectory. In particular, \citet{ahmadi2020safe} employ piecewise-polynomial approximations of the trajectories, together with regularity side information, to construct a data-driven differential inclusion model capable of predicting trajectory evolution. Safety analysis is then carried out using BCs, followed by the synthesis of controllers that deterministically guarantee the safety over a finite horizon. In the continuous-time setting, \citet{nejati2022data} propose a data-driven framework for safety controller design for unknown nonlinear systems with polynomial dynamics that provides deterministic guarantees and can also accommodate input constraints, albeit at the cost of introducing bilinearity.

    The work by~\citet{10804185} presents a data-driven approach for safety controller design for nonlinear systems with polynomial dynamics in the discrete-time setting, providing deterministic safety guarantees.
    The work by \cite{gardner2025trust} introduces \textsf{TRUST}, an open-source tool for synthesizing safety-enforcing controllers (using the notion of CBCs), as well as stabilizing controllers, directly from data. The approach requires only input--state data collected along a single trajectory of a system with unknown linear and polynomial dynamics, in both continuous- and discrete-time settings. Implemented as a user-friendly \textsf{Python} web application, \textsf{TRUST} employs sum-of-squares (SOS) optimization to construct CBCs from data.
    
    While the aforementioned studies cannot handle systems with general nonlinear terms, \citet{samari2025dataHSCC} propose a data-driven framework for designing safety controllers for general nonlinear systems that offers deterministic finite-horizon safety guarantees. More concretely, by minimizing the effect of nonlinearities and capturing them via the parameter $c$ in~\eqref{eq:cbc3}, the safety guarantee becomes finite-horizon, thereby enabling the treatment of general nonlinear systems.
    Moreover, by leveraging the adding-one-integrator approach, \citet{samari2025dataHSCC} account for input constraints without introducing bilinearity; instead, the resulting framework yields dynamic safety controllers rather than conventional static state-feedback controllers. While previous studies primarily consider single-system settings, \citet{esmaeili2025safeSAFEMARRT} propose a data-driven approach to designing safety controllers, specifically tailored to motion planning tasks for homogeneous linear multi-agent systems operating in a shared, obstacle-filled workspace.
    
    The aforementioned studies assume direct measurement of all state variables of the dynamical systems, which may be restrictive in certain practical scenarios, as discussed previously. To address this challenge, \citet{jahanshahi2023synthesis} propose a data-driven framework for designing CBC-based safety controllers with deterministic formal guarantees for continuous-time polynomial dynamical systems that are partially observable. Their approach assumes a polynomial-type estimator with partially unknown dynamics and a known upper bound on the estimation error and relies on collecting a single input--output trajectory from the system together with a single state trajectory from its estimator. In a related direction, for discrete-time linear dynamical systems, \citet{11312441} present a data-driven approach formulated as a quadratically constrained quadratic program, which deterministically guarantees both safety, via the notion of CBFs, and stability using only input--output data, without assuming the availability of an estimator. In particular, an augmented system constructed from historical input--output measurements enables the data-driven formulation of CBFs.

    An almost inevitable scenario in practice, not considered in the studies reviewed above, is that data collected from dynamical systems are corrupted by noise arising from process disturbances or measurement imperfections. Motivated by this practical challenge, several studies have investigated settings in which the available data are subject to noise.
    Within this line of work, building upon the contribution by~\citet{10068731} on computing reachable sets directly from noise-corrupted data for discrete-time dynamical systems, \citet{11312081} propose a data-driven safety verification framework for unknown discrete-time linear dynamical systems subject to process disturbances. The framework leverages matrix zonotopes and BCs to deterministically verify system safety directly from noise-corrupted input--state data.
    For the same class of dynamical systems and based on a single set of noise-corrupted input--state data, \citet{modares2025unifyingDirectandIndirect} propose a data-driven framework for safe controller synthesis, with deterministic guarantees, that unifies direct and indirect learning by representing the closed-loop dynamics through constrained matrix zonotopes. We note that the two previous studies fall within the class of data-driven approaches that derive data-parameterized system representations (cf. Section~\ref{Subsec: Structural Approach}), a categorization that also applies to the literature reviewed hereafter.
    
    Beyond linear dynamical systems, the problem of safe stabilization for partially unknown discrete-time input-affine nonlinear dynamical systems is studied by~\citet{10554661}, where deterministic guarantees are provided. Leveraging a set of noisy input--state data, \citet{10554661} propose a data-driven approach to certify the existence of an (acausal) control policy capable of safely stabilizing all systems consistent with the collected data. Moreover, inspired by the work of~\citet{9763329}, which leverages noise-corrupted data to design controllers for discrete-time linear dynamical systems with deterministic guarantees of robust invariance for a given polyhedral state set, \citet{luppi2024dataEJC} address the safe control problem for unknown input-affine polynomial systems in continuous time subject to process disturbances by designing robustly invariant sets and providing deterministic out-of-sample performance guarantees. Designing safety controllers with deterministic guarantees from noise-corrupted input–state data for discrete-time input-affine nonlinear systems subject to process disturbances, without restricting the nonlinear dynamics to polynomial forms, is also addressed by~\citet{modares2025nonNonConservative}, where safety specifications are expressed using polyhedral sets.

    By leveraging the concept of CBCs, \citet{11312857} propose a data-driven design of safety controllers for discrete-time input-affine dynamical systems with polynomial nonlinearities subject to process disturbances, providing deterministic infinite-horizon safety guarantees.
    For the same class of systems, \citet{ashoori2025physicsTAC} offer a data-driven approach to designing CBC-based safety controllers capable of delivering both deterministic infinite- and finite-horizon safety guarantees. The latter increases the likelihood of finding suitable CBCs and associated safety controllers. Moreover, inspired by the study by~\citet{10318172}, which focuses on constructing robust invariant sets (see also the work by~\citet{WangLCSS2021}) under safety constraints for discrete-time linear dynamical systems, \citet{ashoori2025physicsTAC} incorporate prior knowledge of the system into the data-driven design, resulting in a meaningful reduction in the number of samples required for the design task.
    
    While the aforementioned studies are primarily tailored to delay-free systems, safety controller design for systems with unknown models and time delays remains largely unexplored, with the exception of the recent work by~\citet{akbarzadeh2026data}. In particular, leveraging a single set of input--state data, corrupted by unknown-but-bounded noise, \citet{akbarzadeh2026data} propose a data-driven framework for designing CBC-based safety controllers for discrete-time input-affine polynomial systems subject to both process disturbances and time-invariant delays, providing deterministic infinite-horizon safety guarantees. Notice that, since stability analysis and controller deign are not the primary focus of this survey, we refer interested readers to~\citet{9484756} (and the references therein) for data-driven approaches to designing stabilizing controllers for systems with time delays.

    Although the aforementioned literature exploits certain structural properties of dynamical systems to obtain data-parameterized representations, as discussed in Section~\ref{Subsec: Structural Approach}, another structural property that has recently attracted significant attention is the monotonicity of dynamical systems. In particular, monotone dynamical systems are endowed with the property that their trajectories preserve a natural partial order defined on the state space. As is well known, monotone dynamical systems exhibit highly ordered transient and asymptotic behavior~\citep{1235373,8310901,saoud2024characterization}, making them particularly well suited for data-driven approaches, as their intrinsic structure helps mitigate challenges arising from limited or noisy data. Despite its demonstrated benefits, this important structural property has not yet been thoroughly explored for the data-driven synthesis of safety certificates, with only a few recent exceptions, which we review in the sequel.

    For discrete-time monotone dynamical systems, \citet{11036788} propose a neural-network-based data-driven framework for safety verification using a finite number of samples, providing deterministic out-of-sample performance guarantees. Specifically, by embedding BCs into a higher-dimensional space, a new formulation for safety verification is introduced that is compatible with interval analysis, which is then employed for data-driven safety verification. Moreover, \citet{11036788} establish a connection between embedded BCs and neural networks with positive weights and non-decreasing activation functions, facilitating their learning through monotone neural architectures. The embedded BCs are represented as the difference of two monotone neural networks to enhance flexibility. It is demonstrated by~\citet{11036788} that, primarily due to leveraging this structural monotonicity property, the proposed approach requires significantly less data compared with methods relying on Lipschitz continuity assumptions, \emph{e.g.}, the work by~\citet{nadali2023transferCDC} reviewed previously.

    For the same class of dynamical systems considered by~\citet{11036788}, \citet{11312124} propose a data-driven framework for safety verification with deterministic correctness guarantees. Relying on multiple system trajectories, a family of monotone basis functions is constructed that remains non-increasing along all trajectories. Using this class of basis functions, a sampling-based optimization approach is developed to synthesize BCs and verify system safety with deterministic guarantees, without relying on Lipschitz continuity assumptions.
    
   Most of the existing data-driven literature based on structural property-based methods focuses on constructing \emph{quadratic} CBCs from sufficiently rich data for \emph{specific classes} of nonlinear systems. Motivated by these developments, we propose the following research avenue as a promising direction for future work.
    
    \begin{resp}
    	   \begin{openproblem}
       Given the general nonlinear system $\mathcal{S}_g$ in Definition~\ref{def:dt-GNS}, develop a data-driven approach based on a single non-i.i.d. trajectory of the system to synthesize a more general non-quadratic CBC together with its safety controller while simultaneously enforcing input constraints.
    	\end{openproblem}
    \end{resp}

    Most of the studies grounded in the notion of CBCs surveyed in this subsection primarily provide infinite-horizon safety guarantees, \emph{i.e.}, $c = 0$ is considered in condition~\eqref{eq:cbc3} (cf.~Remark~\ref{rem:c0}). This implies that the value of a CBC should be non-increasing along system trajectories, which may introduce conservatism and become restrictive in certain practical scenarios. As discussed in Section~\ref{Subsec: MB-FCA} and motivated by this challenge, the literature proposes the use of \(k\)-inductive CBCs (or \(k\)-inductive BCs for safety verification), which relax this requirement. We review this notion in the subsequent subsection.

    \subsection{\(k\)-Inductive (Control) Barrier Certificates}\label{Subsec:data_KCBCs}
    To highlight the differences between conventional CBCs and their \(k\)-inductive counterparts, we introduce \(k\)-inductive control barrier certificates in the following definition, adapted from the work by~\citet{anand2021safetyCDC}.

    \begin{definition}\label{def:k_inductive_cbc}
        Consider the system $\mathcal{S}_g$ as in Definition~\ref{def:dt-GNS}, with the initial set $X_0 \subset X$ and the unsafe set $X_u \subset X$. A function $\mathds{B}_{\mathrm{k}} : X \to \R$ is considered as a \(k\)-inductive control barrier certificate ($k$-CBC) for the system $\mathcal{S}_g$ if there exist $k \in \Np$, $c_{\mathrm{k}} \in \Rpz$, and $\eta_{\mathrm{k}}, \gamma_{\mathrm{k}} \in \R$, with $ \eta_{\mathrm{k}} + (k - 1) c_{\mathrm{k}} < \gamma_{\mathrm{k}}$, such that
         \begin{subequations}\label{eq:KIcbc}
             \begin{itemize}
                 \item $\forall x \in X_0$:
                 \begin{align}
                     \mathds{B}_{\mathrm{k}} (x) \leq \eta_{\mathrm{k}},\label{eq:KIcbc1}
                 \end{align}
                 \item $\forall x \in X_u$:
                 \begin{align}
                    \mathds{B}_{\mathrm{k}} (x) & \geq \gamma_{\mathrm{k}}, \label{eq:KIcbc2}
                 \end{align}
                 \item $\forall x \in X, \; \exists u \in U,$ such that
                 \begin{align}
                     &\mathds{B}_{\mathrm{k}}(f(x, u)) \leq \mathds{B}_{\mathrm{k}}(x) + c_{\mathrm{k}}, \label{eq:KIcbc3}\\
                     & \mathds{B}_{\mathrm{k}}(f^k(x, u)) \leq \mathds{B}_{\mathrm{k}}(x). \label{eq:KIcbc4}
                 \end{align}
             \end{itemize}
         \end{subequations}
    \end{definition}
    
    Before comparing CBCs and $k$-CBCs, we first present the following theorem, which provides safety guarantees based on the notion of $k$-CBCs~\citep{anand2021safetyCDC}.
    
     \begin{theorem}\label{thm:safety_KIcbc}
    	Given $\mathcal S_g$ in Definition~\ref{def:dt-GNS}, let $\mathds{B}_{\mathrm{k}}$ be its $k$-CBC in the sense of Definition~\ref{def:k_inductive_cbc}. Then, $\mathcal S_g$ is safe over an infinite-time horizon, meaning that all state trajectories originating from $X_0$ remain outside the unsafe set $X_u$ for all time (i.e., $x_{x_0u}(k) \notin X_u$ for all $x_0 \in X_0$ and all $k \in \N$), under the control input $u$ corresponding to $k$-CBC $\mathds{B}_{\mathrm{k}}$.
    \end{theorem}
    
    As is evident, the most important distinction between CBCs and $k$-CBCs is that the latter can ensure infinite-horizon safety guarantees even when $c_{\mathrm{k}} \neq 0$, whereas the former requires $c = 0$ to provide such guarantees. In this sense, the $k$-CBC notion can be viewed as a relaxation of the conventional CBC. More concretely, to obtain infinite-horizon safety guarantees via CBCs, the value of the CBC should be non-increasing along system trajectories. In contrast, $k$-CBCs relax this requirement by permitting up to $k-1$ one-step increases, while enforcing a non-increasing property over every $k$ consecutive steps. We note that when $k = 1$ and $c_{\mathrm{k}} = 0$, the notion of $k$-CBCs coincides with that of CBCs with $c = 0$. We note that while $k$-CBCs can be leveraged to design safety controllers, $k$-inductive barrier certificates ($k$-BCs) can be employed for safety verification (analogous to the distinct usage of BCs and CBCs).
    
    Despite the advantages offered by $k$-CBCs, their synthesis is generally more demanding than that of conventional CBCs, particularly in data-driven settings where the system dynamics are (partially) unknown. In particular, as in the case of CBCs, the unknown dynamics appear in condition~\eqref{eq:KIcbc3}, but this is not the only challenge. More importantly, due to the structure of the left-hand side in condition~\eqref{eq:KIcbc4}, the problem of synthesizing $k$-CBCs becomes inherently nonconvex, which makes the development of data-driven synthesis approaches particularly challenging. Accordingly, only a limited number of works have addressed the data-driven synthesis of $k$-CBCs. 
    
    Within this context, \citet{9800956} propose a data-driven approach based on an i.i.d. sampling scheme for synthesizing $k$-BCs for discrete-time dynamical systems, providing probabilistic out-of-sample performance guarantees. In particular, for a prescribed confidence level and under certain Lipschitz continuity assumptions, \citet{9800956} formulate a scenario-based program whose solution, grounded in the notion of $k$-BCs, yields probabilistic safety guarantees.
    To move beyond safety verification and enable safety controller synthesis using the notion of $k$-CBCs, \citet{wooding2024learningKCBC} exploit structural properties of the system and propose a data-driven framework for a class of discrete-time input-affine nonlinear dynamical systems (with matched nonlinearities). Specifically, the approach relies on noise-free input--state data collected along a single trajectory to construct a data-parameterized representation of the system, which is subsequently employed for controller design, yielding deterministic out-of-sample performance guarantees (cf. Section~\ref{Subsec: Structural Approach}).
    
    From the above review, it follows that the literature on data-driven synthesis of $k$-BCs and $k$-CBCs remains relatively limited, leaving several promising directions for future research, including the treatment of noise-corrupted data, which is formally stated below.
    
    \begin{resp}
    	\begin{openproblem}\label{OP1}
    		Given the system $\mathcal{S}_g$ as in Definition~\ref{def:dt-GNS}, develop a data-driven approach for synthesizing safety certificates, using the notion of $k$-CBCs, that can accommodate noise-corrupted data while providing formal out-of-sample performance guarantees.
    	\end{openproblem}
    \end{resp}
    
    We note that Research Avenue~\ref{OP1} represents only one possible direction for future research in this area; other potential avenues include accommodating systems subject to process disturbances.
    
     In the next subsection, we review another class of functional certificates that is more naturally suited to properties beyond safety, such as liveness and \(\omega\)-regular specifications~\citep{baier2008principles}.
    
    \subsection{(Control) Closure Certificates}
    In this section, we first aim to clarify the main ideas and key features behind closure certificates (CCs). For simplicity of exposition, we restrict the discussion to BCs and CCs. Broadly speaking, BCs are natural and effective tools for safety verification as they are constructed to characterize inductive state invariants, \emph{i.e.}, sets of states that contain all reachable behaviors while remaining disjoint from unsafe ones. While this state-based viewpoint has also been extended to richer temporal objectives, its use beyond safety, \emph{e.g.}, for liveness or \(\omega\)-regular properties, is inherently conservative~\citep{murali2024closure}. A typical example of such properties is the requirement that a system visit a given set of states infinitely often.
    
    Although safety can be established via an inductive argument, liveness requires the development of a well-foundedness reasoning~\citep{cook2009principles}. In this context, \citet{1319598} introduce the notion of transition invariants as a set of pairs of states such that the second state may be reachable from the first, thereby providing a superset of the transitive closure of the transition relation governing the system evolution. Leveraging these transition invariants, one can develop a well-foundedness argument to refute liveness properties in the context of proving program termination~\citep{1319598}. Moreover, it is demonstrated by~\citet{1319598} that how transition invariants can be utilized to verify programs against \(\omega\)-regular properties, which form an expressive and well-behaved class of formal specifications capturing linear-time properties.
    
    Inspired by this perspective, \citet{murali2024closure} introduce the notion of CCs as functional transition invariants for the verification of dynamical systems against \(\omega\)-regular properties.
    Formally, a CC is a real-valued function defined over pairs of states of the system. Consider the states \(x\), \(x^{\prime}\), and \(x^{\prime\prime}\), with \(x^{\prime}\) being the immediate successor of \(x\). As a base case, the CC is required to be nonnegative for every pair \((x, x^{\prime})\). Moreover, if the CC is nonnegative for the pair \((x^{\prime}, x^{\prime\prime})\), then it should also be nonnegative for the pair \((x, x^{\prime\prime})\). Taken together, these conditions ensure that whenever the state \(x^{\prime\prime}\) is reachable from the state \(x\), the CC takes a nonnegative value on the pair \((x, x^{\prime\prime})\). Additional constraints can then be imposed on this construction in order to verify safety, refute liveness, or establish \(\omega\)-regular properties of interest~\citep{nadali2024neuralAAAI}.
    We note that, whereas CCs are suited to verifying whether a system satisfies a given \(\omega\)-regular property, control closure certificates can be employed for controller synthesis to enforce such properties.
    
    Despite the advantages offered by CCs, their synthesis remains challenging even in the model-based setting; it becomes significantly more demanding when the system models are unknown. This problem has been addressed in two recent studies by~\citet{nadali2024neuralAAAI} and~\citet{11312799}. More specifically, \citet{nadali2024neuralAAAI} propose a data-driven approach tailored to discrete-time unknown dynamical systems, in which a neural network is trained to represent a CC for the unknown system. The proposed approach relies on grid-based sampling (cf. Fig.~\ref{subfig:Lip-based_data}) and exploits certain Lipschitz continuity conditions to ensure the correctness of the learned CC beyond the observed data. For the same class of systems, \citet{11312799} propose a data-driven approach for constructing CCs, offering either PAC guarantees based on i.i.d. sampling (cf. Section~\ref{Subsec: Scenario Approach}) or deterministic correctness guarantees via grid-based sampling, combined with certain Lipschitz continuity assumptions (cf. Section~\ref{Subsec: Lipschitz Approach}). Notably, the former type of guarantee requires fewer samples compared with the approach by~\citet{nadali2024neuralAAAI}.
    
    The majority of the existing data-driven approaches, surveyed in this paper, assume access to \emph{full-state} information. In practice, however, many systems are only partially observable, with measurements available from a limited set of sensors rather than the complete system state. Motivated by this practical scenario, we propose the following research avenue.
    
    	\begin{resp}
    	\begin{openproblem}
        For a dynamical system evolving in either discrete or continuous time, develop a data-driven scheme for constructing either (in)finite abstractions or functional certificates using only input--output data collected from a partially observed system, without requiring access to the full system state.
    	\end{openproblem}
    \end{resp}
    
    To conclude this section, it is worth noting that the data-driven functional certificate approaches surveyed here, as well as the data-driven abstraction-based methods reviewed in Section~\ref{Sec: Deterministic Setting_ABA}, are primarily tailored to single dynamical systems, often of relatively low dimension, and are therefore not directly applicable to large-scale interconnected networks comprising numerous subsystems. This crucial observation naturally raises a valid question: how can such data-driven approaches be extended to accommodate large-scale interconnected networks? In the subsequent section, we focus on solution paradigms proposed in the literature to address this challenge.
	
	\section{Deterministic Setting: Data-Driven Compositional Techniques}\label{Sec: Deterministic Setting_CT}
	In this section, we provide a comprehensive overview of data-driven abstraction-based and functional-certificate-based approaches that can be applied to large-scale interconnected networks. Before reviewing such approaches, we first clarify that why the data-driven studies surveyed previously cannot be directly applied to such networks, thereby highlighting the primary challenges\footnote{We remark that even in the model-based setting, these approaches cannot be directly applied to large-scale networks due to the substantial computational complexity related to high dimensionality.}. We first focus on the first two general data-driven methodologies discussed in Sections~\ref{Subsec: Scenario Approach} and~\ref{Subsec: Lipschitz Approach}. Recall that these two data-driven frameworks rely on i.i.d. sampling and grid-based sampling, respectively. Consequently, both approaches require multiple system initializations, and as the system dimension increases, the number of decision variables to be designed grows (at least linearly). Consequently, applying the i.i.d. sampling framework to large-scale networks would require collecting a substantially large amount of data, which becomes extremely burdensome due to the need for multiple system initializations.
	
	This issue becomes even more challenging when employing approaches that utilize certain Lipschitz continuity conditions (cf. Section~\ref{Subsec: Lipschitz Approach}). Such approaches typically exhibit \emph{exponential} sample complexity with respect to the system dimensions. Consequently, if these methods were to be applied to large-scale networks, the number of required samples, and thus the number of system initializations, would grow prohibitively large, essentially rendering their application to high-dimensional systems impractical.
	
	We further highlight that both i.i.d. sampling and grid-based sampling frameworks, whether used in data-driven abstraction-based or functional-certificate-based methodologies, typically require solving linear programs (\emph{e.g.}, the SCP~\eqref{eq:General_SCP}), whose number of constraints grows with the number of samples. While linear programs are generally scalable, in the case of large-scale interconnected networks, the resulting optimization problems may become computationally intractable due to the excessive number of constraints. Collectively, these considerations indicate that the aforementioned data-driven frameworks cannot be directly applied to networks composed of a large number of subsystems.
	
	In contrast, the third general data-driven framework in Section~\ref{Subsec: Structural Approach}, which exploits non-i.i.d. time-series data collected from the system during a single finite-time experiment (cf.~the data in~\eqref{eq:data_single_trajectory}), allows for the collection of large amounts of data from the entire network without requiring multiple system initializations. However, for large-scale networks, the primary challenge shifts to the solution paradigm, as such approaches typically involve solving data-dependent SOS optimization programs (for systems with polynomial dynamics) or data-dependent semidefinite programs (SDPs). It is well known that SOS-based methods do not scale well and can rarely be applied to systems with more than $10$ state variables. Although SDP-based approaches are generally more scalable than SOS-based ones, they also tend to become computationally intractable for systems with more than a few hundred state variables~\citep{8619019}. Collectively, these observations indicate that even the third general data-driven framework faces severe scalability limitations and thus cannot be directly applied to large-scale interconnected networks.
	
	To address this scalability challenge and extend the previously reviewed data-driven abstraction-based and functional-certificate-based approaches to large-scale interconnected networks, compositional techniques can be employed. Broadly speaking, these techniques decompose an interconnected network into smaller subsystems, collect samples at the subsystem level, and establish data-dependent conditions under which analyses and out-of-sample performance guarantees obtained for the individual subsystems can be systematically lifted to the entire network.
	More specifically, abstraction-based compositional methods aim to obtain an (in)finite abstraction of the entire network by first building each subsystem's abstraction and establishing a formal behavioral relation between every subsystem and its abstraction (\emph{e.g.}, see Definition~\ref{def:SFs-deterministic}) in a data-driven manner. These local data-driven abstractions and relations are then combined to obtain a global abstraction for the network together with its overall behavioral relation. In the compositional variant of functional certificate approaches, the objective is instead to construct a global functional certificate (\emph{e.g.}, a CBC as in Definition~\ref{def:cbc}) and a controller for the entire network by systematically combining the data-driven certificates and local controllers of the subsystems.
	As noted in Section~\ref{Subsec:A-BT}, compositional techniques mainly rely on small-gain arguments~\citep{dashkovskiy2010small,mironchenko2023input} or dissipativity-based reasoning~\citep{arcak2016networks}. 
	
	We now elaborate on how each compositional reasoning operates. To this end, consider a large-scale network composed of $\mathscr{N} \in \Np$ subsystems. Here, we denote a subsystem by $\mathcal{S}_{l_i}$, where the subscript $i \in \{1, \ldots, \mathscr{N}\}$ represents the $i^{\text{th}}$ subsystem. 
	Since subsystems within an interconnected network influence one another, the dynamics in~\eqref{eq:LTI_model} should be modified as
	\begin{align}
		\mathcal{S}_{l_i} : x_i^+ = A_i x_i + B_i u_i + D_i w_i, \label{eq:LTI_model_net}
	\end{align}
	where the term $D_i w_i$ captures the influence of the other subsystems on $\mathcal{S}_{l_i}$. In particular, $D_i$ denotes the internal input matrix, and $w_i$ represents the internal input, \emph{i.e.}, the input capturing the effect of other subsystems within the network.
	Notice that the definitions of unknown matrices $A_i$ and $B_i$, as well as the state vector $x_i$ and control input $u_i$ with their corresponding sets, remain analogous to those in Definition~\ref{def:LTI-system}, but are now specified for each subsystem individually. 
	
	One of the key differences between small-gain and dissipativity-based arguments lies in how the interconnection is represented. Specifically, within the small-gain reasoning, $D_i$ and $w_i$ are typically partitioned as
	\begin{subequations}\label{eq:SGR}
		\begin{align}
			D_i &=
			[D_{i1} \; \ldots \; D_{i(i-1)} \; D_{i(i+1)} \; \ldots \; D_{i\mathscr{N}}],\label{eq:DSGR}
		\end{align}
		and
		\begin{align}
			w_i &= [w_{i1}; \; \ldots ; \; w_{i(i-1)}; \; w_{i(i+1)}; \; \ldots ; \; w_{i\mathscr{N}}]	, \label{eq:IIC_small}
		\end{align}
	\end{subequations}
	where $D_{ij} \in \R^{n_i \times n_j}$ and $w_{ij} = x_j$ if subsystem $j$ influences subsystem $i$; otherwise, $D_{ij} w_{ij} \equiv \Zero_{n_i}$ ($w_{ij} = \Zero_{n_j}$ and $D_{ij} = \Zero_{n_i \times n_j}$). Consequently,
	$D_i \in \R^{n_i \times \zeta_i}$ and $w_i \in W_i \subset \R^{\zeta_i}$, with $W_i$ being the internal input set and $\zeta_i = \sum_{j=1,\, j \neq i}^{\mathscr{N}} n_j$.
	
	Within the dissipativity-type compositional reasoning, however, the partitioning is carried out in a different manner. In particular, under this framework, the internal inputs are constructed as
	\begin{align}
		[w_1 ; \; \ldots ; \; w_{\mathscr{N}}]
		=
		\mathds{M} \, [x_1 ; \; \ldots ; \; x_{\mathscr{N}}] , \label{eq:IIC_dissi}
	\end{align}
	where $\mathds{M} = \{\boldsymbol{\mathrm{m}}_{ij}\}$, $i,j \in \{1, \ldots, \mathscr{N}\}$, with $\boldsymbol{\mathrm{m}}_{ij} \in \R^{\bar \zeta_i \times n_j}$, denotes the coupling block matrix that characterizes the interactions among the subsystems and thus captures the network topology. It follows from this construction of the internal inputs that, in this case, $D_i \in \R^{n_i \times \bar \zeta_i}$ and $w_i \in \bar W_i \subset \R^{\bar \zeta_i}$, with $\bar W_i$ denoting the internal input set, and $\bar \zeta_i \in \Np$.
	
	To elaborate on how an interconnected network is formed, we use, with a slight abuse of notation, the same symbols as in Definition~\ref{def:LTI-system}, emphasizing that they now refer to the network.
	Accordingly, given $\mathscr{N} \in \Np$ subsystems $\mathcal{S}_{l_i}, i \in \{1, \ldots, \mathscr{N}\}$, with the internal input configuration either as in~\eqref{eq:IIC_small} or in~\eqref{eq:IIC_dissi}, their interconnection forms the interconnected network $\mathcal{S}_n$, described by
	\begin{align*}
		\mathcal{S}_n : x^+ = Ax + Bu,
	\end{align*}
	where $x = [x_1; \ldots; x_\mathscr{N}] \in X \coloneq \prod_{i=1}^{\mathscr{N}} X_i$ and $u = [u_1; \ldots; u_\mathscr{N}] \in U \coloneq \prod_{i=1}^{\mathscr{N}} U_i$ are the state and control input of the network, with $X$ and $U$ denoting the state and control input sets. Moreover, $A \in \R^{n \times n}$, with $n \coloneq \sum_{i=1}^{\mathscr{N}} n_i$, is a block matrix, with diagonal blocks $(A_1, \ldots, A_{\mathscr{N}})$ and off-diagonal blocks $D_{ij}$, $i,j \in \{1, \ldots, \mathscr{N}\}, i \neq j$, in the case of partitioning based on small-gain reasoning (see~\eqref{eq:SGR}), or with diagonal blocks of $(A_1 + D_1 \boldsymbol{\mathrm{m}}_{11}, \ldots, A_{\mathscr{N}} + D_{\mathscr{N}} \boldsymbol{\mathrm{m}}_{\mathscr{N}\mathscr{N}})$ and off-diagonal blocks of $D_i \boldsymbol{\mathrm{m}}_{ij}$ in the case of partitioning based on dissipativity-based reasoning (see~\eqref{eq:IIC_dissi}), and $B = \mathsf{blkdiag}(B_1,\ldots,B_\mathscr{N}) \in \R^{n \times m}$, with $m \coloneq \sum_{i=1}^{\mathscr{N}} m_i$.
	
	Since each subsystem is influenced by other (neighboring) subsystems, this interaction should be explicitly taken into account when constructing (in)finite abstractions of the subsystems or synthesizing functional certificates for them. To illustrate this, consider the scenario in which one aims to synthesize a CBC, as in Definition~\ref{def:cbc}, for the subsystem $\mathcal{S}_{l_i}$ in~\eqref{eq:LTI_model_net}. In this case, an additional term should be included on the right-hand side of condition~\eqref{eq:cbc3} to capture the influence of the internal input $w_i$. The specific form of this term differs depending on whether small-gain or dissipativity-based reasoning is employed. 
	
	In particular, under small-gain reasoning, the additional term takes the form $\xi_i \Vert w_i \Vert^2$, where $\xi_i \in \Rp$, typically referred to as the interaction gain. This quantity captures the effect of the other subsystems on the $i^{\text{th}}$ subsystem for all $i \in \{1, \ldots, \mathscr{N}\}$ and essentially provides a degree of robustness against the internal input $w_i$. In contrast, under dissipativity-based compositional reasoning, this term is incorporated as
\begin{align*}
	\begin{bmatrix}
		w_i\\
		x_i
	\end{bmatrix}^{\!\top}\underbrace{\begin{bmatrix}
	\mathds{Z}^{11}_i & \mathds{Z}^{12}_i\\
	\mathds{Z}^{21}_i & \mathds{Z}^{22}_i
	\end{bmatrix}}_{\mathds{Z}_i}
	\begin{bmatrix}
		w_i\\
		x_i
	\end{bmatrix} \! \! ,
	\end{align*}
    \vspace{-0.35cm}
    \newline
	where
	\(
	\mathds{Z}_i
	\)
	is a symmetric block matrix that should be designed.
	Notice that similar modifications are also required in abstraction-based approaches. For instance, in the right-hand side of condition~\eqref{eq:con2-def-SF-discrete}, when employing small-gain reasoning, the term $\xi_i \Vert w_i - \hat{w}_i \Vert^2$ should be included, where $\hat{w}_i$ denotes the internal input of the abstract subsystem.
	
	Subsequently, in the small-gain framework, once the interaction gains $\xi_i$ are computed in a data-driven manner, a gain interconnection matrix is constructed. If the spectral radius of such a matrix is less than one~\citep{dashkovskiy2010small}, the safety certificates and the associated controllers designed for the individual subsystems can be lifted to the entire network. In the dissipativity-based framework, on the other hand, a matrix incorporating both the network topology captured by $\mathds{M}$ and the matrices $\mathds{Z}_i$ for all $i \in \{1, \ldots, \mathscr{N}\}$ is constructed. If such a matrix is negative semidefinite~\citep{arcak2016networks}, then the results obtained for the subsystems can be extended to the whole network.
	
	\subsection{Literature on Data-Driven Compositional Techniques}
	
	In the context of data-driven compositional techniques for abstraction-based approaches, \citet{lavaei2023symbolicCDCDATA} proposes a compositional framework grounded in small-gain reasoning to construct complete finite abstractions of large-scale interconnected networks while providing probabilistic out-of-sample performance guarantees. In particular, the proposed framework relies on i.i.d. sampling and exploits certain Lipschitz continuity conditions to eliminate the violation parameter; however, the resulting guarantee still involves a confidence level due to the i.i.d. sampling (cf. Section~\ref{Subsec: Lipschitz Approach}).
    Furthermore, \citet{samari2024dataNetABSTTAC} present a compositional methodology for constructing complete finite abstractions of large-scale interconnected networks, providing deterministic correctness guarantees by exploiting Lipschitz continuity conditions and relying on grid-based sampling (cf. Section~\ref{Subsec: Lipschitz Approach}). Importantly, unlike the study by~\citet{lavaei2023symbolicCDCDATA}, the proposed approach does not require knowledge of the network topology or the use of classical small-gain conditions.
	
	Despite these advancements, these approaches exhibit exponential sample complexity at the subsystem level when establishing formal out-of-sample performance guarantees. This limitation is characteristic of data-driven approaches that rely on Lipschitz continuity conditions, as discussed earlier. It is worth noting that no data-driven study has yet addressed the construction of infinite abstractions for large-scale interconnected networks, which therefore remains an open challenge.
	
	\begin{resp}
		\begin{openproblem}\label{OP3}
			Consider a large-scale interconnected network consisting of $\mathscr{N}$ subsystems. Develop a data-driven compositional framework for constructing an infinite abstraction of the network, {i.e.}, a ROM of the network, together with a formal SF-based relation between the ROM and the network, while providing out-of-sample performance guarantees.
		\end{openproblem}
	\end{resp}
	
	Moving to data-driven compositional techniques for functional certificate approaches, based on dissipativity reasoning, \citet{noroozi2021dataLCSS} present a data-driven framework for verifying the safety of interconnected networks, using the notion of BCs, with probabilistic guarantees. In particular, the approach relies on i.i.d. sampling and exploits Lipschitz continuity conditions to eliminate the violation parameter, thereby providing safety guarantees for the network that involve only a confidence level (cf. Section~\ref{Subsec: Lipschitz Approach}). In addition, \citet{lavaei2023compositionalTACStochandDeterministic} and \citet{lavaei2022formalCNCStochandDeterministic} develop data-driven frameworks based on dissipativity and small-gain reasoning, respectively, to verify the safety of large-scale networks including autonomous vehicles, while providing probabilistic safety guarantees that involve only confidence levels.
	
	While the previous data-driven studies assume that the number of subsystems is known a priori, the problem of data-driven synthesis of safety certificates through the notion of BCs has also been investigated for infinite networks (\emph{i.e.}, extremely large networks with no fixed upper bound on the number of subsystems). In particular, leveraging Lipschitz continuity conditions and grid-based sampling, \citet{aminzadeh2024compositionaldataECC} and \citet{zaker2025datasafetyautomatica} propose data-driven approaches for constructing BCs for infinite networks while providing deterministic safety guarantees at the network level. Importantly, while~\citet{aminzadeh2024compositionaldataECC} assume that the interconnection topology is known, an assumption also adopted by~\citet{noroozi2021dataLCSS,lavaei2023compositionalTACStochandDeterministic} and~\citet{lavaei2022formalCNCStochandDeterministic}, the work by~\citet{zaker2025datasafetyautomatica} does not require such knowledge, which is particularly suitable for data-driven settings, where requiring less prior knowledge about the network is preferable.
	
    Despite these data-driven advancements, such approaches exhibit \emph{exponential} sample complexity at the subsystem level, motivating the development of frameworks that overcome this limitation. Moreover, they are primarily limited to safety verification for networks and do not address the synthesis of safety controllers. Motivated by these challenges, \citet{akbarzadeh2025formalCBCTAC}  and \citet{akbarzadeh2024dataDissAUTOMATICA} exploit structural properties of networks, composed of input-affine nonlinear subsystems with polynomial dynamics, and derive data-parameterized representations for the subsystems (cf. Section~\ref{Subsec: Structural Approach}). These representations are subsequently utilized to design CBCs and the corresponding safety controllers for each subsystem. Then, through small-gain compositional reasoning, the obtained CBCs and safety controllers are lifted to the network level, resulting in a global CBC and its associated safety controller for the entire network, thereby providing deterministic infinite-horizon safety guarantees. 
	We note that this study requires a single set of input--state data collected from each subsystem, which contrasts with frameworks based on Lipschitz continuity conditions that require multiple subsystem initializations. Moreover, the results by~\citet{akbarzadeh2025formalCBCTAC} can accommodate noise-corrupted data, which previous data-driven compositional studies cannot handle.
	
	\begin{resp}
		\begin{openproblem}
			For a large-scale network composed of $\mathscr{N}$ subsystems, develop a data-driven compositional framework for constructing either abstractions or functional certificates, where the interconnection topology is not only unknown but also dynamic.
		\end{openproblem}
	\end{resp}
	
	Before concluding this section, we note that data-driven compositional methodologies have also been explored in the context of stability analysis for large-scale interconnected networks; see, \emph{e.g.}, the studies by~\citet{zaker2025dataHSCC,samari2025dataSMCTAC,zaker2025dataAUTOMATICA,11347537,lavaei2023dataISSLCSS} and~\citet{nakano2025dissipativityarXiv}. However, since the primary focus of this survey is on formal verification and synthesis with respect to specifications beyond stability, we do not review those studies.
	
    Having reviewed data-driven approaches for formal verification and policy synthesis in deterministic settings, we now turn to their stochastic counterparts. Before doing so, it is worth emphasizing that many of the fundamental ideas developed for deterministic systems, including abstraction-based reasoning, functional certificates such as barrier certificates, and compositional arguments, remain central in the stochastic setting as well. The main distinction is that formal guarantees on system behavior are now expressed in terms of the probability measures induced by the stochastic dynamics. For instance, the satisfaction of a desired specification is quantified as the probability that the corresponding system trajectories satisfy that specification. This notion of probability should not be confused with the confidence levels that arise in data-driven finite-sample guarantees, which instead quantify the reliability of conclusions drawn from finite datasets rather than the likelihood of satisfying a given specification.
	
	\section{Stochastic Setting: Data-Driven Abstraction-based Approaches}\label{Sec: Stochastic Setting_ABA}
	This section provides a comprehensive review of abstraction-based techniques for stochastic dynamical systems. The primary focus is on the construction of finite abstractions, given the extensive body of literature on this topic. To this end, we first introduce the class of discrete-time stochastic dynamical systems considered in this section.
	
	\begin{definition}\label{def:sys_stoch_barrier}
		A discrete-time stochastic dynamical system $\mathcal{S}_{sg}$ evolves according to
		\begin{align}
			\mathcal{S}_{sg} : x^+ = f(x, u, \varsigma), \label{eq:gen_disc_org_stoch}
		\end{align}
		where $x \in X$ denotes the system state, $u \in U$ represents the control input, and $\varsigma$ is a sequence of i.i.d. random variables from a sample space $\Omega$ to the measurable space $(\mathds{V}_\varsigma, \mathds{F}_{\varsigma})$:
		\[
		\varsigma \coloneq \{\varsigma(k) : (\Omega, \mathds{F}_{\Omega}) \to (\mathds{V}_\varsigma, \mathds{F}_{\varsigma}), \quad k \in \N\}.
		\]
		Moreover, $X \subset \R^n$ is the Borel state space of the system, $U \subset \R^m$ is the Borel input space of the system, and $f : X \times U \times \mathds{V}_\varsigma \to \R^n$ is an unknown measurable function characterizing the state evolution of the system. We denote by $\mathbb{U}$ the collection of input sequences $\{ u(k) : \Omega \to U, \; k \in \N \}$, in which $u(k)$ is independent of $\varsigma(z)$ for any $k, \, z \in \N$ and $z \geq k$. For any initial state $x_0 \coloneq x(0)\in X$ and input sequence $u\in\mathbb{U}$, we denote by $\{x_{x_0u}(k)\}_{k\in\mathbb{N}}$ the solution process of $\mathcal{S}_{sg}$ corresponding to the initial state $x_0$ and the input sequence $u$.
	\end{definition}
	
	Having introduced the system under consideration, we first elaborate on the construction of its finite abstraction. For clarity of exposition, we first present the procedure intuitively and then formalize it. To do so, let us first carry out the discussion in the model-based setting and then highlight the associated challenges and outline potential solution frameworks for transitioning to data-driven settings. Similar to the deterministic setting discussed in Section~\ref{Subsec:FA-data}, the construction of a finite abstraction begins by partitioning the state and input spaces into finitely many regions, denoted by \(X = \cup_i \mathsf{X}_i\) and \(U = \cup_i \mathsf{U}_i\), respectively. We then select representative points \(\hat{x}_i \in \mathsf{X}_i\) and \(\hat{u}_i \in \mathsf{U}_i\), which serve as the discrete (\emph{i.e.}, abstract) states and control inputs. Similar to Section~\ref{Subsec:FA-data}, we denote the finite abstraction of the system \(\mathcal{S}_{sg}\) by \(\hat{\mathcal{S}}_{sg}\). Accordingly, the sets of discrete states and inputs are defined as \(\hat{X} \coloneq \{\hat{x}_i \mid i = 1, \ldots, n_{\hat{x}}\}\) and \(\hat{U} \coloneq \{\hat{u}_i \mid i = 1, \ldots, n_{\hat{u}}\}\), respectively.
	
	\begin{figure}[t!]
		\centering
		\includegraphics[width=0.7\linewidth]{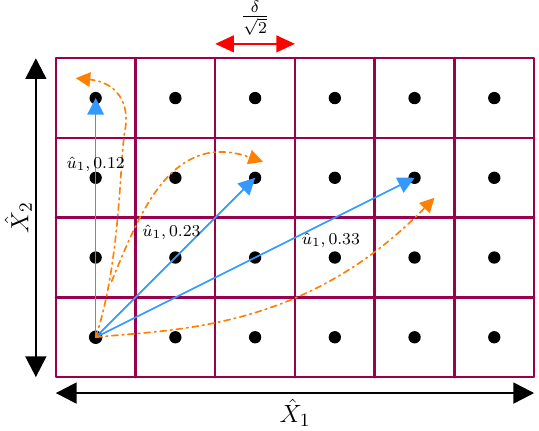}
		\caption{Construction of a finite MDP for a stochastic dynamical system: the state and input spaces are partitioned into finitely many cells (with $\delta$ denoting the state discretization parameter), representative points are selected, and transition probabilities between cells are computed for each state–input pair; the figure only illustrates selected transitions.}
		\label{fig:FA_Stoch}
	\end{figure}
	
	One of the main differences in constructing a finite abstraction of a stochastic system, compared with a deterministic one, is that, due to inherent randomness in the dynamics, the successor state may fall into different partitions even when starting from the same discrete state and applying the same discrete input. Accordingly, one should compute the probability of transitioning to each partition given a discrete state and input, and, by repeating this for all state–input pairs, obtain a transition probability matrix that defines a finite Markov decision process (MDP)~\citep{lavaei2022automated}; see also Fig.~\ref{fig:FA_Stoch}. It is worth noting that the procedure for constructing finite abstractions remains the same for stochastic dynamical systems without control inputs, with the key distinction that the resulting model reduces to a finite Markov chain (MC), rather than an MDP, since transitions depend only on the state and not on inputs~\citep{lavaei2022automated}.
		
\begin{figure}[t!]
		\centering
		\includegraphics[width=0.7\linewidth]{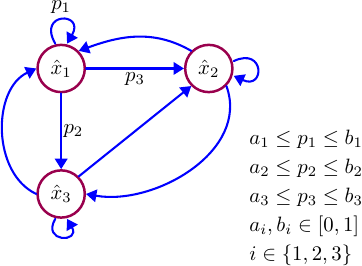}
		\caption{Illustration of an IMC.}
		\label{fig:IMC}
	\end{figure}

    Another approach to constructing finite abstractions of a stochastic system is to build interval Markov decision processes (IMDPs), where lower and upper bounds on transition probabilities are computed instead of exact values. While the construction of IMDPs is generally more computationally expensive than that of finite MDPs, as both lower and upper bounds on transition probabilities should be computed, IMDPs offer notable advantages, particularly when enforcing infinite-horizon specifications~\citep{lavaei2022automated}. Similar to the above discussion, in the absence of control inputs, IMDPs reduce to interval Markov chains (IMCs), an example of which is depicted in Fig.~\ref{fig:IMC}.
	
	Akin to the deterministic setting, finite abstractions of stochastic systems, whether represented as finite MDPs or IMDPs, can be employed for the verification and enforcement of complex specifications. More precisely, they enable the quantification of the probability of satisfying a specification of interest and, when control inputs are available, the synthesis of policies that maximize this probability, typically via dynamic programming techniques based on Bellman equations~\citep{lavaei2020amytissSpringer}.
	
    We now aim to formalize the above discussion. For the system $\mathcal{S}_{sg}$, its finite MDP can be represented as
	\begin{align}
		\hat{\mathcal{S}}_{sg} : \hat x^+  =  \hat f(\hat x, \hat u, \varsigma),\label{eq:FMDP}
	\end{align}
	where $\hat f : \hat X \times \hat U \times \mathds{V}_\varsigma \to \hat X$, with $\hat X$ and $\hat U$ denoting the finite state and input sets of $\hat{\mathcal{S}}_{sg}$. The transition $\hat f$ is characterized as
	\begin{align*}
		\hat f(\hat x, \hat u, \varsigma) = \Pi(f(\hat x, \hat u, \varsigma)),
	\end{align*}
	with $\Pi : X \to \hat X$ denoting the quantization map that assigns to each $x \in X$ a representative point $\hat x \in \hat X$ of the corresponding partition set containing $x$. Similar to the complete abstraction in the deterministic setting, the map $\Pi$ satisfies
	\begin{align*}
		\Vert \Pi(x) - x \Vert  \leq \delta, \quad \forall x \in X,
	\end{align*}
	where $\delta$ is the state discretization parameter, as introduced in Definition~\ref{def:FA_construction}. For simplicity, we assume here that the output of $\mathcal{S}_{sg}$ is the state itself, \emph{i.e.}, $y = Cx$ with $C = \I_n$; the same assumption applies to $\hat{\mathcal{S}}_{sg}$.
	
	Having described the definition of $\hat{\mathcal{S}}_{sg}$, we now elaborate on how to establish a relation between $\mathcal{S}_{sg}$ and $\hat{\mathcal{S}}_{sg}$ that enables the systematic transfer of results derived for $\hat{\mathcal{S}}_{sg}$ to $\mathcal{S}_{sg}$. This relation can be established via the notion of stochastic simulation functions (S-SFs), which serve as the stochastic counterpart of SFs in the deterministic setting, introduced in Definition~\ref{def:SFs-deterministic}~\citep{lavaei2022automated}. For simplicity, we adopt the same notation for S-SFs and their associated parameters as for SFs; however, the intended meaning will be clear from the context.
	
	\begin{definition}\label{def:SSFs}
		Consider a stochastic system $\mathcal{S}_{sg}$ in~\eqref{eq:gen_disc_org_stoch} and its finite MDP $\hat{\mathcal{S}}_{sg}$ in~\eqref{eq:FMDP}.
		A function $\mathds{S} : X \times \hat{X} \to \Rpz$ is
		called a stochastic simulation function (S-SF) from $\hat{\mathcal{S}}_{sg}$ to $\mathcal{S}_{sg}$ if there exist constants $\alpha_s\in \Rp$, $\rho_s, \, \psi_s \in \Rpz$, and $0 < \kappa_s \leq 1 $ such that
		\begin{subequations}
			\begin{itemize}
				\item $\forall x \in  X,\forall \hat x \in \hat{X},$
				\begin{align}
					\alpha_s\Vert x - \hat x\Vert^2\le  \mathds{S}(x,\hat x),\label{eq:con1-def-SSF-discrete}
				\end{align}
				\item $\forall x\in X,\forall\hat x\in\hat{X}, \forall \hat u \in \hat{U}, \exists u \in U,$ such that
				\begin{align}
					\mathds{E}&  \big[ \mathds{S}(f(x, u, \varsigma),  \hat f(\hat x, \hat u, \varsigma)) \mid x, \hat x, u, \hat u \big]\notag \\ & \leq \kappa_s \mathds{S}(x,\hat x) + \rho_s \,\Vert \hat u \Vert^2 + \psi_s. \label{eq:con2-def-SSF-discrete}
				\end{align}
			\end{itemize}
		\end{subequations}
	\end{definition}
	
	Analogous to the notion of SFs, S-SFs establish a (probabilistic) relationship between the output trajectories of two systems, namely $\mathcal{S}_{sg}$ and $\hat{\mathcal{S}}_{sg}$ (recall that here we assume $y = x$ and $\hat y = \hat x$). The expectation operator is applied to the left-hand side of condition~\eqref{eq:con2-def-SSF-discrete} with respect to the stochastic noise, since the future evolution of the system is random, \emph{i.e.}, condition~\eqref{eq:con2-def-SSF-discrete} is required to hold in expectation over one-step transitions.
	
	In order to construct an S-SF, the stochastic system $\mathcal{S}_{sg}$ is required to be endowed with a property analogous to that given in Definition~\ref{def:DISS}. For completeness, we present this property in the following definition~\citep{lavaei2022automated}.
	
	\begin{definition}\label{def:P_DISS}
		The system $\mathcal{S}_{sg}$ is called incrementally input-to-state stable ($\delta$-ISS) if, for some constants $\underline{\alpha}, \, \overline{\alpha} \in \Rp$, $\overline{\rho} \in \Rpz$, and $0 < \overline{\kappa} < 1 $, there exists 
		a function $\mathds{S} : X \times X \to\Rpz$ such that
		\begin{subequations}
			\begin{itemize}
				\item $\forall x, \, x^{\prime}  \in  X,$
				\begin{align}
					\underline{\alpha} \Vert x - x^{\prime} \Vert^2 \le  \mathds{S}(x, x^{\prime}) \leq \overline{\alpha} \Vert x - x^{\prime} \Vert^2,\label{eq:con1-def-P_DISS-discrete}
				\end{align}
				\item $\forall x, \, x^{\prime}  \in  X, \forall  u, \, u^{\prime} \in U,$
				\begin{align}
					\mathds{E}& \big[ \mathds{S}(f(x, u, \varsigma),  f(x^\prime, u^\prime, \varsigma)) \mid x, x^\prime, u, u^\prime \big]\notag \\ & \leq \overline{\kappa} \,  \mathds{S}(x, x^{\prime}) + \overline{\rho}  \,\Vert u - u^{\prime} \Vert^2. \label{eq:con2-def-P_DISS-discrete}
				\end{align}
			\end{itemize}
		\end{subequations}
	\end{definition}
	
    Definition~\ref{def:P_DISS} ensures that the expected distance between the state trajectories, measured by the function $\mathds{S}$, remains bounded with respect to $\overline{\rho} \,\Vert u - u^{\prime} \Vert^2$. This definition is essential for bounding the expected distance between two solution processes, originating from different initial conditions and evolving under different input trajectories, via the notion of S-SFs.
	
	Having introduced S-SFs and the conditions required for their construction, we now present the formal guarantee quantifying the probabilistic closeness between the state trajectories of $\mathcal{S}_{sg}$ and $\hat{\mathcal{S}}_{sg}$~\citep{lavaei2022automated}.
	\begin{theorem}
	 Let $\mathds{S}$ be an S-SF from $\hat{\mathcal{S}}_{sg}$ to $\mathcal{S}_{sg}$. For any input trajectory $\hat u \in \hat{\mathbb{U}}$ that preserves the Markov property of the closed-loop system $\hat{\mathcal{S}}_{sg}$, and for any initial conditions $x_0 \in X$ and $\hat x_0 \in \hat X$, one can construct a corresponding input trajectory $u \in \mathbb{U}$ for $\mathcal{S}_{sg}$ via the interface function associated with $\mathds{S}$, such that
	\begin{align}
		\mathds{P} \Bigl\{  \sup_{0 \leq k \leq T} \Vert  x_{x_0 u}(k) -  \hat x_{\hat x_0 \hat u}(k) \Vert < \upsilon \mid x_0, \hat x_0 \Bigr\} \geq 1 - \varphi, \label{eq:ssf_pb}
	\end{align}
	where, if $0< \kappa_s < 1$:
	\begin{subequations}\label{eq:ssf_error}
		\begin{align}
			\varphi \coloneq \begin{cases} \! 1 \! -\! (1\! -\! \frac{\mathds{S}(x_0, \hat x_0)}{\alpha_s \upsilon^2})(1 \!-\!\frac{\phi_s}{\alpha_s \upsilon^2})^{T} \! , & \!\!\!\! \text {if } \alpha_s \upsilon^2 \! \geq \! \frac{\phi_s}{1-\kappa_{s}}, \\ \! (\frac{\mathds{S}(x_0, \hat x_0)}{\alpha_s \upsilon^2}) \kappa_{s}^{T}\! + \!(\! \frac{\phi_s}{(1-\kappa_{s}) \alpha_s \upsilon^2}\!)(1\! -\! \kappa_{s}^{T}), & \!\!\!\! \text {if } \alpha_s \upsilon^2 \! < \! \frac{\phi_s}{1-\kappa_{s}},\end{cases} \label{eq:ssf_b1}
		\end{align}
		whereas, if $0 < \kappa_{s} \leq 1$:
		\begin{align}
			\varphi \coloneq \frac{\mathds{S}(x_0, \hat x_0) + \phi_s T}{\alpha_s \upsilon^2}, \label{eq:ssf_b2}
		\end{align}
		with $\phi_s \coloneq \rho_s \Vert \hat u \Vert_\infty^2 + \psi_s$ and $\upsilon \in \Rpz$.
	\end{subequations}
	\end{theorem}
	We provide three important remarks on this result. First, although the bound in~\eqref{eq:ssf_b1} is slightly tighter than that in~\eqref{eq:ssf_b2}, the latter is often more attractive in practice, as it allows $\kappa_{s} = 1$, thereby facilitating the satisfaction of condition~\eqref{eq:con2-def-SSF-discrete}. Second, the guarantee provided in~\eqref{eq:ssf_pb} holds over a finite time horizon $[0, T]$, making it particularly suitable for finite-horizon specifications. If $\phi_s = 0$, the guarantee can be extended to the infinite-horizon case, with $\varphi \coloneq \frac{\mathds{S}(x_0, \hat x_0)}{\alpha_s \upsilon^2}$, \emph{i.e.},
	\begin{align}
		\mathds{P} \Bigl\{  \sup_{0 \leq k < \infty} \! \Vert  x_{x_0 u}(k) \! - \!  \hat x_{\hat x_0 \hat u}(k) \Vert \! < \! \upsilon \! \mid \! x_0, \hat x_0 \! \Bigr\} \!\! \geq \! 1 \! - \! \frac{\mathds{S}(x_0, \hat x_0)}{\alpha_s \upsilon^2}. \label{eq:ssf_pb_inf}
	\end{align}
	Third, the guarantees in~\eqref{eq:ssf_pb} and~\eqref{eq:ssf_pb_inf} are specification-free, meaning that, regardless of the specification satisfied by the finite abstraction, these bounds provide a probabilistic guarantee on the closeness between the state (or, more generally, output) trajectories of the two systems, either over a finite-time horizon or over an infinite-time horizon.
	
	To leverage such guarantees in a data-driven setting, one should first construct the finite abstraction from data, which requires computing the transition probability matrix directly from data; this is the first main challenge in developing data-driven abstraction-based approaches for stochastic systems. Moreover, once such an abstraction is constructed, one should still establish an S-SF as in Definition~\ref{def:SSFs} from data, since the system dynamics are unknown. While this difficulty already arises in the deterministic setting, the stochastic nature of the system introduces an additional layer of complexity in the data-driven case due to the expectation operator appearing in~\eqref{eq:con2-def-SSF-discrete}. Addressing this issue typically requires replacing the expectation term with its empirical approximation computed from data. Nevertheless, to preserve formal guarantees, it is necessary to rigorously quantify the discrepancy between the true expectation and its empirical approximation, typically using concentration inequalities such as Chebyshev's inequality~\citep{hernandez2001chebyshev}. This, in turn, introduces an additional probabilistic layer, corresponding to the confidence level under which the bound on the discrepancy between the true expectation and its empirical approximation holds.
	
	The first challenge, \emph{i.e.}, constructing finite (I)MDPs from data, can be addressed using approaches based on Chernoff bounds~\citep{chernoff1952measure} or by employing maximum likelihood estimation methods~\citep{myung2003tutorial} to estimate the parameters of the underlying probability distributions from observed data~\citep{9815301}. We primarily focus on how the second and third challenges can be addressed in a data-driven manner, while stressing that alternative approaches have also been proposed in the literature to tackle the first challenge.
	
    To elaborate on how the literature typically addresses the second and third aforementioned challenges, we present a general solution framework that can be suitably adapted to align with either of the two data-driven approaches described in Sections~\ref{Subsec: Scenario Approach} and~\ref{Subsec: Lipschitz Approach}.
	For simplicity of exposition, we restrict our attention to the verification setting (\emph{i.e.}, neither the system nor its finite abstraction is endowed with control inputs).
	We first aim to formulate the problem of constructing an ROP.
	To this end, we fix a parameterized structure for the S-SF as $\mathds{S}(x,\hat x, \mathrm{q}) = \sum_{j = 1}^{\mathrm{r}} \mathrm{q}_j \mathrm{p}_j (x, \hat x)$, where each $\mathrm{p}_j(x, \hat x)$ is a basis function, and $\mathrm{q} = [\mathrm{q}_1 ~\ldots ~ \mathrm{q}_{\mathrm{r}}]^\top \in \R^{\mathrm{r}}$ denotes the unknown coefficients to be designed.  Accordingly, the construction of the S-SF as in Definition~\ref{def:SSFs} can be formulated as the following robust optimization program (ROP):
	\begin{mini!}|s|[2]<b>
		{[\mu; d]}{\mu}
		{\label{eq:RCP_SSF}}{}
		\addConstraint{ \max_j \{ h_j(x, \hat x, d) \} \! \leq \! \mu,  j \! \in \! \{1, 2\},   \forall x \! \in \! X,  \forall \hat x \! \in \! \hat X }{ \label{eq:RCP_SSF_ST} }
		\addConstraint{d = [\alpha_s;\psi_s;\kappa_{s};\mathrm{q}_1; \mathrm{q}_2;\ldots;\mathrm{q}_{\mathrm{r}}] \in \R^{\mathrm{r}+3}}{ \notag}
		\addConstraint{\alpha_s \in \Rp, \; \mu \in \R, \; \psi_s \in \Rpz, \; \kappa_{s} \in (0,1],}{ \notag}
	\end{mini!}
	where
	\begin{subequations}\label{eq:hSSF}
		\begin{align}
			h_1(x, \hat x, d) & = \alpha_s\Vert x - \hat x\Vert^2  -  \mathds{S}(x,\hat x, \mathrm{q}),\\
			h_2(x, \hat x, d) & = \mathds{E} \big[ \mathds{S}(f(x, \varsigma),  \hat f(\hat x, \varsigma), \mathrm{q}) \mid x, \hat x \big] \! - \! \kappa_s \mathds{S}(x,\hat x, \mathrm{q}) \notag \\ & ~~~ - \psi_s.
		\end{align}
	\end{subequations}
	To address the second challenge, arising from the presence of unknown dynamics in condition~\eqref{eq:con2-def-SSF-discrete}, we collect sample pairs of the form $(x^{z_i}, f(x^{z_i}, \varsigma))$, with each $x^{z_i} \in X$, for all $i \in \{1, \ldots, N\}$, corresponding to a scenario. Consequently, the scenario optimization program (SOP) corresponding to the ROP~\eqref{eq:RCP_SSF} can be formulated as
	\begin{mini!}|s|[2]<b>
		{[\mu; d]}{\mu}
		{\label{eq:SCP_SSF}}{}
		\addConstraint{ \max_j \{ h_j(x^{z_i}, \hat x, d) \}   \leq   \mu,  j   \in   \{1, 2\} }{ \label{eq:SCP_SSF_ST} }
		\addConstraint{\forall x^{z_i}   \in   X, \forall i \in \{1, \ldots, N\},  \forall \hat x   \in   \hat X}{ \notag}
		\addConstraint{d = [\alpha_s;\psi_s;\kappa_{s};\mathrm{q}_1; \mathrm{q}_2;\ldots;\mathrm{q}_{\mathrm{r}}] \in \R^{\mathrm{r}+3}}{ \notag}
		\addConstraint{\alpha_s \in \Rp, \; \mu \in \R, \; \psi_s \in \Rpz, \; \kappa_{s} \in (0,1],}{ \notag}
	\end{mini!}
	with functions $h_1(x^{z_i}, \hat x, d)$ and $h_2(x^{z_i}, \hat x, d)$ as in~\eqref{eq:hSSF} ($x$ is replaced by $x^{z_i}$). Given that the constraint associated with $h_2(x, \hat{x}, d)$ is nonconvex, one can fix $\kappa_s$, after which the resulting optimization problems can be solved.
	
	While the SOP~\eqref{eq:SCP_SSF} addresses the second challenge, it still faces an additional difficulty, as a closed-form expression for the expected value in $h_2(x, \hat{x}, d)$ with respect to $\varsigma$ is still required. To address this, one can take $\hat{N}$ i.i.d. samples from $\varsigma$, denoted by $\varsigma_{l}$, $l \in \{1, \ldots, \hat N\}$, for each $x^{z_i}$. Then, according to Chebyshev's inequality~\citep{hernandez2001chebyshev}, one has
	\begin{align}
		&\mathds{P} \Big\{    \big \vert   \mathds{E} \big[ \mathds{S}(f(x^{z_i}, \varsigma),  \hat f(\hat x, \varsigma), \mathrm{q}) \big] \notag  \\  & ~~~~ -  \frac{1}{\hat{N}}   \sum_{l=1}^{\hat{N}}   \mathds{S}(f(x^{z_i}, \varsigma_l),  \hat f(\hat x, \varsigma_l), \mathrm{q}) \big \vert   \leq   \bar{\epsilon}   \Big\}    \geq   1  -  \bar{\varepsilon}_2,\label{eq:ch_SSF}
	\end{align}
	where $\bar{\epsilon} \in \Rp$ and $\bar{\varepsilon}_2 \in (0, 1]$ denote the approximation error and the confidence level corresponding to the empirical approximation, respectively.
	
	Consequently, the last challenge can be addressed by substituting the expected value in $h_2(x^{z_i}, \hat x, d)$ with its empirical approximation, as in~\eqref{eq:ch_SSF}. The SOP~\eqref{eq:SCP_SSF} can be now updated by replacing $h_2(x^{z_i}, \hat x, d)$ with $\bar h_2(x^{z_i}, \hat x, d)$, which is defined as
	\begin{align}
		\bar h_2(x^{z_i}, \hat x, d) & \coloneq   \frac{1}{\hat{N}}   \sum_{l=1}^{\hat{N}}   \mathds{S}(f(x^{z_i}, \varsigma_l),  \hat f(\hat x, \varsigma_l), \mathrm{q}) \notag\\ & ~~~ - \kappa_s \mathds{S}(x^{z_i},\hat x, \mathrm{q}) - \psi_s + \bar{\epsilon}. \label{eq:barh_SSF}
	\end{align}
	It is worth noting that, since~\eqref{eq:barh_SSF} incorporates the term $\bar{\epsilon} \in \Rp$, it introduces additional conservatism compared with $h_2(x^{z_i}, \hat x, d)$, which is required to account for the error induced by the empirical approximation. Due to the introduction of a confidence parameter arising from the use of an empirical approximation of the expectation term, the guarantee in~\eqref{eq:ssf_pb} or~\eqref{eq:ssf_pb_inf} (without control inputs in the described approach) in the data-driven setting holds with a confidence level of at least $1 - \varepsilon_2$, \emph{i.e.},
	\begin{align}
		\mathds{P} \Big\{\eqref{eq:ssf_pb} \text{ or \eqref{eq:ssf_pb_inf} holds}\Big\} \geq 1-\varepsilon_2. \label{eq:GMDP}
	\end{align}
	Importantly, $\varepsilon_2$ should account for $\bar{\varepsilon}_2$, which originates from the empirical approximation, but it may also include additional terms depending on the data-driven approach or the sampling strategy employed, \emph{i.e.}, whether the $N$ sample pairs are obtained via an i.i.d. sampling approach (cf. Fig.~\ref{subfig:scenario_data}) or a grid-based sampling approach (cf. Fig.~\ref{subfig:Lip-based_data}).
	
	Equipped with an understanding of the challenges and the potential solution frameworks for developing data-driven abstraction-based methodologies for stochastic dynamical systems, we now proceed to provide a comprehensive review of the corresponding literature.
	
	\subsection{Literature on Data-Driven Abstraction-based Approaches for Stochastic Systems}
	In recent years, a variety of indirect and direct data-driven techniques have been proposed for constructing finite abstractions of stochastic dynamical systems, addressing both verification and synthesis problems. Within the category of indirect data-driven approaches, \citet{jackson2021strategyHSCC} propose a framework for control synthesis of discrete-time, partially unknown, switched stochastic systems. In the proposed methodology, the unknown dynamics are first learned via GP regression, after which an IMDP abstraction is constructed and used for controller synthesis to maximize the probability of satisfying a given LTL specification over finite traces. Subsequently, \citet{10349682} and \citet{11108000} propose replacing standard GP regression with deep kernel learning for modeling the system dynamics, enhancing scalability and representational capacity for complex nonlinear stochastic systems, while similarly constructing an IMDP abstraction. In addition, \citet{schon2024dataADHS} propose a framework for unknown discrete-time stochastic systems without control inputs that, rather than relying on standard GPs, exploits the so-called binary-tree GP, whose piecewise-constant posterior mean and covariance naturally induce a partition of the state space. This feature facilitates the construction of IMCs for the verification of infinite-horizon reachability specifications while providing probabilistic guarantees. More recently, \citet{10678810} propose an indirect framework based on parametric identification and robust simulation relations, where Bayesian regression is employed to compute a credible parameter set with a prescribed confidence level.
	
	In the context of direct data-driven approaches, \citet{9095986} propose a reinforcement-learning-based framework for the control synthesis of unknown discrete-time stochastic control systems, aiming to maximize the probability of satisfying a given specification, expressed as a syntactically co-safe LTL formula, over a bounded time horizon. The proposed method abstracts the unknown system into a finite MDP with unknown transition probabilities, synthesizes a control strategy over this finite MDP, and subsequently refines it for the original system with approximate optimality guarantees. In addition,~\citet{9815301} propose a data-driven method, grounded in an SCP, for constructing finite MDPs of unknown discrete-time stochastic systems, providing probabilistic closeness guarantees in the form of~\eqref{eq:GMDP} for finite-horizon specifications. The proposed methodology relies on i.i.d. sampling and requires certain Lipschitz continuity conditions to ensure the validity of the guarantees (cf. Section~\ref{Subsec: Lipschitz Approach}).
	
	While the previous two studies are primarily tailored to finite-horizon specifications, the frameworks proposed by~\citet{pmlr-v283-nazeri25a} and~\citet{11312907} can be employed to enforce both finite- and infinite-horizon specifications. More precisely, \citet{pmlr-v283-nazeri25a} introduce a data-driven abstraction-based technique for unknown discrete-time Lipschitz continuous dynamical systems subject to additive stochastic noise. The approach leverages samples of the system dynamics to learn the enabled actions and transition probabilities of the abstraction, represented as an IMDP, while providing PAC guarantees (cf. Section~\ref{Subsec: Scenario Approach}) on the probability of satisfying a specified control objective. In a similar vein, the work by~\citet{11312907} extends the previous framework by relaxing certain assumptions, as it does not require noise-free samples of the system dynamics and considers a broader class of stochastic systems in which the noise is not restricted to be additive. In a recent effort, \citet{sauglam2026incrementalAAAI} propose a direct data-driven abstraction-based approach for unknown discrete-time stochastic systems that combines online learning with incremental game-solving. Specifically, \citet{sauglam2026incrementalAAAI} learn under- and over-approximations of reachable sets from noisy data, under Lipschitz continuity and known bounded noise support, and use these to construct finite stochastic game graphs as abstractions for synthesizing policies that satisfy infinite-horizon temporal objectives almost surely.
	
	A common requirement across several direct data-driven studies reviewed here is the availability of Lipschitz constants, particularly for the stochastic system~\citep{pmlr-v283-nazeri25a,11312907,sauglam2026incrementalAAAI}. In this direction, \citet{pmlr-v238-zhang24i} develop theoretical results based on nonparametric estimation to compute asymptotic upper bounds for the Lipschitz constants of discrete-time stochastic systems. Building on these bounds, \citet{pmlr-v238-zhang24i} construct IMDP abstractions and employ them to verify temporal-logic specifications, while also allowing synthesis in the presence of inputs.
	
    Despite the rich literature reviewed here, all these studies focus on constructing finite abstractions of discrete-time stochastic systems; a natural future direction is therefore to investigate the data-driven construction of infinite abstractions, \emph{i.e.}, ROMs, for stochastic dynamical systems.
	
	\begin{resp}
		\begin{openproblem}\label{OP:SROM}
			Develop a data-driven approach for constructing infinite abstractions (ROMs) of stochastic systems, together with their S-SFs, while providing out-of-sample performance guarantees.
		\end{openproblem}
	\end{resp}
	
   We note that while the work by~\citet{pmlr-v288-nadali25a} can potentially be viewed as a step toward Research Avenue~\ref{OP:SROM}, it does not directly address it. Specifically, this work proposes a data-driven framework for transferring control between stochastic systems by learning a stochastic neural simulation function and an interface map, assuming a given lower-dimensional source system rather than constructing one, while providing infinite-horizon guarantees (cf.~\eqref{eq:GMDP}) under grid-based sampling and Lipschitz continuity conditions.
	
	Having discussed various aspects of data-driven abstraction-based approaches for stochastic dynamical systems, we now proceed to the next section, where we review data-driven functional certificate approaches for stochastic systems.
	
	\section{Stochastic Setting: Data-Driven Functional Certificate Approaches}\label{Sec: Stochastic Setting_FCA}
	This section presents functional certificate-based approaches for stochastic systems, with a primary focus on stochastic control barrier certificates (S-CBCs). When the objective is verification rather than controller synthesis, the corresponding notion is stochastic barrier certificates (S-BCs). The use of S-CBCs yields a quantitative guarantee expressed as a lower bound on the probability that the system trajectory remains within the safe region. In fact, safety guarantees derived from S-CBCs and S-BCs are intrinsically probabilistic, stemming from the noise present in the system dynamics.
	
	To enable the discussions in this section, let us again consider the system described in Definition~\ref{def:sys_stoch_barrier}.
	We now proceed to present the notion of S-CBCs. To this end, with a slight abuse of notation, we adopt the same symbols as those used for CBCs in Definition~\ref{def:cbc} to denote their stochastic counterparts.
	Nevertheless, the main differences between the definition of an S-CBC and that of a CBC are twofold: \emph{(i)} in contrast to a CBC, an S-CBC is required to be a nonnegative function (essential for~\eqref{eq:non_neg}), \emph{i.e.}, $\mathds{B} : X \to \Rpz$, which in turn implies that $\eta, \gamma \in \Rpz$ with $\gamma > \eta$, and \emph{(ii)} condition~\eqref{eq:cbc3} should be reformulated to account for the stochastic dynamics, whereas conditions~\eqref{eq:cbc1} and~\eqref{eq:cbc2} remain unchanged.
	In particular, for the stochastic system $\mathcal{S}_{sg}$ introduced in Definition~\ref{def:sys_stoch_barrier}, condition~\eqref{eq:cbc3} is modified as
	\begin{itemize}
		\item $\forall x \in X,$ $\exists u \in U$, such that
		\begin{align} 
			\mathds{E}\big[ \mathds{B}(f(x, u, \varsigma)) \mid x, \, u \big] \leq \kappa_{\varsigma} \, \mathds{B}(x) + c,\label{eq:scbc3}
		\end{align}
		for some constants $c \in \Rpz$ and $0 < \kappa_{\varsigma} \leq 1$.
	\end{itemize}
	
	With these modifications in place, one can derive a lower bound on the probability that $\mathcal{S}_{sg}$ remains within the safe region over a finite time horizon~\citep{lavaei2022automated}.
	In particular, for any initial condition $x_0 \in X_0$, the probability that the solution process of $\mathcal{S}_{sg}$, under the input signal $u$ associated with the S-CBC $\mathds{B}$, does not reach the unsafe set $X_u$ over the time interval $[0,\, T]$ is lower bounded by $1 - \vartheta$, \emph{i.e.},
	\begin{align}
		\mathds{P} \bigl\{ x_{x_0 u}(k) \notin X_u \; \text{for all} \; k \in [0,\, T] \mid x_0 \in X_0 \bigr\} \geq 1 - \vartheta, \label{eq:scbc_pb}
	\end{align}
	where, if $0 < \kappa_{\varsigma} < 1$:
	\begin{subequations}\label{eq:non_neg}
		\begin{align}
			\vartheta \coloneq \begin{cases}1-(1-\frac{\eta}{\gamma})(1-\frac{c}{\gamma})^{T} \! , & \text { if } \gamma \geq \frac{c}{1-\kappa_{\varsigma}}, \\ (\frac{\eta}{\gamma}) \kappa_{\varsigma}^{T}+(\frac{c}{(1-\kappa_{\varsigma}) \gamma})(1-\kappa_{\varsigma}^{T}), & \text { if } \gamma<\frac{c}{1-\kappa_{\varsigma}},\end{cases} \label{eq:scbc_b1}
		\end{align}
		whereas, if $0 < \kappa_{\varsigma} \leq 1$:
		\begin{align}
			\vartheta \coloneq \frac{\eta + cT}{\gamma}. \label{eq:scbc_b2}
		\end{align}
	\end{subequations}
    Akin to the discussion in Section~\ref{Sec: Stochastic Setting_ABA}, the bound in~\eqref{eq:scbc_b1} is less conservative than that in~\eqref{eq:scbc_b2}, as it provides a tighter probabilistic guarantee; however, the latter is more broadly applicable, since for some systems and dynamics there may not exist a constant $\kappa_{\varsigma} < 1$ satisfying condition~\eqref{eq:scbc3}, which is required in~\eqref{eq:scbc_b1}.
	
	The underlying safety guarantees can be extended to infinite-time horizons if $c = 0$, where $\vartheta \coloneq \frac{\eta}{\gamma}$. More concretely, in this case, the guarantee~\eqref{eq:scbc_pb} is modified to
	\begin{align}
		\mathds{P} \bigl\{ x_{x_0 u}(k) \notin X_u \; \text{for all} \; k \in [0,\, \infty) \mid x_0 \in X_0 \bigr\} \geq 1 - \frac{\eta}{\gamma}. \label{eq:scbc_g2}
	\end{align}
	We note that enforcing condition~\eqref{eq:scbc3} with $c = 0$ is typically restrictive and, in certain scenarios, may preclude the existence of a valid S-CBC~\citep{salamati2024dataAUTOMATICA}. Allowing $c > 0$ relaxes this condition and increases the likelihood of constructing a valid S-CBC as a $c$-martingale certificate, albeit at the cost of restricting the safety guarantee to a finite-time horizon.
	
	To leverage the guarantee in~\eqref{eq:scbc_pb} or~\eqref{eq:scbc_g2}, it is necessary to construct an S-CBC for the system $\mathcal{S}_{sg}$. However, condition~\eqref{eq:scbc3} explicitly depends on the unknown system dynamics. 
	In addition, the stochastic setting introduces a further layer of difficulty due to the presence of the expectation operator in~\eqref{eq:scbc3}.
	As detailed in Section~\ref{Sec: Stochastic Setting_ABA}, addressing this challenge typically requires estimating the expectation term via empirical approximations, while employing concentration inequalities such as Chebyshev's inequality~\citep{hernandez2001chebyshev} to formalize this approximation at the cost of introducing a confidence level under which it holds.
	
	Hereafter, following the approach outlined in Section~\ref{Sec: Stochastic Setting_ABA}, we describe a systematic data-driven procedure to address these challenges, while emphasizing that alternative data-driven approaches may also be employed. For simplicity of exposition, and in line with Section~\ref{Sec: Stochastic Setting_ABA}, we focus on the synthesis of an S-BC. Inspecting the RCP~\eqref{eq:RCP_BC}, and fixing the structure of an S-BC in a manner analogous to Section~\ref{Subsec:DD-CBC}, it follows that the ROP corresponding to the synthesis of an S-BC can be formulated as
	\begin{mini!}|s|[2]<b>
		{[\mu; d]}{\mu}
		{\label{eq:RCP_SBC}}{}
		\addConstraint{ \max_j \{ h_j(x, d) \} \leq \mu, \; j \in \{0,\ldots, 4\},  \; \forall x \in X }{ \label{eq:RCP_SBC_ST} }
		\addConstraint{d = [\eta;\gamma;c;\kappa_{\varsigma};\varkappa;\mathrm{q}_1; \mathrm{q}_2;\ldots;\mathrm{q}_{\mathrm{r}}] \in \R^{\mathrm{r}+5}}{ \notag}
		\addConstraint{\varkappa \in \R \backslash \Rpz, \; \mu \in \R, \; \eta,\gamma, c \in \Rpz, \; \kappa_{\varsigma} \in (0,1],}{ \notag}
	\end{mini!}
	where $h_1(x,d)$--$h_3(x,d)$ are defined analogously to those in~\eqref{eq:h_CBC}, while $h_0(x, d)$ enforces the nonnegativity of the S-BC and $h_4(x, d)$ corresponds to condition~\eqref{eq:scbc3}, \emph{i.e.},
	\begin{align}
		\begin{split}\label{eq:h_SBC}
			h_0(x, d) & = - \mathds{B}(x, \mathrm{q}),\\
			h_4(x, d) & = \mathds{E}\big[ \mathds{B}(f(x, \varsigma), \mathrm{q}) \mid x \big] - \kappa_{\varsigma} \, \mathds{B}(x, \mathrm{q}) - c.
		\end{split}
	\end{align}
	By collecting sample pairs of the form $(x^{z_i}, f(x^{z_i}, \varsigma))$, where each $x^{z_i}$, for all $i \in \{1, \ldots, N\}$, one can formulate the corresponding SOP as
	\begin{mini!}|s|[2]<b>
		{[\mu; d]}{\mu}
		{\label{eq:SCP_SBC}}{}
		\addConstraint{ \max_j \{ h_j(x, d), h_4(x^{z_i}, d) \} \leq \mu, \; j \in \{0,1,2,3\} }{ \label{eq:SCP_SBC_ST} }
		\addConstraint{\forall x \in X, \; \forall x^{z_i}   \in   X, \; \forall i \in \{1, \ldots, N\}}{ \notag}
		\addConstraint{d = [\eta;\gamma;c;\kappa_{\varsigma};\varkappa;\mathrm{q}_1; \mathrm{q}_2;\ldots;\mathrm{q}_{\mathrm{r}}] \in \R^{\mathrm{r}+5}}{ \notag}
		\addConstraint{\varkappa \in \R \backslash \Rpz, \; \mu \in \R, \; \eta,\gamma, c \in \Rpz, \; \kappa_{\varsigma} \in (0,1].}{ \notag}
	\end{mini!}
	To resolve the second challenge regarding the expected value in SOP~\eqref{eq:SCP_SBC},  for each $x^{z_i}$, we draw $\hat{N}$ i.i.d. samples of $\varsigma$, denoted by $\varsigma_{l}$, $l \in \{1, \ldots, \hat{N}\}$.
	Consequently, by invoking Chebyshev's inequality, one obtains\vspace{-0.2cm}
	\[
	\mathds{P} \Big\{ \!\! \big \vert   \mathds{E}\big[ \mathds{B}(f(x^{z_i}, \varsigma), \mathrm{q}) \big] \! -  \frac{1}{\hat{N}} \! \sum_{l=1}^{\hat{N}} \! \mathds{B}(f(x^{z_i}, \varsigma_{l}), \mathrm{q}) \big \vert \! \leq \! \bar{\epsilon} \! \Big\} \!\! \geq \! 1  -  \bar{\varepsilon}_2, \vspace{-0.15cm}
	\]
	where $\bar{\epsilon} \in \Rp$ denotes the approximation error and $\bar{\varepsilon}_2 \in (0, 1]$ represents the confidence level associated with the empirical estimate. Accordingly, one can formulate an alternative SOP, analogous to~\eqref{eq:SCP_SBC}, by replacing $h_4(x^{z_i}, d)$ in~\eqref{eq:SCP_SBC_ST} with its empirical counterpart $\bar{h}_4(x^{z_i}, d)$, defined as
	\begin{align}
		\bar{h}_4(x^{z_i}, d) \! \coloneq \! \frac{1}{\hat{N}} \sum_{l=1}^{\hat{N}} \! \mathds{B}(f(x^{z_i}, \varsigma_{l}), \mathrm{q})   - \kappa_{\varsigma} \, \mathds{B}(x^{z_i}, \mathrm{q}) - c + \bar{\epsilon}.\label{eq:barh_SBC}
	\end{align}
	Observe that, due to the presence of the term $\bar{\epsilon} \in \Rp$ in~\eqref{eq:barh_SBC}, $\bar{h}_4(x^{z_i}, d)$ is more conservative than $h_4(x^{z_i}, d)$, so as to account for the approximation error introduced by replacing the expectation with its empirical estimate.
	
	Having addressed both challenges, it follows that, upon solving the SOP incorporating $\bar{h}_4(x^{z_i}, d)$ and constructing an S-BC, the resulting safety guarantee takes the form
	\begin{align}
		\mathds{P} \Big\{\mathds{P}\big\{\mathcal{S}_{sg} \vDash \text{safety}\big\} \geq 1-\vartheta\Big\} \geq 1-\varepsilon_2, \label{eq:2lP_SBC}
	\end{align}
	where $\varepsilon_2 \in (0, 1]$ denotes the overall confidence level. As detailed in Section~\ref{Sec: Stochastic Setting_ABA}, this quantity necessarily includes the term $\bar{\varepsilon}_2$ arising from the empirical approximation of the expectation, and may incorporate additional contributions depending on the specific data-driven methodology and sampling scheme employed. We note that the inner probability in~\eqref{eq:2lP_SBC} reflects the stochastic nature of the system, which is present even in model-based settings, whereas the outer probability captures the confidence level induced by the use of data. While the details and descriptions presented here focus on S-BCs and S-CBCs, the same arguments extend naturally to other functional certificate frameworks.
	
	Another approach to constructing S-CBCs is to employ data-driven frameworks that exploit structural properties of systems to derive data-parameterized representations, as presented in Section~\ref{Subsec: Structural Approach}, with suitable modifications. More specifically, for $\mathcal{S}_{sg}$ in Definition~\ref{def:sys_stoch_barrier}, the system can be initialized at a given initial condition and driven by a sequence of arbitrary inputs over the time horizon $[0, \mathcal{T}]$, where $\mathcal{T} \in \Np$ denotes the experiment horizon, yielding
	\begin{subequations}\label{eq:data_single_trajectory_stoch}
		\begin{align}
			\mathcal{O}^{i} & \coloneq \begin{bmatrix}
				x(0) & ~~ x^{i}(1) & ~~ \ldots & ~~ x^{i}(\mathcal T - 1)
			\end{bmatrix} \!\! ,\\
			\mathcal{I} & \coloneq \begin{bmatrix}
				u(0) & ~~ u(1) & ~~ \ldots & ~~ u(\mathcal T - 1)
			\end{bmatrix} \!\! ,\\
			\mathcal{O}^{+^i} & \coloneq \begin{bmatrix}
				x^{i}(1) & ~~ x^{i}(2) & ~~ \ldots & ~~ x^{i}(\mathcal T)
			\end{bmatrix} \!\! ,\\
			\mathbb{Z}^i & \coloneq \begin{bmatrix}
				\varsigma^{i}(0) & ~~\varsigma^{i}(1) & ~~ \ldots & ~~ \varsigma^{i}(\mathcal T - 1)
			\end{bmatrix} \!\! ,
		\end{align}
	\end{subequations}
	where $i \in \{1, 2, \ldots, \bar{N}\}$, with $\bar{N} \in \Np$, indexes trajectories corresponding to different noise realizations $\mathbb{Z}^i$. We note that all trajectories in this setting originate from the same initial condition $x(0)$ and evolve under the same input sequence $\mathcal{I}$~\citep{lavaei2025dataSOLETAC}, which ensures their comparability and enables consistent capture of the noise effect via its empirical mean. We emphasize that $\mathbb{Z}^i$ is unknown and not measured. This modification enables the extension of the approach discussed in Section~\ref{Subsec: Structural Approach} to stochastic settings.
	
	It is worth noting that, in the \emph{continuous-time} setting, the expectation operator in~\eqref{eq:scbc3} is replaced by an infinitesimal generator of the underlying stochastic system. This generator is a partial differential operator that encodes rich information about the evolution of the stochastic system. However, in data-driven settings where the system dynamics are unknown, the infinitesimal generator should be estimated from data. Similar to the discrete-time case, this estimation is accompanied by a confidence level, thereby introducing an additional probabilistic layer in the resulting safety guarantees. We refer the interested reader to the work by~\citet{9805799} for a data-driven estimation of the infinitesimal generator from data. Owing to the need to estimate the infinitesimal generator from noisy measurements while accounting for the associated estimation error, the data-driven analysis and synthesis of \emph{continuous-time} stochastic systems is considerably more challenging, motivating the following research avenue.
	
		\begin{resp}
		\begin{openproblem}
			For a stochastic dynamical system evolving in continuous time and subject to both Brownian motion and Poisson processes, develop a data-driven framework exploiting system structural properties and a single trajectory of the system for constructing either (in)finite abstractions or functional certificates while formally estimating the corresponding infinitesimal generator from noisy measurements and providing probabilistic guarantees.
		\end{openproblem}
	\end{resp}
	
	\subsection{Literature on Data-Driven Design of Functional Certificates for Stochastic Systems}
	In recent years, there has been growing interest in data-driven functional certificate methodologies for stochastic dynamical systems. In this context, an indirect data-driven approach for synthesizing stochastic control barrier functions for continuous-time stochastic dynamical systems with unknown diffusion terms is proposed by~\citet{9992657}. The approach relies on i.i.d. sampling and yields infinite-horizon probabilistic safety guarantees. Indirect data-driven approaches for learning system dynamics via GP regression have also been considered in the literature for the synthesis of S-BFs~\citep{reed2025errorAAAI,10886850} and S-CBCs~\citep{pmlr-v168-wajid22a}. These works rely on i.i.d. sampling and provide probabilistic safety guarantees over finite-time horizons. Moreover, \citet{pmlr-v202-wang23as} propose an indirect data-driven safe reinforcement learning framework for unknown continuous stochastic systems, where a neural generative model of the environment is learned from trajectory data and subsequently used to synthesize S-BFs that enforce safety constraints over finite horizons.
	
	Several studies fall within the first two categories of direct data-driven approaches discussed in Section~\ref{Sec: NaP}. In this regard, \citet{SalamatiIFAC} propose a data-driven approach, based on the notion of S-BCs, for the formal safety verification of unknown discrete-time stochastic dynamical systems. The method relies on i.i.d. sampling and imposes certain Lipschitz continuity assumptions (cf. Section~\ref{Subsec: Lipschitz Approach}) in order to provide probabilistic finite-horizon safety guarantees.
	Subsequently, \citet{salamati2024dataAUTOMATICA} extend the results of the aforementioned study and propose a data-driven approach that addresses both verification and synthesis problems. We note that the guarantees provided in both studies are expressed in the form of~\eqref{eq:2lP_SBC}. In addition, \citet{salamati2022safetyLCSS}, building on repetitive scenario design~\citep{7484339}, and \citet{pmlr-v168-salamati22a}, based on the wait-and-judge approach \citep{campi2018waitMP}, propose data-driven methods that reduce sample complexity while providing probabilistic finite-horizon safety guarantees in the form of~\eqref{eq:2lP_SBC}.
	A common limitation of the preceding four studies is the need to fix the certificate template a priori. To overcome this, \citet{11312074} propose a data-driven approach for safety controller synthesis in discrete-time stochastic systems with unknown dynamics, where both the S-CBC and the controller are represented by neural networks learned from data. By incorporating Lipschitz-based validity conditions into training, the approach ensures generalization beyond the samples and provides probabilistic finite-horizon safety guarantees with a desired confidence level.
	
	As another relevant work, \citet{10644349} use conditional mean embeddings~\citep{klebanov2020rigorous} to verify safety of discrete-time stochastic systems via S-BCs by reformulating the conditions as a data-driven optimization problem and deriving distributionally robust guarantees. The resulting SOS-based approach, combined with a GP envelope and i.i.d. sampling, provides probabilistic finite-horizon safety guarantees (similar to~\eqref{eq:2lP_SBC}).
	Within this line of work, \citet{casablanca2026lucid} propose \textsc{Lucid}, a tool for certifying the safety of discrete-time unknown stochastic systems from finite data, which uses a finite Fourier kernel expansion to reformulate the problem as a tractable linear program, providing distributionally robust probabilistic finite-horizon safety guarantees under i.i.d. sampling. More recently, \citet{schon2026kernel} extend this line of work from safety verification to control synthesis using S-CBCs, addressing temporal logic specifications via Streett supermartingales~\citep{abate2024stochasticCAV} and providing probabilistic finite-horizon guarantees.
	
	    As is evident from the earlier discussion, existing data-driven approaches for stochastic systems are limited and typically consider only process noise with i.i.d. properties. To enhance the practical applicability of these methods, we propose the following research avenue.
	
	\begin{resp}
		\begin{openproblem}
			For a stochastic dynamical system subject to both process and measurement noise, where the noise processes may be non-i.i.d., develop a data-driven framework for constructing either (in)finite abstractions or functional certificates while providing probabilistic out-of-sample performance guarantees.
		\end{openproblem}
	\end{resp}
	
	Within the third category of data-driven approaches, \emph{i.e.}, those exploiting structural properties to obtain data-parameterized representations (cf. Section~\ref{Subsec: Structural Approach}), \citet{10101826} design probabilistic safe controllers for uncertain linear discrete-time systems directly from data, ensuring safe-set invariance in probability. Extending this perspective beyond linear dynamics, \citet{10771977} propose a direct data-driven risk-averse safe control framework for stochastic nonlinear parameter-varying systems, where gain-scheduling and nonlinear control components are learned from finite data to enforce probabilistic set invariance with robustness to stochastic disturbances. The recent study by~\citet{lavaei2025dataSOLETAC} introduces a data-driven framework for designing safety controllers via the notion of S-CBCs for discrete-time stochastic dynamical systems with polynomial dynamics, providing probabilistic finite-horizon safety guarantees directly from data collected along $N$ different system trajectories (cf. data in~\eqref{eq:data_single_trajectory_stoch}).  In contrast to the previous two studies, the framework by~\citet{lavaei2025dataSOLETAC} does not require rendering the entire safe set invariant in probability, but instead allows the initial set to be chosen as a subset of the safe set, which can lead to reduced conservativeness.
	
	\section{Stochastic Setting: Data-Driven Compositional Techniques}\label{Sec: Stochastic Setting_CT}
	In this section, we provide an overview of data-driven compositional techniques for high-dimensional interconnected networks of stochastic dynamical subsystems, covering both data-driven abstraction-based and functional-certificate-based approaches. The underlying principle of such compositional techniques is largely analogous to that discussed in Section~\ref{Sec: Deterministic Setting_CT}. However, a key distinction arises due to the stochasticity present at the subsystem level: the corresponding analyses are probabilistic, which in turn induces probabilistic guarantees at the network level.
	
	In the realm of data-driven compositional techniques for abstraction-based approaches, \citet{10304199} propose a reinforcement-learning-based framework that implicitly abstracts each discrete-time stochastic subsystem into a finite MDP with unknown transition probabilities. In a subsequent stage, following an assume--guarantee paradigm, control strategies are synthesized for finite MDPs using reinforcement learning and then refined to the original continuous-space network with approximate optimality guarantees. Accordingly, the final guarantee is expressed as a lower bound on the satisfaction probability of the overall network, obtained by combining the probabilities of individual subsystems together with additional factors, including certain Lipschitz constants.
	
	Moreover, \citet{10384089} presents a data-driven compositional approach to constructing finite MDPs of unknown large-scale interconnected networks of discrete-time stochastic subsystems. The proposed framework exploits dissipativity properties of both the subsystems and their corresponding finite MDPs by leveraging the notion of stochastic storage functions. The approach relies on i.i.d. sampling (cf. Fig.~\ref{subfig:scenario_data}), together with certain Lipschitz continuity conditions (cf. Section~\ref{Subsec: Lipschitz Approach}), and the resulting guarantees take the form of~\eqref{eq:GMDP} for finite-horizon specifications.
	Importantly, while the types of guarantees provided by~\citet{10304199} and~\citet{10384089} differ, the latter can, in general, yield less conservative guarantees for networks with a large number of subsystems, primarily due to the effectiveness of dissipativity-based compositional reasoning.
	
	Within the context of data-driven compositional techniques for functional certificate approaches, \citet{lavaei2023compositionalTACStochandDeterministic} propose a data-driven framework for verifying the safety of large-scale networks using dissipativity-based compositional reasoning. More precisely, \citet{lavaei2023compositionalTACStochandDeterministic} collect i.i.d. data from each unknown continuous-time stochastic subsystem and construct stochastic storage certificates for each subsystem. These certificates are then composed to obtain S-BCs for the overall interconnected network, which can be used to verify network safety while providing finite-time-horizon safety guarantees in the form of~\eqref{eq:2lP_SBC}, whose out-of-sample correctness (with a predetermined confidence level) is ensured by leveraging certain Lipschitz continuity conditions (cf. Section~\ref{Subsec: Lipschitz Approach}).
	
	In a similar vein, \citet{lavaei2022formalCNCStochandDeterministic} propose a data-driven methodology for the formal estimation of collision risks in autonomous vehicles (AVs) with discrete-time unknown stochastic dynamics, within a multi-agent framework. The approach involves collecting data from each AV using an i.i.d. sampling scheme, based on which, and by leveraging certain Lipschitz continuity conditions to ensure out-of-sample correctness (with a specified confidence level), a stochastic sub-barrier certificate is constructed for each agent. These certificates are then composed, using small-gain-based compositional reasoning, to synthesize the S-BC for the network of AVs, thereby enabling a formal estimation of the collision risk over a finite time horizon in the form of~\eqref{eq:2lP_SBC}.
	
	\section{Concluding Discussion}\label{Sec: Conclusion}
	In this survey, we have provided, to the best of our knowledge, the first unified and in-depth overview of data-driven formal methods for complex dynamical systems, with particular emphasis on methodologies that provide rigorous formal guarantees for verification and controller synthesis against complex specifications beyond stability. We have organized the rapidly growing literature around three main methodological pillars, namely abstraction-based techniques, functional certificate approaches, and compositional methods, and have discussed them for both deterministic and stochastic systems. In parallel, we have classified the main data-driven guarantees into three broad categories, namely \emph{(i)} PAC/scenario-based guarantees, \emph{(ii)} guarantees relying on Lipschitz continuity properties, and \emph{(iii)} guarantees exploiting structural properties of the underlying systems, such as data-parameterized representations or monotonicity. Through this organization, we have aimed to clarify the common principles, advantages, and limitations of a literature that is broad, technically diverse, and often fragmented across different communities, while also highlighting several avenues for future research.
	
	The survey also highlights a recurring trade-off underlying nearly all existing approaches. On the one hand, methods providing PAC guarantees are attractive due to their generality and flexibility. Nevertheless, they require i.i.d. sampling with multiple system initializations, and their guarantees allow violations over a small subset of the state space. On the other hand, approaches based on Lipschitz continuity conditions can provide stronger guarantees, in particular deterministic ones in non-stochastic settings, yet they often suffer from \emph{exponential} sample-complexity growth. Methods exploiting structural properties can alleviate some of the limitations of the previous two categories, for instance, by reducing data-collection burdens, and can, in certain cases, provide deterministic guarantees from remarkably limited data; however, this comes at the cost of restricting the system class and often requiring stronger prior knowledge. A similar tension appears between abstraction-based and functional certificate methods. Abstraction-based approaches are highly expressive and naturally compatible with algorithmic synthesis for rich logical specifications, whereas functional certificate methods avoid state-space discretization and can be more scalable, but are often conservative for rich temporal properties. Moreover, while compositional techniques are indispensable for large-scale interconnected systems, their current data-driven developments remain limited in scope and still leave major generality challenges unresolved.
	
	Overall, the present survey indicates that data-driven formal methods have matured into a promising and rapidly advancing research area, while a few fundamental problems remain open. We hope that this survey helps make the area more accessible to researchers new to the area, clarifies the current state of the art, and contributes to shaping a coherent research agenda toward reliable, scalable, and practically deployable data-driven formal methods.
	
	\bibliographystyle{agsm}
	\bibliography{biblio}

\end{document}